\documentclass[12pt]{article}
\usepackage{palatino,achicago,amssymb,amsfonts,amsmath,amsthm,latexsym,setspace}
\usepackage{graphicx,xcolor,booktabs,microtype,array,rotating}
\usepackage[T1]{fontenc}
\usepackage[detect-all]{siunitx}
\usepackage{epstopdf}
\usepackage{subcaption, float, ragged2e, lscape, longtable}
\usepackage{threeparttablex} 
\usepackage{makecell}
\usepackage{booktabs, tabularx}        
\newcolumntype{C}[1]{>{\centering\arraybackslash}p{#1}}
\newcolumntype{L}[1]{>{\centering\arraybackslash}p{#1}}
\usepackage{hyperref}
\usepackage{cleveref}
\newcommand{\cmmnt}[1]{}
\definecolor{DoepkeBlue}{rgb}{.2,0,.8}
\hypersetup{
    breaklinks=true,   
    pdfusetitle=true,  
    colorlinks=true,
    linkcolor=DoepkeBlue,
    filecolor=DoepkeBlue,
    urlcolor=DoepkeBlue,
    citecolor=DoepkeBlue,
}

\allowdisplaybreaks

\newtheorem{prp}{Proposition}

\newenvironment{prf}{{\bf Proof:} }{\hfill $\Box$}
\newenvironment{prfof}{{\bf Proof of} }{\hfill $\Box$}

\makeatletter

\renewcommand\section{\@startsection{section}{1}{0cm}{-1.5ex \@plus
-.2ex \@minus -.2ex}%
{.5ex \@plus.2ex} {\normalfont\large\bfseries}}

\renewcommand\subsection{\@startsection{subsection}{1}{0cm}{-1.5ex \@plus
-.2ex \@minus -.2ex}%
{.5ex \@plus.2ex} {\normalfont\bfseries}}

\renewcommand\subsubsection{\@startsection{subsubsection}{1}{0cm}{-1.5ex \@plus
-.2ex \@minus -.2ex}%
{.5ex \@plus.2ex} {\normalfont}}

\makeatother

\begin{document}

\begin{titlepage}
\begin{singlespacing}
\title{Educational Inequality\footnote{This chapter has been prepared for the Handbook of the Economics of Education Volume 6. We thank Kwok Yan Chiu, Ricard Grebol, and Ashton Welch for excellent research assistance and the editors, John Jerrim, Sandra McNally, Anders Björklund, Jonas Radl, Martin Hällsten, and Christopher Rauh for helpful comments. Financial support from the National Science Foundation (grant SES-1949228), the Comunidad de Madrid and the MICIU (CAM-EPUC3M11, H2019/HUM-589, ECO2017-87908-R) is gratefully acknowledged. This paper uses data from the National Educational Panel Study (NEPS): Starting Cohort 4–9th Grade, doi:10.5157/NEPS:SC4:1.0.0, Next Steps: Sweeps 1-8, 2004-2016. 16th Edition. UK Data Service SN: 5545 http://doi.org/10.5255/UKDA-SN-5545-8, Longitinal Studies of Australian Youth (LSAY): 15 Year-Olds in 2003, doi:10.4225/87/5IOBPG,  Education Longitudinal Study (ELS), 2002:  Base Year. https://doi.org/10.3886/ICPSR04275.v1. Blanden: School of Economics, University of Surrey, Stag Hill, Guildford, UK and Centre for Economic Performance, London School of Economics, UK; (e-mail: J.Blanden@surrey.ac.uk). Doepke: Department of Economics,
Northwestern University, 2211 Campus Drive, Evanston, IL 60208
(e-mail: doepke@northwestern.edu). Stuhler: Department of Economics, Universidad Carlos III de Madrid (e-mail: jan.stuhler@uc3m.es).}}
\author{Jo Blanden \and Matthias
Doepke \and Jan Stuhler}
 \date{April 2022}

\maketitle

\vspace{-1cm}

\begin{abstract}

This chapter provides new evidence on educational inequality and reviews the literature on the causes and consequences of unequal education. We document large achievement gaps between children from different socio-economic backgrounds, show how patterns of educational inequality vary across countries, time, and generations, and establish a link between educational inequality and social mobility. We interpret this evidence from the perspective of economic models of skill acquisition and investment in human capital. The models account for different channels underlying unequal education and highlight how endogenous responses in parents' and children's educational investments generate a close link between economic inequality and educational inequality. Given concerns over the extended school closures during the Covid-19 pandemic, we also summarize early evidence on the impact of the pandemic on children's education and on possible long-run repercussions for educational inequality.

\end{abstract}


\end{singlespacing}
\thispagestyle{empty}
\end{titlepage}


\clearpage

\section{Introduction}

In modern economies, people's livelihoods are based in large part on skills acquired through education. The importance of such skills has steadily increased over time. Whereas in the early nineteenth century few children received any formal education at all, large fractions of recent cohorts in high-income countries continue their studies through higher education and spend a substantial part of their lives enrolled in school and university. The benefits of education extend not just to higher earnings in the labor market and more secure employment, but also include wider advantages such as better health \cite{lleras2005relationship}, higher life satisfaction \cite{Powdthaveeetal2015}, reduced criminal behavior \cite{lochner2020education}, and greater civic participation \cite{lochner2011non}.\footnote{See \citeN{gunderson2020returns} for a survey of economic returns to education and \citeN{oreopoulos2011priceless} for an overview of non-pecuniary benefits.}

The essential economic role of education implies that unequal education can be a driver of unequal outcomes between different groups in society. What is more, educational inequality is at the root of low social mobility across generations. If only the children of wealthy and successful parents have access to the best educational opportunities, inequality will be more persistent across generations compared to a society where education is less dependent on family background. Understanding the nature and determinants of educational inequality is therefore crucial to the study of overall economic inequality and of the distribution of economic opportunity in society.

This chapter reviews the literature on educational inequality and presents new evidence on the extent to which family background is associated with differences in educational outcomes. We also discuss the mechanisms that underlie socio-economic gaps and examine how economic conditions, institutions, and policies shape these gaps. Lastly, given the importance of education inequality for social mobility, we endeavor to understand what the future may hold: will socio-economic gaps in education close, or are they likely to become even more marked in the future? 

We start by documenting test score gaps by family background using internationally comparable data from the OECD's Programme for International Student Assessment (PISA). We show that in all countries considered, there are large achievement gaps between students from families of higher versus lower socioeconomic status. In addition, achievement gaps within countries are wide compared to the observed variation in average achievement across countries. Using longitudinal data for a smaller set of countries, we document similar socioeconomic gaps in terms of educational attainment. We then discuss how socioeconomic gaps vary between countries, over time, and across generations, as well as how they relate to other aspects of economic inequality. We emphasize, in particular, two prominent empirical findings in the recent literature, both of which are central to our discussion of the mechanisms underlying educational inequality.

The first finding is the so-called ``Great Gatsby Curve,'' whereby countries or regions with high economic inequality tend to have low intergenerational mobility in income (\citeNP{hassler2007inequality}, \citeNP{corak2013income}, \citeNP{blanden2013cross}). Educational inequality is a potential source of the Great Gatsby Curve: if higher inequality increases the gap in educational achievements between children from richer and poorer families, lower social mobility is likely to follow. Accordingly, we examine the empirical relationship between income inequality and inequality by family background in educational outcomes. In terms of educational achievements in school, this link is fairly weak. That is, more unequal countries generally do not have wider test score gaps between students at the top and bottom ends of the socioeconomic scale. In contrast, there is a strong and robust relationship between income inequality and the intergenerational correlation in educational attainment: the ``Educational Great Gatsby Curve.'' The observation that income inequality matters much for attainment but little for achievement, as measured by test scores at school, helps shed light on the channels underlying the overall link between economic inequality and social mobility. In particular, mechanisms that generate socio-economic gaps in educational attainment conditional on achievement---such as financial constraints in higher education or different educational aspirations between families of different backgrounds---are likely to play a role.

The second finding is that educational inequality is surprisingly persistent across multiple generations. Simply extrapolating observed parent-child correlations would imply substantial regression to the mean when considering social mobility between grandparents and children, and little persistence in the economic status of different families over three or four generations. Yet, recent empirical evidence shows that differences in economic status across families instead persist over many generations (e.g., \citeNP{Clark2014book}, \citeNP{LindahlPalme2014_IGE4Generations}). One potential explanation for this puzzle is that conventional measures of social mobility from parents to children may understate persistence because educational advantages cannot be fully captured by simple summary measures such as years of schooling. For example, horizontal stratification in the learning process, such as variation in the quality of the educational institutions attended by children from richer and poorer families, may have an additional impact on intergenerational persistence. Similarly, a comparison of distant kins suggests that conventional measures also understate the contribution of assortative mating to educational inequality and low social mobility (\citeNP{ColladoOrtunoStuhler2019aa}).

After reviewing this evidence, we lay out models to understand the mechanisms that drive educational inequality. We first consider the role of parental investments, public investments, and neighborhood and peer effects in determining children's educational achievements throughout their school years.  We then consider the roles of ability, financial constraints, and uncertainty in young adults' decisions to go to and complete college. In a last step, we consider simple models of intergenerational transmission in an effort to explain the sources of high multi-generation persistence. The models frame our discussion of the related theoretical and empirical literatures.

A main insight from our model of skill acquisition during childhood is that the central role of parents in shaping their children's education generates a link between different sources of educational inequality. Parents invest in their children's skills directly, from talking and playing with them in the early years, to helping them with homework and studying later on. They are, however, constrained in these choices by inequalities in time, skills, and money, and their investments furthermore depend on other inputs such as the quality of public schools. Parents also shape peer and neighborhood experiences by choosing where to live and in which schools and extra-curricular activities to enroll their children. Their decisions in these matters add to inequality in school inputs, particularly in settings such as the United States where public school quality varies considerably and there are expensive private school options. 

Parental decisions also underlie interconnections between inequality in the economy at large, educational inequality, and social mobility. The economic approach to parenting envisions parental decisions as being informed by concern over children's welfare or economic success. If economic conditions are such that returns to formal education are high, parents worry more about the quality of schools that their children attend, push their children harder towards educational achievement, and attempt to endow them with preferences and aspirations that favor high educational attainment. But not everyone is able to make the same investments: higher inequality also implies a wider resource gap between richer and poorer parents in terms of both money and time. Hence, a more unequal economic environment results in greater educational inequality and lower intergenerational mobility: the ``Great Gatsby Curve'' arises. 

Socioeconomic differences can arise both from what parents ``do,'' namely differing kinds of investment in children's education, and from what they ``are,'' as captured by the notion of endowments in the classic \citeN{BT79} model of intergenerational transmission. The descriptive evidence in Section~\ref{sec:evidence} reflects both of these influences, as do our models. Endowments can include not only parents' initial wealth, educational attainment, and genetic determinants of ability, but also factors such as aspirations, values, and social norms, as long emphasized in sociological studies (\citeNP{erikson2019does}) as well as in recent economic work (\citeNP{BursztynJensen2015}). That said, our analysis in \Cref{sec:model_lr} of multi-generation transmission suggests that empirical findings based on two generations and focusing on standard measures of education may miss some types of endowments and could consequently understate the wider transmission of advantages and disadvantages. This argument aligns with earlier results based on the comparison of intergenerational and sibling correlations (\citeNP{bjorklund2011education}, \citeNP{BjorklundJaentti2012aa}). 

Broader family endowments---beyond income, wealth, and educational attainment---that are transmitted strongly from generation to generation might also explain high multi-generational persistence. Such persistent endowments are unlikely to primarily consist of genetic characteristics, as persistence across generations would be low unless assortative mating on genetic ability were extraordinarily strong. A more probable candidate would be a persistent family culture, capturing, for example, how a family views its position in society. While economic models of the intergenerational transmission of values and attitudes exist (e.g., \citeNP{bive01}, \citeNP{dozi08b}), an exploration of their ability to account for high multi-generational persistence has yet to be pursued.

The central role of publicly provided inputs in education, together with the prospect of lower social mobility due to educational inequality, has given rise to a number of policy questions. Should the government do more to guarantee equal access to, or even success in, education for children from different backgrounds? If so, what specific policy measures are likely to be successful? We use our theoretical models to discuss how the literature has approached these questions. Policy interventions are especially desirable if there are inefficiencies in the level of educational investments or their distribution across children of varying socio-economic circumstances. Possible sources of inefficiencies include human capital externalities in production, spillovers such as peer effects in the classroom (see \citeNP{epro11} for a review), informational frictions, and incomplete financial markets that make it difficult for poorer families to afford investments in education even if the returns are high. We use the examples of bottlenecks in the school system and financial constraints in access to higher education to illustrate the role of such inefficiencies. In addition, we review the evidence on a range of specific policy issues, including school funding, teacher quality, class size, and instruction time. 

At the time of writing, the world is still in the grip of the coronavirus pandemic. School closures have been a highly visible aspect of the public health response. Our model of children's skill acquisition demonstrates the important role played by schools in equalizing educational opportunities between children from different backgrounds. Several recent papers assess the implications of pandemic school closures for educational attainment and inequality (\citeNP{JangYum2020}; \citeNP{Fuchs_et_al_2020}; \citeNP{ADSZ2021}). We consider the potential impact of the Covid-19 pandemic on educational inequality in Section~\ref{sec:covid}, where we discuss this literature together with insights from empirical work. Given public education's role as ``the great leveler,'' widespread school closures are likely to have profound effects, leading to larger educational inequality among affected cohorts and consequent economic repercussions far into the future.

Our discussion builds on the contributions of \citeN{HanushekWoss2011HB} and \citeN{bjorklund2011education} in an earlier volume of this Handbook series. We focus on socio-economic gradients and do not explore inequality in educational outcomes by race or gender, although these dimensions are clearly important.  Our discussion is informed by a human capital model that emphasizes differences in investment and skill development due to unequal resources and peer effects. It therefore accounts for some of the sources of racial differences but not others, such as discrimination. Recent surveys assessing racial and gender inequality and their underlying mechanisms include \citeN{blau2017gender} and \citeN{lang2020race}. Other topics not addressed in detail here include the political determinants of educational systems, aspects of educational inequality specific to developing countries, the relationship between family structure and educational inequality, and the macroeconomic repercussions of educational inequality \cite{gaze93}. 

We conclude our review with a consideration of open research questions. While the literature on educational inequality has made tremendous progress, the nature of the subject also poses unique empirical challenges. The central role of parenting decisions makes it difficult to design randomized interventions, meaning that empirical evidence is primarily based on observational data that can be hard to interpret. Even with well-identified research designs, relevant outcomes (such as children's future earnings and family decisions) may be realized only decades later (or generations later when analyzing long-run mobility). Furthermore, parenting and education decisions occur in a tremendous range of institutional and cultural contexts, which vary not only between but even within countries. Though not insurmountable, these issues imply that much has yet to be learned. 

In the following section, we present new evidence on the extent of educational inequality in a set of high-income economies. In Section~\ref{sec:model} we examine different sources of educational inequality from the perspective of a model of child development. Section~\ref{sec:model_high_ed} extends this analysis to higher education, including issues such as student loans. Section~\ref{sec:model_lr} discusses mechanisms that can give rise inequalities in education and economic outcomes that extend across many generations. Section \ref{sec:covid} considers the implications of the Covid-19 pandemic on inequality and Section~\ref{sec:conc} concludes.

\section{Evidence on Educational Inequality}

\label{sec:evidence}

This section presents new evidence on the extent of educational inequality by family background in high-income economies. Inequality can be documented using different measures  (e.g., educational attainment and test scores) and at various life stages. We start by looking at evidence on test scores from the OECD's Programme for International Student Assessment (PISA), which allows us to construct measures of educational inequality at the high school level that are comparable across countries (\citeNP{PISA2015Info}). To document socio-economic gaps in higher education, we use longitudinal surveys for Australia, England, Germany, and the United States that provide information on family background, test scores, and educational attainment. Finally, we assess the contribution of educational inequality to the persistence of economic status over multiple generations through a review of recent evidence in the literature on intergenerational mobility.

\subsection{Socio-Economic Gaps in Test Scores}

\begin{figure}[t]
\centering
\captionsetup[sub]{}
\begin{subfigure}[]{0.49\linewidth}
    \centering
     \includegraphics[width=\linewidth]{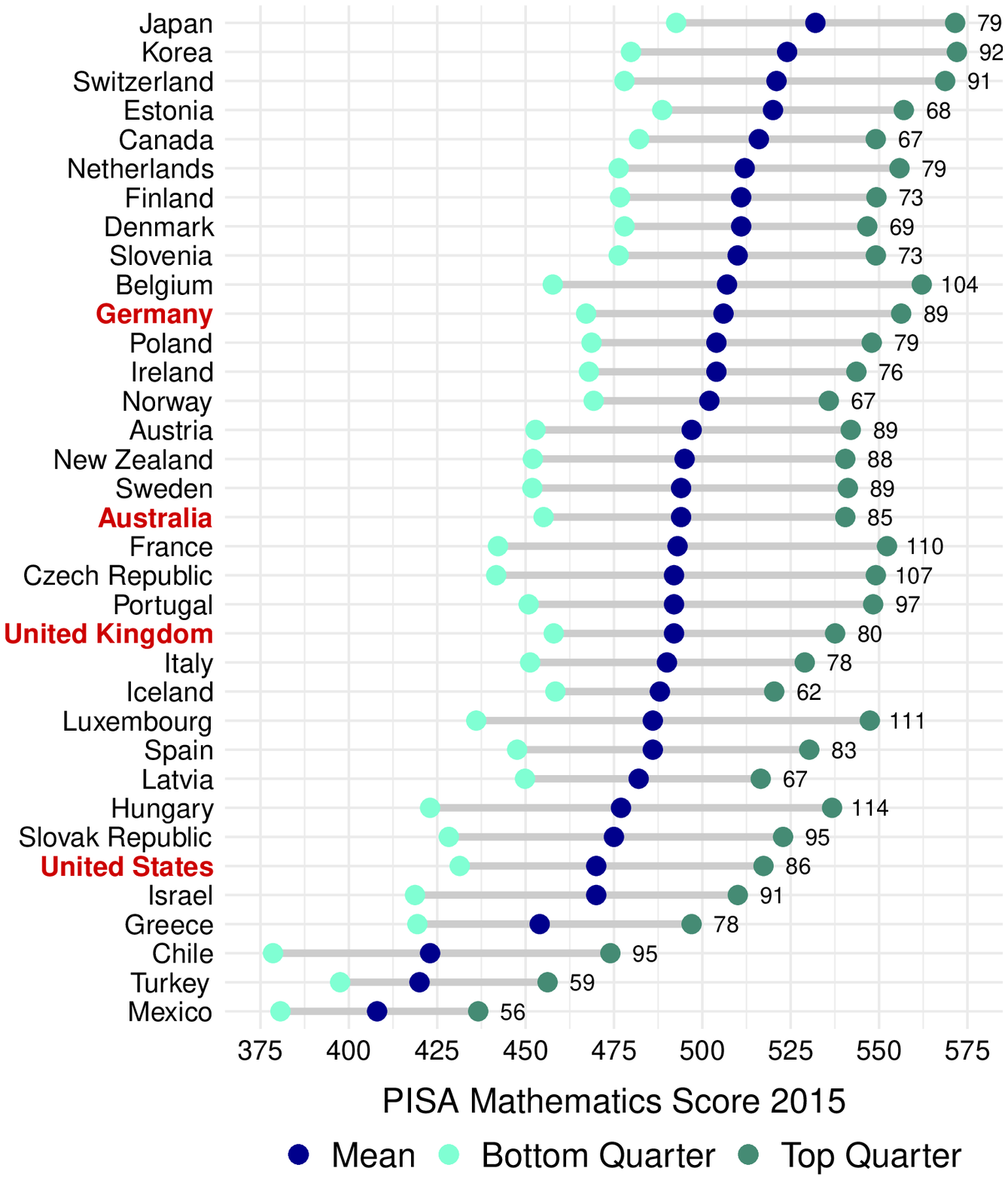}
    \caption{Mathematics}
    \label{fig:PISA_Math} 
\end{subfigure}
\hfill
\begin{subfigure}[]{0.49\linewidth}
    \centering
     \includegraphics[width=\linewidth]{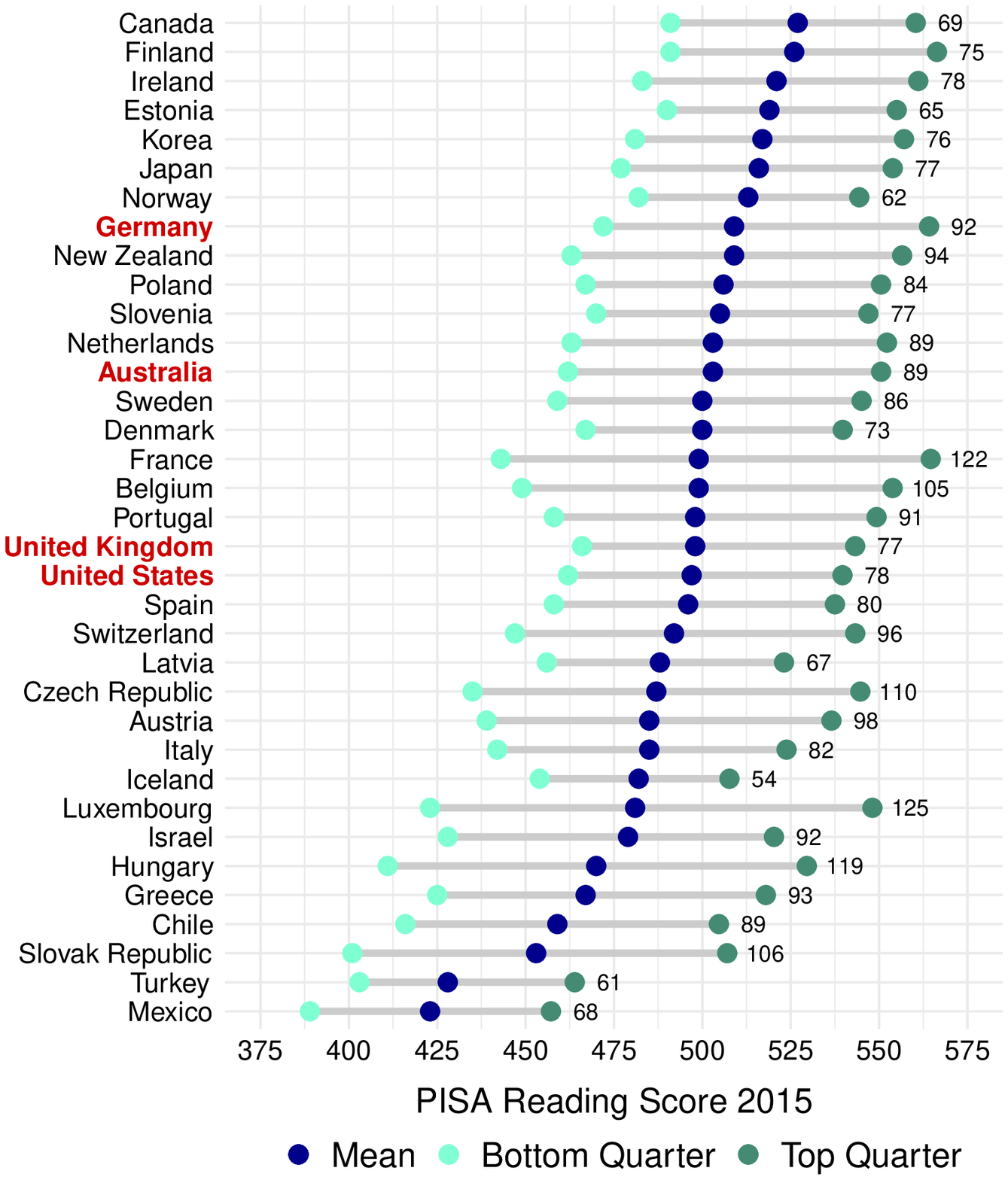}
    \caption{Reading}
    \label{fig:PISA_Reading} 
\end{subfigure}
\centering
\caption[Pisa results]{PISA scores by country and socio-economic background.\\\\
\footnotesize
\emph{Notes:} The figure reports the mean PISA 2015 results for OECD countries and the mean scores in the top and bottom quarters of the PISA index of economic, social, and cultural status (ESCS). The numbers refer to the gap between the mean scores in the top and bottom quarters for each country. Source: \citeN{PISA2015Info}}
\label{fig:PISA} 
\end{figure}

Differences in educational achievements appear early in life and are large in all stages of educational attainment.  \Cref{fig:PISA} provides a snapshot of achievement gaps in high school, comparing PISA scores at age 15.\footnote{See also \citeN{HanushekWoss2011HB}, who provide a comprehensive survey of economic research on differences in educational achievement based on earlier PISA waves, as well as the Trends in International Mathematics and Science Study (TIMSS), and the Progress in International Reading Literacy Study (PIRLS).} For each country, the figure plots the average score on the 2015 PISA assessment in mathematics (left panel) and reading (right panel), the average scores within the bottom and top quarters of the PISA index of socio-economic status (ESCS), as well as the gap between the two.   

\captionsetup[figure]{}
  
Two observations stand out. First, the gap in test scores between the top and bottom quarters of socio-economic status is pronounced in each of the 35 considered countries. Second, these socio-economic gaps are large compared to the overall differences in achievement between countries. Even in the best-performing countries such as Finland or Canada, the achievement of students from disadvantaged backgrounds is below the OECD average of 500 points. Furthermore, while the reported mean gap between countries rarely exceeds 50, the average gap by family background is 84 for reading and 86 for mathematics, corresponding to nearly one standard deviation.   

How do these differences in test scores translate into differences in knowledge? While such conversions are conceptually problematic, a number of studies have estimated the grade equivalence of PISA points \cite{PISA2015Info}. Moreover, learning gains in national and international tests during a given year generally amount to between one-quarter and one-third of a standard deviation \cite{Woessmann2016}. The evidence in Figure \ref{fig:PISA} therefore suggests that by the age of 15, children in the bottom quarter in terms of socio-economic background are more than two years behind their more advantaged peers. A strong connection between family background and student achievement has been well documented in the literature. However, the strength of this relation varies across countries, and can in part be explained by institutional differences in education systems \cite{HanushekWoss2011HB}, as we further discuss in Section \ref{sec:compensating_investments}. Whether the magnitudes of these socio-economic gaps have changed over time remains more controversial, a point we return to in Section \ref{subsec:inequalitytime}.

\subsection{Socio-Economic Gaps in Educational Attainment}

Beyond test scores, socio-economic differences also extend to educational attainment. Children from high-income families are more likely to continue on to post-secondary programs, and conditional on attending they are more likely to complete their studies and obtain a degree. Moreover, these gaps do not arise solely because children from well-off families perform better in school (as documented in the previous section). Socio-economic differences are pronounced even conditional on intermediate measures of achievement, such as test scores during high school. 

To shed light on these patterns, we report results from four different data sets that contain detailed information on educational careers and family background. For England, we use Next Steps: the Longitudinal Study of Young People in England (LSYPE) \cite{UCL_next_steps}; for the United States, the Education Longitudinal Study (ELS) \cite{ELS_US}; for Australia, the Longitudinal Surveys of Australian Youth (LSAY) \cite{LSAY_Aus}; and for Germany, the National Education Panel Study (NEPS, \citeNP{BlossfeldMaurice2011}).

\begin{table}[p]
\begin{scriptsize}  

\begin{longtable}[thp]
    					{l L{.22\textwidth}
                      L{.15\textwidth}
                      L{.15\textwidth}
                      L{.15\textwidth}
                      L{.15\textwidth}} 
\caption{\textcolor{black}{Description of Longitudinal Datasets}\label{table:1}}\\
\\
\toprule 
&\RaggedRight  \textbf{Country} & \Centering  \textbf{England} &\Centering   \textbf{United States} &\Centering  \textbf{Australia} &\Centering  \textbf{Germany}  \\ \cmidrule(l){3-6} 
&\RaggedRight  Name &\RaggedRight  Next Steps (formerly LSYPE) &\RaggedRight   Educational Longitudinal Study 2002 &\RaggedRight  Longitudinal Survey of Australian Youth 2003 &\RaggedRight  National Education Panel Study (NEPS), SC 4  \\
&\RaggedRight  Birth cohort &\RaggedRight  1989-1990 &\RaggedRight  $\sim$ 1987-1988
(grade sampled) &\RaggedRight  1988 &\RaggedRight  $\sim$ 1995-1996 (grade sampled) \\
&\RaggedRight  Starting sample size &\RaggedRight 16000 &\RaggedRight  17591  &\RaggedRight 12500 &\RaggedRight  16425 \\
&\RaggedRight  Grade and age in Wave 1 of Survey &\RaggedRight 9\textsuperscript{th} grade, age 14 &\RaggedRight 10\textsuperscript{th} grade, standard age 15/16  &\RaggedRight  Age 15 (70\% in 10\textsuperscript{th} grade) &\RaggedRight  9\textsuperscript{th} grade, standard age 14/15 \\
&\RaggedRight  Parents with ``higher education'' (one parent more than secondary) (\%) &\RaggedRight  50.5\newline[8494] &\RaggedRight   72.8\newline[15612] &\RaggedRight 54.3\newline[6536] &\RaggedRight  79.6\newline[8487] \\
&\RaggedRight  Single parents (\%) &\RaggedRight  22.2\newline[8450] &\RaggedRight  23.6\newline[13592] &\RaggedRight  21.6\newline[6593] &\RaggedRight  17.1\newline[6148] \\
&\RaggedRight  Correlation between parents' years of education &\RaggedRight  0.453\newline[5738] &\RaggedRight 0.590\newline[10298] &\RaggedRight 0.470\newline[6234] &\RaggedRight 0.518\newline[7834] \\
&\RaggedRight  Correlation between math and reading test scores &\RaggedRight 0.785\newline[7861] &\RaggedRight 0.759\newline[15244] &\RaggedRight  0.760\newline[10370] &\RaggedRight 0.512\newline[8434] \\
&\RaggedRight  Parent expects child to attend university (\%) &\RaggedRight 58.6\newline[15513] &\RaggedRight  77.6\newline[12877] &\RaggedRight N/A &\RaggedRight 56.8\textsuperscript{\emph{a}}\newline[5725]  \\
&\RaggedRight Student expects to attend university (\%) &\RaggedRight 65.1\newline[15431] &\RaggedRight   72.2\newline[15273] &\RaggedRight 63.4\newline[10356] &\RaggedRight  66.2\newline[5023] \\
&\RaggedRight  Studying for a degree at age 20 (\%) &\RaggedRight 37.6\newline[8478] &\RaggedRight   43.1\newline[16162] &\RaggedRight 33.0\newline[6609] &\RaggedRight  46.1\newline[8309] \\
&\RaggedRight  Definition of selective university &\RaggedRight Russell Group (42 research intensive universities) &\RaggedRight Four-year institutions with average test scores in top 20\%   &\RaggedRight  ``Group of 8'' research intensive universities &\RaggedRight  University program with minimum entry grade requirements \\
&\RaggedRight Studying at selective university at age 20 (\%) &\RaggedRight 9.4\newline[8576] &\RaggedRight  19.5\newline[12226] &\RaggedRight 7.4\newline[6609]  &\RaggedRight  28.6\newline[8309] \\
&\RaggedRight  Degree obtained by age 25 (\%) &\RaggedRight 26.8\newline[7569] &\RaggedRight   32.9\newline[16197] &\RaggedRight 47.3\newline[3700] &\RaggedRight  N/A \\
&\RaggedRight  Attended at age 20 but no degree by age 25 (\%) &\RaggedRight 36.2\newline[3539]  &\RaggedRight  34.8\newline[9253] &\RaggedRight 15.0\newline[1598] &\RaggedRight  N/A \\
\bottomrule
\end{longtable}

\centering
\vspace{.2cm}

\begin{minipage}[t]{\columnwidth}
\footnotesize
 \emph{Notes:} Square brackets report the sample sizes upon which the calculations are based. We restrict our sample to those who participated in or after wave 9 (NEPS) or to those for whom we have information on whether they started university (other samples). Variables are weighted using panel entry weights (NEPS) or the first wave of the sample in which the variable is observed (other samples).\textsuperscript{\emph{a}} Question relates to wishes rather than expectations. 
\end{minipage}
\end{scriptsize}
\end{table}

As a simple measure of family background, we use an indicator for whether at least one parent has obtained an level of education beyond high school.\footnote{For Germany, our definition of higher education includes intermediate school-leaving (\textit{Mittlere Reife}) with vocational qualifications; for England this reflects having more than a GCSE qualification (obtained upon leaving school at age 16); and for the United States and Australia this means having a qualification higher than a high school diploma.} As shown in Table \ref{table:1}, this share varies between 50.5 (England) and 79.6 (Germany) percent, partially due to differences in the structure of the educational systems. 

In Table \ref{table:2a}, we regress the student’s standardized test scores in high school on our indicator for socio-economic background, for Australia, the United States, England and Germany. Note that these estimates reflect the overall importance of family background, for which parental education is a proxy, and not the causal effect of parental education itself.\footnote{Based on a review of the literature and an application to Swedish data, \citeN{HolmlundLindahlPlug2011} conclude that intergenerational schooling associations are largely driven by selection rather than direct causal effects. \citeN{BjoerklundJaentti2020} note that the pattern is qualitatively similar for income, with estimates of the causal effect of parent on child income being much smaller than the corresponding descriptive associations.} The results are similar across the four countries and across the school subjects: the average test score of children of less educated parents is between 0.4 and 0.6 standard deviations lower than that of children with highly educated parents, in both mathematics and reading.\footnote{See also Section 4.2 in \citeN{HanushekWoss2011HB} for a comprehensive review of international comparisons of socio-economic gaps in test scores.} As shown in Table \ref{table:1}, the correlation in scores between the two subject areas is high in all four countries. 

\begin{table}[htb]
\begin{small}    
    \begin{longtable}[ht!]
    					{l L{.11\textwidth}
                      L{.11\textwidth}
                      L{.11\textwidth}
                      L{.11\textwidth}
                      L{.11\textwidth}} 
\caption{\textcolor{black}Associations between Test Scores and Parental Education\label{table:2a}}\\
\toprule 
      \multicolumn{1}{C{.11\textwidth}}{}    &
      \multicolumn{1}{C{.11\textwidth}}{\textbf{England}}    &
      \multicolumn{1}{C{.11\textwidth}}{\textbf{US}}    &
      \multicolumn{1}{C{.11\textwidth}}{\textbf{Australia}} &
      \multicolumn{1}{C{.11\textwidth}}{\textbf{Germany}}    \\\cmidrule(l){2-5} 
        \multicolumn{5}{@{}l}{\textbf{Standardised scores on parental higher education}} \\ \midrule 
    Mathematics & \makecell{$0.606$\\$(0.030)$}
    & \makecell{$0.549$\\$(0.024)$} & \makecell{$0.395$\\$(0.027)$} & \makecell{$0.534$\\$(0.028)$} \\
    Reading &  \makecell{$0.593$\\$(0.028)$}
    & \makecell{$0.559$\\$(0.022)$} & \makecell{$0.437$\\$(0.029)$} & \makecell{$0.487$\\$(0.033)$}  \\ 
    Sample size & $7876$ & $16197$ & $10131$ & $7672$ \\ 
    \bottomrule
\end{longtable}

\centering
\vspace{.2cm}

\begin{minipage}[t]{\columnwidth}
\footnotesize
\emph{Notes:} All models are weighted using panel entry or longitudinal weights. Parental higher education is an indicator for whether at least one parent has obtained education beyond high school.
\end{minipage}

\end{small}
\end{table}

In Panel (a) of Table \ref{table:2b}, we show socio-economic gaps in university attendance, reporting both unconditional estimates and estimates that condition on test scores in high school at around age 14 or 15 (depending on data source).\footnote{Attendance is defined as attending a professional academy, university of applied sciences, or university in Germany, as studying for a bachelors' degree in England and Australia, or a four-year college degree in the United States.}  The probability of attending university at age 20 is between 18 (Australia) and 28 (United States) percentage points higher for children of highly educated parents, as defined above. To illustrate the size of these effects compared to the baseline attendance rate (see Table \ref{table:1}) we also report odds ratios, which suggest that the odds of attending university--the proportion of students attending over those non-attending---are up to 3.4 times higher for children of highly educated parents.

Attainment gaps in higher education partially reflect achievement gaps in high school, and therefore narrow when controlling for test scores in math and reading. Differences in test scores explain about half of the gap in university attendance in the United States, Germany, and Australia. That said, the gaps remain large even conditional on test scores. For example, in Germany, students from more advantaged backgrounds are still 11 percentage points more likely to attend university than their peers with comparable test scores at age 14--15, compared to an overall attendance rate of 46 percent. Notably, conditional gaps are particularly large in the expensive US system and small in England, suggesting that costs and credit constraints may in part drive these differences.\footnote{Note that the English data refers to the period before fees were increased to their current relatively high level \cite{jerrim2012socio}. Though, \citeN{murphy2019end} show that the new student finance arrangement has not led to a rise in socio-economic gaps in participation.} 

To summarize, attainment gaps in higher education are large even conditional on observed achievement gaps in secondary school. Panel (b) in Table 3 shows qualitatively similar results for degree attainment by age 25.\footnote{These estimates are not computed for Germany as completion by age 25 is relatively less common.} 

Test scores are a noisy measure of achievement (\citeNP{jacobrothstein2016}), such that the positive coefficient on family background conditional on test scores may still reflect differences in abilities (i.e., an omitted variable bias). The extent to which prior achievement can explain socio-economic differences in university attendance continues to be debated \cite{jerrimvignoles2015}. One way to investigate this concern is to control for a more extensive set of ability measures and test scores. In the German sample, for example, the coefficients on parent education decrease slightly but still remain large when accounting for these additional controls. 

\begin{table}[p]
\begin{scriptsize} 

    \begin{longtable}[ht!]
    					{l L{.11\textwidth}
                      L{.11\textwidth}
                      L{.11\textwidth}
                      L{.11\textwidth}
                      L{.11\textwidth}} 

\caption{\textcolor{black}Associations between University Attendance and Parental Education\label{table:2b}}\\
      \toprule      
      \multicolumn{1}{C{.11\textwidth}}{}    &
      \multicolumn{1}{C{.11\textwidth}}{\textbf{England}}    &
      \multicolumn{1}{C{.11\textwidth}}{\textbf{US}}    &
      \multicolumn{1}{C{.11\textwidth}}{\textbf{Australia}} &
      \multicolumn{1}{C{.11\textwidth}}{\textbf{Germany}}    \\\cmidrule(l){2-5} 
        \multicolumn{5}{@{}l}{\textbf{(a) Attending university at age 20 on high parental education}} \\ \midrule 
        \multicolumn{5}{c}{\emph{Unconditional}} \\ 
    Regression Coefficient  &
    \makecell{$0.226$\\$(0.012)$} &  \makecell{$0.277$\\$(0.012)$} & \makecell{$0.177$\\$(0.012)$} & \makecell{$0.251$\\$(0.016)$} \\
    Odds-Ratio & \makecell{$2.727$\\$(0.151)$} & \makecell{$3.441$\\$(0.207)$} & \makecell{$2.265$\\$(0.149)$} & \makecell{$2.949$\\$(0.234)$} \\ 
    \multicolumn{5}{c}{\emph{Conditional on maths and reading scores}} \\ 
    Regression coefficient  &  
    \makecell{$0.074$\\$(0.011)$}& 
    \makecell{$0.142$\\$(0.011)$} & \makecell{$0.087$\\$(0.01)$} & \makecell{$0.114$\\$(0.016)$}   \\
    Odds-ratio   &  
    \makecell{$ 1.431$\\$(0.089)$}& 
    \makecell{$2.204$\\$(0.145)$} & \makecell{$1.577$\\$(0.119)$} & \makecell{$1.865$\\$(0.171)$}  \\ 
    Sample size & $7861$ & $15612$ & $6487$ & $6869$ \\ \bottomrule
        \multicolumn{5}{@{}l}{\textbf{(b) Obtaining a degree by age 25 on high parental education}} \\ \midrule 
        \multicolumn{5}{c}{\emph{Unconditional}} \\ 
    Regression Coefficient  & 
    \makecell{$0.166$\\$(0.011)$} & \makecell{$0.222$\\$(0.011)$} & \makecell{$0.217$\\$(0.02)$} \\
    Odds-Ratio & \makecell{$2.375$\\$(0.149)$} & \makecell{$3.142$\\$(0.202)$} & \makecell{$2.481$\\$(0.226)$} \\ 
    \multicolumn{5}{c}{\emph{Conditional on maths and reading scores}} \\ 
    Regression coefficient  & 
    \makecell{$0.058$\\$(0.011)$}& 
    \makecell{$0.107$\\$(0.010)$} & \makecell{$0.124$\\$(0.02)$} \\
    Odds-ratio   &
    \makecell{$1.354$\\$(0.090)$}& \makecell{$1.976$\\$(0.138)$} & \makecell{$1.817$\\$(0.174)$}  \\  
    Sample size & $7023$ & $15612$ & $3685$ \\
    \bottomrule
        \multicolumn{5}{@{}l}{\textbf{(c) Attending selective university at age 20 on high parental education}} \\ \midrule 
        \multicolumn{5}{c}{\emph{Unconditional}} \\
    Regression Coefficient  & \makecell{$0.093$\\$(0.007)$} & \makecell{$0.179$\\$(0.010)$} & \makecell{$0.073$\\$(0.006)$} & \makecell{$0.157$\\$(0.014)$} \\
    Odds-Ratio & 
    \makecell{$4.129$\\$(0.436)$} & \makecell{$4.600$\\$(0.489)$} & \makecell{$3.310$\\$(0.403)$} & \makecell{$2.384$\\$(0.219)$} \\ 
    \multicolumn{5}{c}{\emph{Conditional on maths and reading scores}} \\ 
    Regression coefficient  & \makecell{$0.036$\\$(0.006)$} & \makecell{$0.076$\\$(0.009)$} & \makecell{$0.048$\\$(0.006)$} & \makecell{$0.069$\\$(0.014)$}  \\
    Odds-ratio   & 
    \makecell{$1.735$\\$(0.194)$} & 
    \makecell{$2.463$\\$(0.283)$} & \makecell{$2.255$\\$(0.285)$} & \makecell{$1.615$\\$(0.158)$}  \\ 
    Sample size & $7861$ & $11780$ & $6487$ & $6869$ \\ 
      \bottomrule
\end{longtable}

\centering
\vspace{.2cm}

\begin{minipage}[t]{\columnwidth}
\footnotesize
\emph{Notes:} All models are weighted using panel entry or longitudinal weights. Parental higher education is an indicator for whether at least one parent has obtained education beyond high school. 
\end{minipage}
\end{scriptsize}
\end{table}

Why might parental background affect attendance even conditional on ability?\footnote{See also \citeN{boudon1975education} and the sociological literature on primary effects, namely gaps in actual academic performance, and secondary effects, which include social origin influences that operate over and above academic performance. For example, \citeN{jackson2007primary} find that secondary effects account for at least one quarter, and possibly up
to one-half, of class differentials as measured by odds ratios in England and Wales.} In Section 4, we present a model of dynamic human capital investments that sheds light on the role of financial resources. Under perfect financial markets (as for example in \citeNP{BT79}), university attendance should not vary with family background, once we condition on acquired skills at the end of secondary school. However, in the presence of borrowing constraints, attendance increases in the financial assets of the parents, conditional on skill. Even if borrowing constraints are not binding, attendance will increase in financial assets if higher education is a risky endeavour, as the disutility of risk is greater for families with low financial resources.

\subsection{Socio-Economic Gaps within Higher Education}

Higher education institutions vary in quality. Likewise, there are substantial differences in the rigor of and economic returns offered by different majors and programs of study in a given university. Hence, additional socio-economic gaps may be present in terms of where students from richer and poorer families study and which courses they take.

Indeed, socio-economic gaps in attendance rates are greater when we restrict our analysis to selective universities (as defined in Table~\ref{table:1}), which tend to attract better students. For example, in Australia, the odds ratio conditional on test scores of attending any university is 1.6, but increases to 2.3 for the eight leading public universities. Panel (c) of Table \ref{table:2b} provides further details on this result, where we see that in the other countries as well, children of highly educated parents are much more likely to study at a selective institution at age 20. Access to elite institutions can be particularly unequal. For example, \citeN{ChettyFriedmanSaezTurner2017} find that in the United States, children whose parents are in the top 1 percent of the income distribution are 77 times more likely to attend an Ivy League college than those whose parents are in the bottom income quintile.\footnote{Such attainment gaps can also generate direct intergenerational spillovers. For example, \citeN{Barrios-Fernandez:2022wc} show that parents' admission to an elite college program causally changes their children’s educational paths, making them more likely to attend an elite private school or college themselves.}

In addition to the quality of the institution, socio-economic gaps can also be observed in the field of study. For example, \citeN{hallsten2018horizontal} find that conditional on previous achievement, about 25 percent of the variation in tertiary field choices in Sweden can be explained by parental background (parents’ education, occupation, income, and wealth). These results illustrate that educational attainment varies not only in a quantitative sense, but also qualitatively, in terms of the achievement conditional on the time spent on schooling \cite{blanden2016educational}. These differences are in turn an important source of economic inequality and immobility: different fields of study are associated with considerably different payoffs \cite{KimetalHorizontal2015}, even after accounting for sorting and variance in the quality of institutions or peer groups (\citeNP{dale2014estimating}, \citeNP{KirkeboenLeuvenMogstad2016}).\footnote{Moreover, children often choose the same field as their parents, contributing to the intergenerational persistence of educational inequality. Using a regression discontinuity design, \citeN{Altmejd2021} shows that a large share of this association is causal, with children being particularly likely to follow their parents' choice in high-paying degrees such as medicine, business, law, and engineering.}

These and other forms of horizontal stratification have long been highlighted in the sociological literature (\citeNP{gerber2008horizontal}; \citeNP{Torche2011}) and have increasingly also been the subject of economic research. As college shares are stabilizing in many countries, the relative importance of qualitative dimensions of stratification in reproducing inequalities could well be on the rise. Accordingly, the differentiation of educational systems in advanced industrialized societies may increasingly warrant a multidimensional approach that classifies education not only hierarchically by level of education but also by horizontal characteristics, such as field of study (\citeNP{AndradeThomsen2017}). That said, quantifying and comparing the extent of horizontal stratification to socioeconomic gaps in the vertical dimension poses a challenge \cite{hallsten2018horizontal}. 

Summary measures of educational inequality and intergenerational mobility tend to focus on years of schooling, abstracting from achievement gaps between students attending the same grade, or from horizontal segregation in institutional quality or field of study. They consequently might understate not only the extent of educational inequalities in the cross-section, but also their persistence across generations. We return to this theme below.

\subsection{Economic Inequality and Intergenerational Mobility}\label{subsec:mobilityinequality}

Gaps in student achievement as documented here have implications for the intergenerational transmission of advantages from parents to children. Education is considered to be the key mediator in the intergenerational persistence of socio-economic status in both the economics and sociology literatures \cite{Goldthorpe2014}. Economic models in the tradition of \citeN{BT79} interpret educational attainment as an investment decision, subject to financial constraints and affected by market conditions. In such models, a rise in economic inequality can make financial constraints more binding for low-income parents, and hence reduce social mobility. A more recent class of theories models the dynamics of human capital investments over multiple stages in the life cycle and allows for parental investments in terms of both money and time, educational investments at school, and neighborhood and peer effects (see \Cref{sec:model}). In such models, additional links between economic and educational inequality arise, implying that cross-sectional inequality is a key determinant of intergenerational persistence.

Given that income inequality has increased considerably in many developed countries, recent research on intergenerational mobility has devoted much attention to this potential feedback from economic to educational inequality. Does greater economic inequality actually lower intergenerational mobility? A stylized fact consistent with this hypothesis is the observation that nations with high cross-sectional inequality tend to have low intergenerational mobility in income, an observation often referred to, as mentioned above, as the ``Great Gatsby Curve'' (\citeNP{corak2013income}; \citeNP{blanden2013cross}; \citeNP{oecd2018broken}). This association implies a double disadvantage for poor families when inequality is high: not only are the income gaps larger, but their children are also more likely to remain poor themselves. The robustness of this cross-country evidence remains, however, debated, partly due to considerable uncertainty regarding the available estimates of income mobility (\citeNP{mogstad2021family}).\footnote{Standard measures of income mobility are subject to potentially large attenuation and life-cycle biases (\citeNP{NybomStuhlerJHR2017}, \citeNP{MazumderDeutscher2021}). \citeN{mogstad2021family} further argue that a cross-country association between inequality and income mobility is not clear, in that it appears to be driven by differences between just three clusters of countries: developing countries with high inequality and low mobility; a small set of Nordic countries with low inequality and high mobility; and the majority of the OECD countries with intermediate levels of income inequality and mobility.}

One intriguing question is whether the negative relation between inequality and mobility also holds for educational mobility.\footnote{Empirically, the two dimensions of mobility are closely related: the correlation between estimates of income mobility and educational mobility is around 0.7 across developed countries (\citeNP{Stuhler2018JRC}).}  The answer matters for two reasons. First, the data requirements for measuring educational outcomes are lower than for income, and the estimates likely more comparable across countries, in particular when based on standardized international assessments such as PISA. The resulting evidence may therefore be more robust to mismeasurement. Second, the association between income inequality and educational gaps is directly related to the key investment mechanism in standard economic models, as considered in Section \ref{sec:model}.\footnote{\citeN{durlauf2021great} describe a wider class of theories and mechanisms to explain the Great Gatsby Curve, including models involving social interactions and segregation.} Evidence on this association could therefore be indicative of the mechanisms underlying the Great Gatsby Curve in income.

\captionsetup[figure]{}
\begin{figure}[t]
\centering
\captionsetup[sub]{}
\begin{subfigure}[]{0.495\linewidth}
    \centering
     \includegraphics[width=\linewidth]{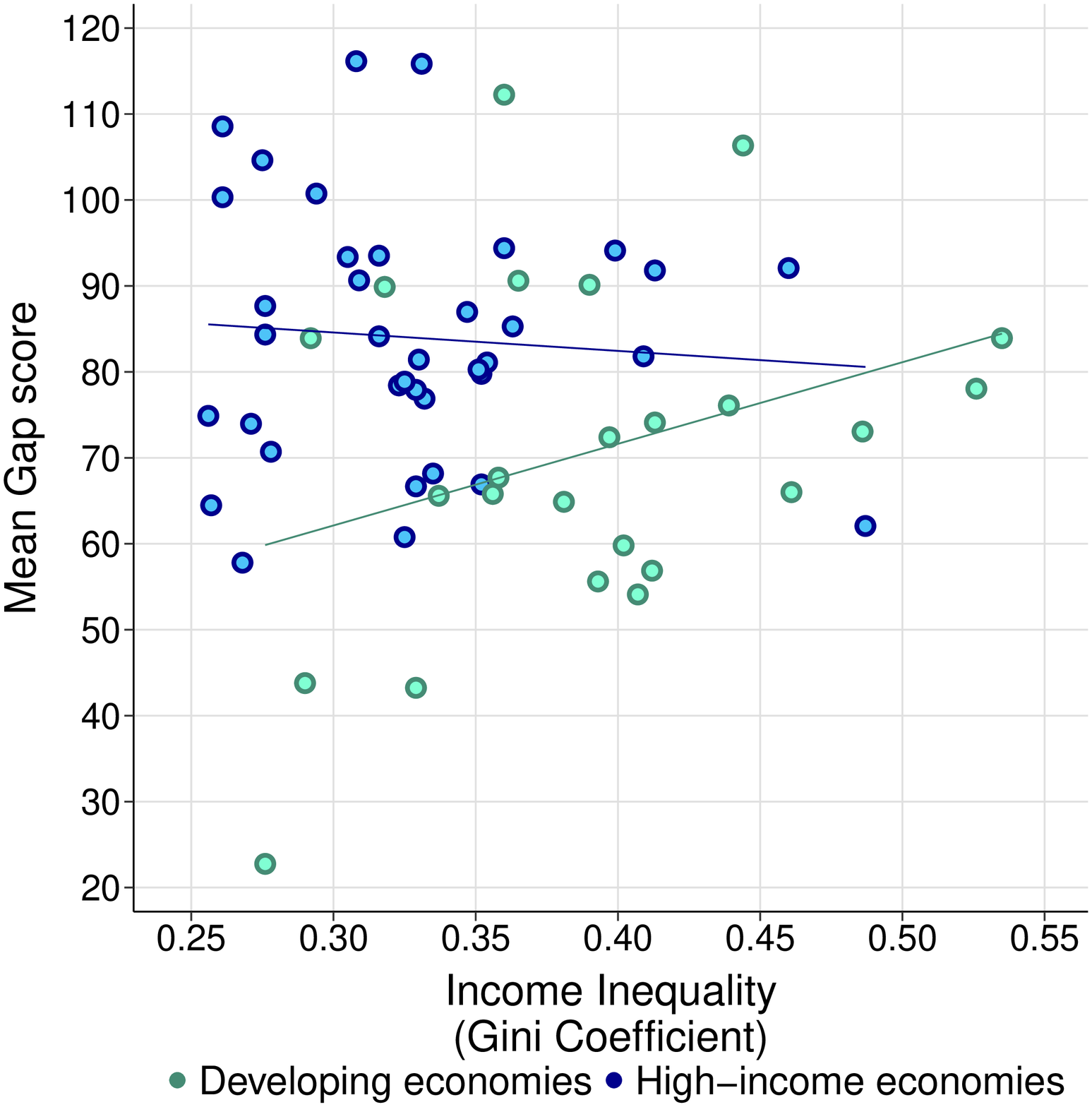}
    \caption{PISA gap scores}
    \label{fig:GG_Pisa} 
\end{subfigure}
\hfill
\begin{subfigure}[]{0.495\linewidth}
    \centering
     \includegraphics[width=\linewidth]{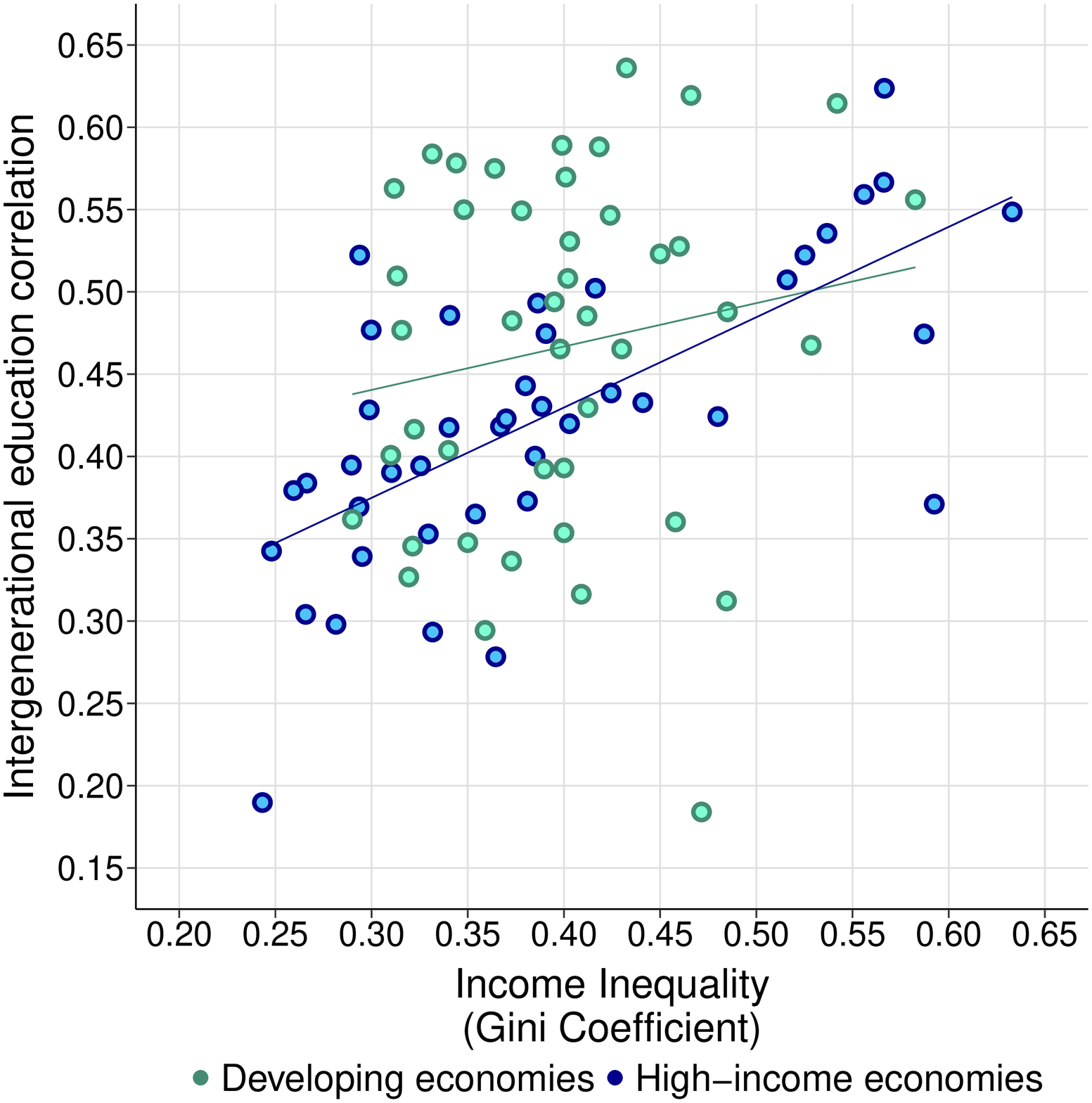}
    \caption{Years of schooling}
    \label{fig:GG_YrsSchooling} 
\end{subfigure}
\centering
\caption[]{The Educational Great Gatsby Curve. \\\\
\footnotesize
    \emph{Notes:} Scatter plot of the 2012 World Bank Gini (or nearest available year) against the gap in average 2015 PISA scores in reading and mathematics between the top and bottom quarters of socio-economic background (Source:  \citeNP{PISA2015Info}) in Panel (a) and of the intergenerational correlation in parents' highest and child's years of schooling (Source: Global Database on Intergenerational Mobility, \citeNP{GDIM2018}) in Panel (b).}
\label{fig:GG_Edu} 
\end{figure}

Following this reasoning, we explore the ``Educational Great Gatsby Curve'' in \Cref{fig:GG_Edu}. The left panel plots the gap in test scores between the bottom and top quarters of the PISA index of socio-economic status (as shown in \Cref{fig:PISA}) against a measure of income inequality (the Gini Coefficient in pre-tax income). Overall, the relationship is weak. We do not observe any clear relationship between income inequality and mobility in student performance in high-income countries, and only a weak positive relationship for developing economies. This result aligns with the work of \citeN{CarneiroToppeta2021} who, using data for younger children from the Second Regional Comparative and Explanatory Study (SERCE), do not find a relationship between income inequality and socio-economic gradients in test scores across Latin American countries.\footnote{The strength of the relationship may, however, vary with alternative measures of mobility or inequality.  \citeN{esping2004untying}, for example, documents a strong positive cross-country relation between inequality and immobility in cognitive skills as measured in the International Adult Literacy Survey (IALS). \citeN{godinhindriks2018} find a similar positive relation in PISA test scores, which may in part be due to tracking: the segregation of pupils based on academic ability is not only associated with greater inequality in test scores \cite{hanushekwoessmann2005} but also with lower mobility \cite{godinhindriks2018} and greater inequality of opportunity \cite{ferreira2014measurement}.}

The right panel in \Cref{fig:GG_Edu} plots a measure of immobility in educational attainment, namely the intergenerational correlation in years of schooling against income inequality.\footnote{In documenting educational inequality, a reliance on attainment gaps as reported in Table~\ref{table:2b} can be sensitive to the share of individuals achieving a given level of education or the relative size of the groups being compared. Using the intergenerational correlation between parents and children in years of schooling offers a more robust alternative. However, years of schooling may capture only part of the overall variation in skills \cite{hanushek2020education}, an issue we return to below. See also \citeN{Hertzetal2008} for additional evidence on intergenerational correlations in schooling for a large set of countries.} 
We observe a clear positive relationship, both within developing and developed countries.\footnote{See also \citeN{NEIDHOFER2018}, who find a pronounced positive relationship between income inequality and educational persistence across 18 Latin American countries, and \citeN{blanden2013cross} who shows that the Great Gatsby Curve also holds for intergenerational persistence in education.} These results are in line with the findings of \citeN{JerrimMacmillan2015} who, using data from the International Assessment of Adult Competencies (PIAAC), study the role of education in explaining the Great Gatsby Curve. They show that not only the intergenerational correlation in schooling but also the returns to schooling increase with income inequality, with both channels contributing to the observed Great Gatsby Curve in income. 

The cross-country relationship between income inequality and intergenerational mobility is therefore stronger for educational attainment than for educational performance. The model provided in Section~\ref{sec:model} provides one potential explanation for this pattern: if financial markets are imperfect, educational investments increase with parental resources, even conditional on a student's academic performance. The contrasting pattern in \Cref{fig:GG_Edu} could therefore be indicative of differences in educational investments and choices net of performance in high school being an important driver of the Great Gatsby Curve in income.\footnote{This interpretation is furthermore consistent with the observation that the share of public expenditures in schooling is negatively correlated with income inequality \cite{Rauh2017}.} That said, comparisons based on PISA are still hampered by data limitations \cite{zieger2020conditioning}, and more research is needed to confirm the relation between socio-economic gaps in test scores and income inequality.

Drawing firm conclusions from international comparisons is, of course, challenging given the myriad ways in which countries differ \cite{durlauf2018understanding}. Following \citeN{chetty2014land}, recent studies have instead estimated intergenerational mobility at the local level within particular countries, so as to assess whether inequality and mobility are related when other institutional aspects are held constant.\footnote{See \citeN{DeutscherMazumder2020Labour} and \citeN{MazumderDeutscher2021} for Australia, \citeN{Leone2019} for Brazil, \citeN{ConnollyHaeckLapierre2019}, \citeN{ConnollyCorakHaeck2019JOLE} and \citeN{Corak2017Landscapes} for Canada, \citeN{fan2021rising} for China, \citeN{ERIKSEN2020109024} for Denmark, \citeN{BellBlundellMachin} for England, \citeN{Dodinetal2021} for Germany, \citeN{GuellPellizzariPicaMora2018} and \citeN{Violante2020} for Italy, \citeN{ButikoferDallaZuannaSalvanes2018} and \citeN{Risa:2019aa} for Norway, \citeN{LlanerasAtlas2020} for Spain, \citeN{Heidrich:2017aa}, \citeN{Branden2019} and \citeN{NyStu2021Geography} for Sweden, and \citeN{Aydemir2019160} for Turkey.}  Generally, this data confirms the Great Gatsby hypothesis of a relation between income inequality and income mobility. For example, comparing regional estimates from Italy, Spain, Sweden, and the Netherlands, \citeN{FregoniHavariStuhler2021} report a pronounced negative relation for all four countries. A fruitful direction for future work would be to explore whether these within-country Great Gatsby Curves are more pronounced for educational mobility than for income mobility, as appears to be the case for between-country comparisons.

\subsection{Economic Inequality and Educational Inequality over Time} \label{subsec:inequalitytime}

In most high-income countries, economic inequality has risen substantially in recent decades (see for example \citeNP{KRUEGER20101}). If the cross-sectional relationship between inequality and social mobility also held within countries over time, we would expect social mobility to have declined over the same period. However, the evidence to this regard is mixed. It is useful here to distinguish between trends in educational inputs and those in educational outcomes. In terms of inputs, studies indeed show widening inequality and hence lower mobility over time. \citeN{ramey10} document, for example, that more educated parents in the United States have increased the time spent on child care much more than less educated parents, and monetary investments are likewise diverging between families at different places along the income scale (\citeNP{corak2013income}; \citeNP{KornrichFurstenberg2012}; \citeNP{SchneiderHastingsBriola2018}; see also \Cref{subsec:econ_edu_inequality} for a more detailed discussion). 

The picture is less clear regarding educational outcomes. In an influential study, \citeN{Reardon2011} argues that achievement gaps in the United States have grown considerably over the last 50 years, and are 30 to 40 percent larger among children born in 2001 than among those born twenty-five years earlier. In contrast, \citeN{hanushek2019unwavering} estimate that achievement gaps between the top and bottom quarters of the SES distribution have remained remarkably constant. Their contrasting findings can be traced to differing data sources and definitions of family background. While \citeN{Reardon2011} combines many different achievement tests and considers parental income, \citeN{hanushek2019unwavering} restrict their analysis to sources with more comparable variable definitions and use information on parental education and home possessions rather than income. Meanwhile, in combining 30 international large-scale assessments over 50 years, representing 100 countries, \citeN{chmielewski2019global} finds that achievement gaps have increased in most countries.  \citeN{jerrim2012socio} instead documents a narrowing of PISA achievement gaps in England and several other OECD countries. 

The evidence regarding educational attainment is also mixed. In a broad analysis of educational mobility in 42 countries, \citeN{Hertzetal2008} find that the intergenerational correlation in years of schooling remained stable over the late 20th century, while the corresponding regression coefficient decreased markedly.\footnote{The regression coefficient is directly scaled by the variance of schooling, and thus can change rapidly in response to compulsory schooling requirements, educational expansions, or other policy changes. See, for example, \citeN{NybomStuhlerTrends2014} and \citeN{karlson2021making}.} In an updated and extended analysis, \citeN{narayan2018fair} report that the correlation coefficient has decreased slightly in high-income countries but remained stable or increased in the developing world. Other studies have instead focused on absolute attainment gaps. For example, \citeN{duncan2017} show that in the United States, the gap in educational levels between children from richer and poorer families grew from the 1970s to the 2000s, and argue that most of this rise can be attributed to increasing income inequality. Others look at intermediate outcomes, such as primary or secondary school completion. Such outcomes can be measured at an earlier age and are therefore observable in standard household data, without the need to link parents and children across households. For example, \citeN{Dodinetal2021} exploit the fact that in Germany only the highest secondary school track (\emph{Abitur}) grants direct access to the university system, such that adolescent track choices are predictive of economic opportunities later in life. 

The choice of inequality measure matters, as educational expansions are associated with systematic changes in both the average level and variance of schooling \cite{ram1990educational}, which affects some inequality measures more than others. For example, secular trends in the probability of obtaining a primary or secondary degree could also influence the intergenerational measures that are based on them.\footnote{See, for instance, \citeN{Dodinetal2021}, who show that in the context of rising \emph{Abitur} shares in Germany, absolute gaps by parental income have remained stable while the Q5/Q1 ratio---the share of children with a higher secondary school degree from the top quintile of the parental income distribution divided by the corresponding share in the bottom quintile---has decreased. Similarly, \citeN{blanden2016educational} show that much of the reduction in inequality in educational attainment among recent cohorts in the UK is explained by the rising share of young people attaining educational thresholds previously only achieved by a minority.} In particular, correlation estimates that abstract from changes in the variance of schooling tend to be more stable over time than those that do not, such as regression estimates \cite{Hertzetal2008}.  

Overall, evidence for a systematic decrease in relative educational mobility in response to recent increases in income inequality is mixed. This observation stands in contrast to the evidence on increasing inequalities in educational inputs, the more robust cross-sectional relation between inequality and mobility, and the theoretical models discussed in the next section. It is possible that definitions of student achievement and socio-economic gaps measured at different points in time may not be sufficiently comparable. Either way, addressing the contrast between cross-sectional and time-series empirical results and theoretical predictions remains a challenge for further research on the question.

\subsection{How Persistent Are Educational Inequalities?}\label{subsec:Sec2multigenerational}

Knowledge about the extent to which educational advantages and disadvantages persist is mainly based on simple summary statistics, such as the parent-child correlation in years of schooling (as shown in the right panel of \Cref{fig:GG_Edu}). 
These measures of mobility over a single generation (from parent to child) would, if they applied independently to each successive generation, imply that educational mobility over multiple generations is very high: the descendants of poor and rich families who lived, say, 150 years ago should have roughly the same average education and income today. 

Nevertheless, there are several reasons why conventional parent-child measures may not capture the full extent to which inequalities are transmitted across generations. First, the descriptive association between parent and child education is only indirectly related to the structural mechanisms through which educational inequalities are transmitted and may therefore not be informative about the extent to which educational inequalities persist in the long run, across multiple generations.\footnote{Relatedly, the causal effect of parents’ on child’s schooling is much smaller than the descriptive association between the two (\citeNP{HolmlundLindahlPlug2011}, \citeNP{BjoerklundJaentti2020}).} Second, socio-economic gaps arise in all stages of educational attainment, such that years of schooling are only a coarse proxy of learning and educational achievements (as noted earlier). Parent-child correlations might therefore not only understate long-run persistence, but also the persistence of educational inequalities across even just one generation, from parents to children.

\begin{figure}[htp]{}
    \centering
    \includegraphics[width=0.7\linewidth,trim=1 1 1 1,clip]{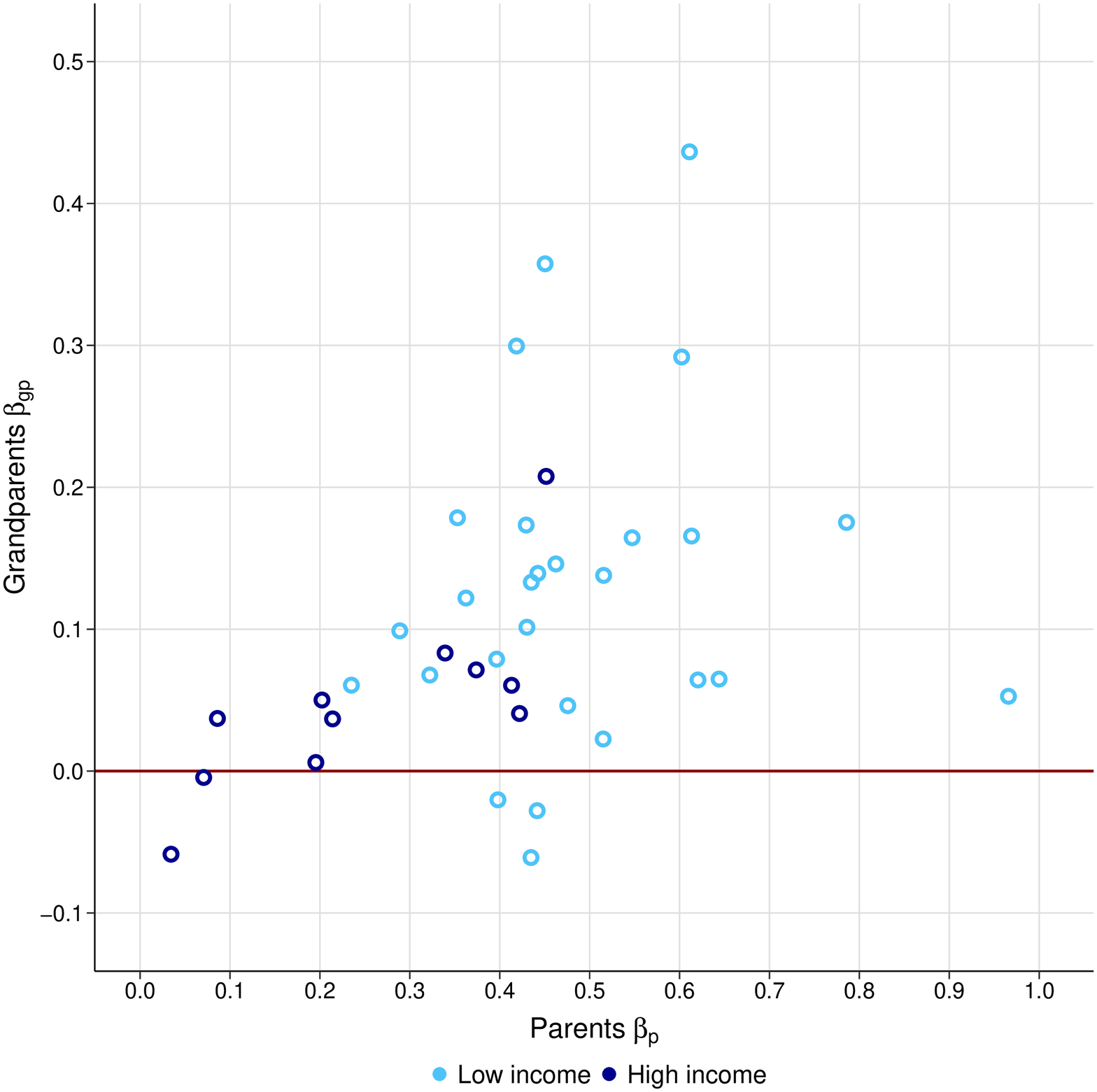}
    \caption[Three generations]{Educational Mobility across Three Generations.\\ 
    \\
    \emph{Source:} Global Database on Intergenerational Mobility, \citeN{GDIM2018}. The figure reports the estimated coefficients on the average education of parents and grandparents in samples with at least 200 observations.}
    \label{fig:3gScatter} 
\end{figure}

Recent studies confirm that parent-child associations are indeed missing part of the picture, and show that mobility across multiple generations is lower than a naive extrapolation from conventional parent-child measures would suggest. One way to demonstrate this point is to note that educational outcomes of ancestors remain predictive of child education, even after conditioning on parent education (e.g., \citeNP{LindahlPalme2014_IGE4Generations}, \citeNP{BraunStuhler2018}, \citeNP{colagrossi2020like}). Specifically, in the regression of years of schooling $y_{it}$ in family $i$ in generation $t$ on parent and grandparent schooling,
\begin{equation}
\label{eq:reg3g} y_{it}=\alpha + \beta_p y_{it-1} + \beta_{gp} y_{it-2} + \varepsilon_{it},
\end{equation}
the coefficient on grandparent schooling $\beta_{gp}$ tends to be positive and, often, sizable.\footnote{The observation that $\beta_{gp}$ is positive is equivalent to the observation that multigenerational correlations decay at less than the geometric rate, such that a simple iteration of the parent-child correlation would understate multigenerational persistence \cite{BraunStuhler2018}.} To illustrate this, Figure \ref{fig:3gScatter} plots estimates of the coefficients $\beta_p$ and $\beta_{gp}$ from equation (\ref{eq:reg3g}) as reported in the Global Database for Intergenerational Mobility (\citeNP{GDIM2018}). With few exceptions, the coefficient estimates of $\beta_{gp}$ are positive. While some of these estimates are based on small and potentially selective (e.g., co-resident) samples, the resulting distribution appears representative of the range of estimates researchers find in more targeted studies. Summarizing the results from 40 different studies, \citeN{AndersonSheppardMonden2017} report that estimates of the coefficient $\beta_{gp}$ tend to be one quarter the size of the coefficient $\beta_p$, as is the case for the estimates reported in \Cref{fig:3gScatter}. 

\captionsetup[figure]{font=footnotesize,labelfont=footnotesize}

That parent-child correlations may understate the role of family background has already been observed by \citeN{bjorklund2011education} in an earlier volume of the Handbook of the Economics of Education. These authors relate sibling to intergenerational correlations and compare their size across countries, concluding that parental schooling accounts for but a minor part of the factors that siblings share, and that neighborhood effects explain only some of the remainder. Indeed, \citeN{BjorklundJaentti2012aa} argue that intergenerational correlations represent only the ``tip of the iceberg'' of family background effects. An inability of conventional parent-child measures to accurately capture the extent to which advantages are transmitted across generations is also consistent with recent studies relying on name-based estimators. Following \citeN{Clark2014book}, such work examines the persistence of educational and socio-economic inequalities on surname levels. For example, using historical census data from Florence, \citeN{BaroneMocetti2016} find persistence in earning advantages over nearly six centuries.\footnote{See \citeN{SantavirtaStuhler2021} for a review of different name-based estimators and a list of recent contributions.}
 
Different methods therefore point to the same conclusion: traditional parent-child correlations understate the transmission of advantages and the persistence of educational inequalities across multiple generations. Why is this the case? The literature has focused on two broad classes of explanations: (i) educational advantages or their underlying determinants might be imperfectly observed in the data, and (ii) grandparents or extended family may have an independent causal influence on child education, over and above that of the child's parents. While these two interpretations are different, both imply that researchers could gain a deeper understanding of the persistence of educational inequality by considering broader kinship, as opposed to the direct connection from parent to child. We return to this argument in Section \ref{sec:model_lr}.

\section{A Model of Skill Acquisition and Educational Inequality}

\label{sec:model}

Our empirical analysis highlights various dimensions of educational inequality among children and young adults. The observation of pervasive educational inequality gives rise to a number of questions: What specific mechanisms are responsible for unequal outcomes? How do institutions, macroeconomic conditions, and inequality in other variables affect the degree of educational inequality? Are policy interventions that push back against educational inequality needed, and if so, which might hold the most promise?

To organize our discussion of the literature addressing these questions, we present a model of human capital accumulation that captures multiple stages of skill acquisition and a variety of inputs (as in \citeNP{CH07}, \citeNP{CHS10}, and \citeNP{AW16}, among others).\footnote{The importance of different stages of child development has long been recognized in the child development literature \cite{E50}, and has recently been widely adopted in the economics literature.} 

\subsection{Setup}

Our model considers a family comprising a mother $f$ (female) and father $m$ (male) with skills $S_f$ and $S_m$ who are raising one child. The child's skills evolve over two periods, early and late childhood. The skills of children and adults can be multi-dimensional vectors and include both cognitive and non-cognitive skills. The initial skills or endowments of the child at the beginning of childhood $s_0$ are a function of parental influences and luck:
\[
s_0=f_0(S_f,S_m,\epsilon_0),
\]
where $\epsilon_0$ is a random term, and $f_0(\cdot)$ is increasing in each of the arguments. 

In early childhood, the child's skills evolve according to the skill accumulation technology:
\begin{equation}\label{eq:ed_tech1}
s_1=f_1(s_0,I_0,E_0,N_0,\epsilon_1)
\end{equation}
Here $I_0$ are parental investments, $E_0$ are non-parental inputs such as preschool, $N_0$ are environmental influences such as peer effects, $\epsilon_1$ is a random term, and $f_1(\cdot)$ is  increasing in each argument.

In late childhood, the skill accumulation technology takes the form:
\begin{equation}\label{eq:ed_tech2}
s_2=f_2(s_1,I_1,E_1,N_1,\epsilon_2)
\end{equation}
where $s_2$ are the child's skills at the beginning of adulthood and $\epsilon_2$ is a random term. Once again, we assume that skills  $f_2(\cdot)$ are an increasing function of each of the arguments.

Parental utility depends on consumption and leisure and on the skill accumulation of the child. We model the way the child's skills enter the parent's utility function as:
\[
W(s_2,X)
\]
where $X$ captures economywide variables that capture the importance of skills, such as returns to education. The parents agree on their objectives and maximize a joint utility function. The parents' full expected utility is given by:
\[
V=E\left[U(C_0,C_1,L_0,L_1)+z W(s_2,X)\right]. 
\]
Here $C_t$ and $L_t$ are parental consumption and leisure in the two periods $t\in\{0,1\}$, and $z$ measures altruism, i.e., the weight at which the child's welfare enters the parents' decision problem. The parents maximize utility subject to the constraint set:
\[
g(C_0,C_1,L_0,L_1,I_0,I_1,S_f,S_m,Y_0,Y_1,X)=0.
\]
Here $Y_0$ and $Y_1$ represent parental income in the two periods. The constraint set can include regular budget constraints for monetary investments in children, time constraints, and also knowledge or ability constraints that depend on parental skills.

\subsection{Sources of Educational Inequality}

\label{sec:sources}

Our setup allows for a number of sources of educational inequality. Inequality can arise in terms of the actual skills acquired at different ages ($s_0$, $s_1$, and $s_2$) and in terms of the inputs and influences from parents, educational institutions, and the environment. When focusing on inequality in the overall skills acquired by the end of childhood $s_2$, we can identify the following sources of inequality:
\begin{enumerate}
    \item \textbf{Inequality in parental skills}. Parents are heterogeneous in initial skills $S_f$ and $S_m$. For given other inputs, inequality in parental skills will result in inequality in children's skills.

    \item \textbf{Assortative mating}. A second source of educational inequality is assortative mating. If, for given inequality in the mother's skill $S_f$ and other inputs, there is a positive correlation between the two parents' skills, educational inequality will be higher.  

    \item \textbf{Inequality in parental inputs}. Educational inequality can also arise from inequality in parental investments $I_0$ and $I_1$. These investments are choices; inequality in investments are a proximate cause of educational inequality, but there are also ultimate causes for why different parents make different choices. Relevant factors include:
    \begin{enumerate}
        \item \textbf{Economic inequality.} Inequality in parental income $Y_0$ and $Y_1$ determines how affordable parenting investments are for parents of greater and lesser means. These income differences, in turn, may depend on educational, wage, and wealth inequality across parents. They can also arise from factors such as single parenthood, which may limit parental resources.
        \item \textbf{Inequality in parental ability and skills.} The parental skills $S_f$ and $S_m$, in addition to having a direct impact on the child's initial skills $s_0$, may also affect parents' relative ability to provide effective educational investments.
        \item \textbf{Inequality in preferences and aspirations.} Parents may differ in their overall degree of altruism towards children $z$, which translates into their willingness to invest in them. Moreover, there may also be inequality in parents' aspirations for their children; for example, how they weigh economic success versus other aspects of their children's quality of life. Such differences would be captured by variation in the function $W(\cdot)$.
        \end{enumerate}
        The importance of these factors for inequality in parental investments may also depend on aggregate conditions $X$. For example, if market returns to education are high, parents will generally be more motivated to make investments in their children's skills. This, in turn, can make economic constraints that vary across parents more binding, accentuating educational inequality.
    
    \item \textbf{Inequality in educational inputs}. There can also be variation in the inputs $E_0$ and $E_1$ provided by educational institutions. In settings where some of these inputs are paid for directly by parents (i.e., private schools and preschools), the same determinants that also matter for parental inputs apply here as well. If these inputs are publicly provided, the organization and financing of the public education system matter for educational inequality. 

    \item \textbf{Inequality in environmental influences}. Lastly, educational inequality may arise from variation in the environmental influences $N_0$ and $N_1$. This variation may again depend in part on choices made by parents, such as which neighborhood to live in and which school to send their children to. Public policy also matters, for example when it comes to policies regarding school choice, housing policy, and taxation (e.g., the extent to which educational expenditures are paid for via local property taxes, as in the United States). 

\end{enumerate}

Few, if any, channels work independently of the others. Perhaps the strongest case for an independent channel behind educational inequality is the ``nature'' aspect of ability, i.e., genetic variation among children as one determinant of educational ability and achievement. But even this genetic variation is subject to economic mechanisms, such as the extent of assortative mating among parents. The choice of who to have children with is arguably in part determined by how a potential partner's characteristics matter for children's future success. For example, if returns to education are high, people should be more motivated to find a partner with high academic ability, which would make sorting by academic ability more assortative. \citeN{mare2016educational} and \citeN{lundberg_et_al_2016} argue that a mechanism of this kind is behind the rising marriage gap by education in the United States, where college graduates are now both more likely to marry each other and to get married in the first place. 

The interdependence of the various channels begs further reflection on the deeper drivers of educational inequality. To this end, we first examine the impact of economic inequality (i.e., income and wealth inequality) on educational outcomes. We then focus specifically on neighborhood and peer effects and on the implications of the  dynamic multi-stage skill acquisition technology \eqref{eq:ed_tech1}--\eqref{eq:ed_tech2} for the effects of early versus late policy interventions. Lastly, we discuss evidence on the determination and importance of parental investments $I_0$ and $I_1$ and on the role of school inputs $E_0$ and $E_1$ for overall educational inequality.

\subsection{Economic Inequality and Educational Inequality}\label{subsec:econ_edu_inequality}

The model highlights that parental inputs arise from decisions that depend on the economic benefit that children will derive from education. This decision problem implies that changes in the economic environment will feed back into parental choices. A possibility that is particularly relevant for educational inequality is that economic inequality can shape the inequality in investments of parents with different socio-economic backgrounds.

To illustrate this possibility, consider a special case of the model where skills are determined entirely by the parental investment $I_1$ in late childhood. There is a single parent with skill $S$ and only two skill levels: low and high, $S,s_2\in\{L,H\}$. The parent derives period utility from consumption according to felicity $U(C_1)$, and the only constraint for the parent is a budget constraint $C_1=Y_1(S,X)-I_1$, where the dependence of income $Y_1$ on skill $S$ and aggregate conditions $X$ captures the impact of skill returns on the income of parents with low and high skill. The variable $X$ here corresponds to the premium for high-skill labor. Accordingly, we set the income of workers with low and high skill to $Y(L,X)=\bar{w}$ and $Y(H,X)=\bar{w}+X$, respectively. Finally, the parent's concern for their child's future outcome $W(s_2,X)$ takes the form of a warm-glow preference derived from the child's future wage, which depends on the returns to skill in the same way as adult income. Hence, if the child ends up with low skill, $s_2=L$, we have $W(L,X)=\bar{w}$, and for a high-skill child $s_2=H$ the parent derives utility 
$W(H,X)=\bar{w}+X$ from the child. We let the probability that the child will end up with $s_2=H$ be given by $\min\{ I_1,1\}$.

The parent's decision problem of choosing the investment $I_1$ can then be written as:
\[
\max_{I_1}\left\{  U(C_1)+z \left(\bar{w}+ I_1 X  \right)  \right\}
\]
where $U(\cdot)$ is an increasing and strictly concave utility function and the maximization is subject to 
\[
C_1=Y_1(S,X)-I_1.
\]
The first-order condition for this decision problem is:
\[
U'(C_1)=z X,
\]
that is, the parent equates the marginal utility of consumption (i.e., the disutility of reduced consumption from investing more in their child's education) to the marginal benefit of a greater probability of the child reaching a high level of human capital, which depends on returns to education $X$. Plugging in the values of $C_1$ and income $Y_1(S,X)$ and writing the optimal investment of a parent with skill $S\in\{L,H\}$ as $I_1(S,X)$ we get:
\begin{align}
\label{eq:focL}U'(\bar{w}-I_1(L,X))=z X,\\    
\label{eq:focH}U'(\bar{w}+X-I_1(H,X))=z X.
\end{align}

We can now establish the following relationship of aggregate returns to education, parental investments, and educational inequality:

\begin{prp}[Impact of Returns to Education on Educational Inequality] \label{prop:ec_ed_ineq}
If the solution for parental investment $I_1(S,X)$ is interior for both low- and high-skill parents, a marginal increase in returns to education $X$ will:
\begin{itemize}
    \item Raise the educational investment $I_1(S,X)$ of all parents.
    \item Increase the difference $I_1(H,X)-I_1(L,X)$ between the (higher) educational investment of high-skill parents and the (lower) investment of low-skill parents.
\end{itemize}
\end{prp}

\begin{prf}
The first part follows directly from the first-order conditions \eqref{eq:focL} and \eqref{eq:focH}: if $X$ increases on the right-hand side, marginal utility has to increase on the left-hand side, which implies that $I_1(S,X)$ rises. The second part holds because an increase in $X$ raises the income of high-skill relative to low-skill parents. The first-order condition \eqref{eq:focH} shows that this results in an additional increase in the investment $I_1(H,X)$ of high-skill parents by the increase in $X$.
\end{prf}

The intuition for the positive impact of the return to education on parental investment is straightforward; the parent cares about the child, and if an increase in returns makes education more valuable, parents will work harder to give their children an extra push. The same effect would arise if an altruistic parent cares about the full utility, rather than just the income, of the child (see for example \citeNP{dozi17_ECTA}). The effect on educational inequality arises from the parental budget constraint. If returns to education rise, the income of highly educated parents increases relative to that of others with less education. This rise in income increases consumption, lowers the marginal utility of consumption, and hence lowers the marginal cost of investing in children's education. 

The results in Proposition~\ref{prop:ec_ed_ineq} provide a possible explanation for the general trends in parental investments in children's education that have been observed in a number of countries since the 1970s. As discussed in Section~\ref{sec:evidence}, in most high-income economies, economic inequality, including education wage premia, increased throughout the 1980s and 1990s, with particularly large changes in the United States and the United Kingdom. Meanwhile, time use studies show that parents today spend many more hours caring for their children compared to the 1970s, particularly when this concerns time spent on education-oriented activities, such as helping children with homework \cite{dozi19}. Our model of parental investments in children's education suggests that at least part of this shift can be seen as a response to a changed economic environment where succeeding in formal education is much more highly rewarded than in earlier times. 

Likewise, in countries where economic inequality has risen quickly, there is clear evidence of increasing inequality in educational inputs. For the United States, \citeN{ramey10} show that in recent decades more educated parents (those with at least some college education) have increased the time spent on child care much more than less educated parents. In the 1970s, there was little difference in child-care time between these groups, but by the 2010s college-educated parents spent more than three additional hours each week interacting with their children. Trends for monetary investments in children are similar. \citeN{corak2013income}, \citeN{KornrichFurstenberg2012}, and \citeN{SchneiderHastingsBriola2018} observe a large increase in the gap in spending on children between high- and low-income households from the 1970s to the 2000s. The overall rise in spending is driven by households in the top quarter of the income distribution, whereas spending has actually declined recently among households in the bottom quarter of the income distribution. These trends mirror household income itself, which has for decades stagnated for households below the median, while simultaneously growing quickly for well-educated households at the top of the income distribution. 

Another facet of parental investment that responds to economic inequality is parenting style. \citeN{dozi17_ECTA} and \citeN{doepke_et_al_2019} develop models of parenting similar to the setup considered here where parents also face a choice between different parenting styles. The models emphasize potential conflicts and disagreements between parent and child. For example, the parent may want the child to put more effort into homework and studying, whereas the child may prefer to go play with friends. Parenting style refers to how such conflicts are resolved. A permissive parent gives much freedom, allowing the child to choose for herself; an authoritarian parent exerts control and demands obedience; and an authoritative parent aims to convince the child to adopt the parent's preferred behavior. 

The models of \citeN{dozi17_ECTA} and \citeN{doepke_et_al_2019} imply that the choice between these styles depends on inequality. Specifically, when income inequality is high (including a large wage gap between those with greater and lesser educational attainment), parents are more concerned about their children's success, and hence are more likely to adopt a more intensive parenting style (i.e., authoritative or authoritarian) that pushes children towards high achievement in education. In line with these predictions, \citeANP{dozi17_ECTA} (\citeyearNP{dozi17_ECTA}, \citeyearNP{dozi19}) use evidence from the World Values Survey to show that in countries with high income inequality, there are more authoritarian and authoritative parents and fewer permissive ones. Parenting styles also respond to changing inequality within countries. Namely, if income inequality goes up over time, so too does the fraction of parents who employ one of the intensive styles of parenting.

The impact of economic inequality on educational inequality extends beyond parental investments. As spousal correlations tend to be high \cite{fero00}, assortative mating is also an important driver of educational inequality.  In the longitudinal data sets used in \Cref{sec:evidence}, the spousal correlation in years of schooling varies between 0.45 and 0.6, and is similarly high in many other countries \cite{fegu01}.\footnote{Moreover, a comparison of spouses and more distant in-laws implies that assortative mating in educational advantages is even stronger than that captured solely by spousal correlation in years of schooling \cite{ColladoOrtunoStuhler2019aa}. We return to this question in \Cref{sec:model_lr}, where we discuss recent evidence on the persistence of educational advantages across multiple generations.} Notably, marital sorting is stronger in the United States (spousal correlation of 0.6) than in England, Germany, and Australia, consistent with the idea that such sorting may be strongest in high inequality environments. Indeed, \citeN{fegu01} show that the spousal correlation in years of schooling tends to be greater in countries where skill premia and income inequality are high. Moreover, recent evidence suggests that mating has become more assortative over time in the United States as inequality has risen (\citeNP{Greenwoodetal2014}, \citeNP{Greenwood2016}, \citeNP{Chiapporietal2020}). 

The feedback from aggregate inequality to educational inequality can be further amplified if there are bottlenecks in the education system. To illustrate this possibility, consider a variant of the model above in which the number of children who are able to attain high skill is fixed, say because of a set number of slots at university. Higher parental effort $I_1$ still increases the chance that a child will become high-skill, but if all parents increase their parenting effort, this will raise the bar for success. Consider an economy with $P$ parents of each type. The probability that a child will achieve high skill is now given by $\psi I_1$, where $\psi$ is taken as given by parents, and in equilibrium is equal to:
\begin{equation}
\label{eq:psi}
\psi=\frac{\phi}{P\left(I_1(L,X)+I_1(H,X)\right)},
\end{equation}
where $\phi$ is the fixed number of children who end up with a high level of education (the bottleneck). In this setting, the parenting investments of one group of parents impose an externality on other parents, because the more they invest the higher the bar for success. We can now show:

\begin{prp}[Impact of Returns to Education on Educational Inequality with Bottlenecks] \label{prop:ec_ed_ineq_bottleneck}
In the model with a fixed number of children who can attain high skill, if the solution for parental investment $I_1(S,X)$ is interior for both low- and high-skill parents, a marginal increase in returns to education $X$ will:
\begin{itemize}
    \item Raise the educational investment $I_1(H,X)$ of high-skill parents.
    \item Increase the difference $I_1(H,X)-I_1(L,X)$ between the (higher) educational investment of high-skill parents and the (lower) investment of low-skill parents.
    \item Increase the probability that a child of a high-skill parent will attain high skill, and lower the probability that a child of a low-skill parent will attain high skill.
\end{itemize}
\end{prp}

\begin{prf}
The first-order conditions for investment \eqref{eq:focL} and \eqref{eq:focH} now read:
\begin{align*}
U'(\bar{w}-I_1(L,X))&=z \psi X,\\    
U'(\bar{w}+X-I_1(H,X))&=z \psi X,
\end{align*}
where the endogenous productivity of investing in skill $\psi$ now appears on the right-hand side (in addition to the altruism factor $z$). Given that the utility function is strictly concave, $U'$ can be inverted and we can write optimal investment at an interior solution as:
\begin{align*}
I_1(L,X)&= \bar{w}- (U')^{-1}( z \psi X),\\    
I_1(H,X)&=\bar{w}- (U')^{-1}( z \psi X)+X.
\end{align*}
Hence, the difference between the investment of high-skill and low-skill parents is given by $I_1(H,X)-I_1(L,X)=X$ and thus increases in $X$. The third part of the proposition follows as given the bottleneck, any increase in the probability of the child of a high-skill parent attaining high skill has to be matched by a corresponding decrease in the probability that the child of a low-skill parent will attain high skill. Lastly, the first part of the proposition follows because if both types of parents lowered their investment, given \eqref{eq:psi} $\psi$ would rise, but then given the results of Proposition~1 (which continue to apply with $\psi X$ taking the role of $X$ in Proposition~1) all parents would like to invest more. Hence, at least high-skill parents have to increase their investment. The absolute change in the investment of low-skill parents is ambiguous, because even though $\psi$ has to decrease, the product $\psi X$ may either rise or fall.
\end{prf}

Hence, a rise in economic inequality not only increases inequality in parenting, but results in an absolute decrease in the chance that a child from a poorer family will succeed. For poorer families, the increase in parenting investments of rich families leads to a discouragement effect, which lowers their incentive to invest and further widens educational inequality. 

In reality, bottlenecks are not as narrow as in this example, where the number of children who achieve high skills is fixed independently of parents' investments and children's achievements. Yet, aggregate feedback mechanisms of this kind arise not just from fixed slot constraints, but also from general equilibrium effects on the price of skill and on the cost of education. For example, if one of group of parents redoubles their investment and demand for college education among their children consequently rises, all else equal college tuition will rise and returns to college will decline once the increased number of college graduates enters the labor market. Both changes have a negative impact on other parents' incentives to make educational investments.

Moreover, actual bottlenecks do exist in certain areas, such as admission slots in top universities (as discussed in Section~2). \citeN{ramey10} cite competition for college admission as one reason for the widening disparity in parenting investments observed in the United States. Notably, few offered slots at top universities can be artificial, in that they may reflect a strategic choice on the part of the institution rather than true scarcity. \citeN{BlairSmetters21} document that even though the total number of students in college more than doubled in recent decades in the United States, elite schools have not expanded enrollment, choosing instead to admit a smaller fraction of applicants. They argue that this outcome reflects an inefficient competition among colleges for prestige, with the end result being that few students get to attend a top institution. In other countries, high-ranking schools have among the largest enrollments (e.g., the University of Toronto in Canada), and hence competition for admission is less fierce. In places such as China and South Korea, a similar bottleneck is created by national entrance exams that govern admission to the best public universities. Entrance to these institutions goes hand in hand  with high parenting investments (such as paying for ``cramming schools''), leading to concerns over a lack of opportunity for students from disadvantaged backgrounds whose parents cannot afford such investments. 

In a more general model, the impact of a rise in $X$ would also depend on the functional form through which parental investment $I_1$ affects children's skills $s_2$ (which is linear in our example). If returns to parental investments are strongly diminishing, a rise in $X$ could result in a rising gap in parental investment (as in Propositions 1 and 2) but little or no change in the skills gap between children of high- and low-skill parents. Such a model could help address the observation in Section~\ref{subsec:inequalitytime} that, while there is clear evidence of larger gaps in parental investments as economic inequality has risen, the evidence for the impact of rising inequality on gaps in skills and attainment is more ambiguous. 

\subsection{From Educational Inequality to the Great Gatsby Curve}

The results from Propositions 1 and 2 provide a possible rationale for the empirical phenomenon of the Great Gatsby Curve discussed in Section~\ref{subsec:mobilityinequality}. A rise in the return to education $X$ would increase the gap in parental investments between low- and high-skill parents, and thereby contribute to widening the gap in outcomes among their children. Moreover, the direct impact of $X$ on the income gap between workers with less and more skill would further accentuate the income gaps between workers of different education levels. 

Several recent studies use richer versions of the model of skill acquisition presented here to examine such a link between educational inequality and social mobility. \citeN{herrington2015}, for example, develops a model with multiple stages of human-capital investment to analyze differences in social mobility between the United States and Norway, two countries with particularly high (United States) and low (Norway) income inequality. Consistent with the Great Gatsby curve, social mobility is higher in Norway, a fact that is matched by the quantitative model. Herrington uses the model to examine the role of taxation and public education spending policies in generating the Great Gatsby curve, and finds that these policy dimensions account for about one-third of the difference in income inequality and 14 percent in social mobility. Spending on early childhood education turns out to have a powerful impact on educational inequality, and also amplifies the effect of other policies.

The model of \citeN{lee_seshadri_2019} features parental investments in early and late childhood, a college decision, and additional accumulation of skills on the job. The model is consistent with empirically observed intergenerational elasticities in terms of lifetime earnings and college attainment. Similar to \citeN{herrington2015}, the authors find that parental background has a particularly large impact on early investments in young children. Policy interventions at this early stage, when poorer parents face considerable constraints, are predicted to increase social mobility.

In a similar setting, \citeN{daruich2018} notes that the impact of such early policy interventions may increase across generations. In addition to the direct impact on children receiving more education, the future earnings generated by this additional education will also induce this cohort to invest more in their own children. Hence, the beneficial effects are propagated to the grandchild generation.

These studies agree on the basic insight that economic inequality and social mobility are closely linked through the determination of investments in the skill acquisition process, along the lines of our stylized model. In addition, they all point to the importance of the timing of investments in skills and public education funding, which we will further discuss in Sections~\ref{sec:peersneighborhood}, \ref{sec:early_late}, and \ref{sec:compensating_investments} below. 

\subsection{The Role of Peers and Neighborhoods}
\label{sec:peersneighborhood}

Thus far, we have focused on a stripped-down version of our model, where economywide inequality affects educational inequality exclusively through effects on parental income $Y$ and parental investments $I_1$. In the general version of our model, additional feedback effects can arise through the other inputs in child development. In particular, the technology for skill acquisition \eqref{eq:ed_tech1}--\eqref{eq:ed_tech2} also allows for educational inputs $E_0$ and $E_1$, which stand for factors such as the quality of schooling, and for environmental influences $N_0$ and $N_1$. Like the direct parental investments $I_0$ and $I_1$, these additional factors respond at least in part to aggregate economic conditions and overall inequality. 

The environmental inputs $N_0$ and $N_1$ capture neighborhood and peer influences in school and beyond. Inequality in these inputs is related to inequality in schooling inputs: if parents choose to put their children in private school or move to an expensive neighborhood because of the availability of high-quality public schools, this will also affect which peers and local role models children are exposed to. Research by \citeANP{CH18a} (\citeyearNP{CH18a}, \citeyearNP{CH18b}) suggests that such local inputs do matter, as prospects for upward mobility differ systematically across neighborhoods and regions in the US data.\footnote{As the degree of segregation is itself a function of the level of cross-sectional inequality \cite{durlauf2018understanding}, the choice of school or neighborhood is likely less important in countries characterized by lower spatial inequality. For example, \citeN{Heramnsenetal2019} find that school and neighborhood correlations are small and declining in Norway.}

Beyond the choice of neighborhood and school, parents can also more directly intervene in the peer group formation of their children, for example by signing them up for specific extracurricular activities or by actively discouraging them to socialize with specific peers. As noted by \citeN{ADSZ2020}, these choices can be interpreted as parents actively weighing the potential benefits and risks of their children interacting with particular peer groups. Learning can be affected by peers both directly, such as children studying together and learning from one another, and indirectly, via their influence on aspirations and social norms. For example, \citeN{BursztynJensen2015} show that students in low-performing peer groups try to avoid social stigma by reducing their educational efforts when those efforts are observable by their peers. Endogenous peer group formation can amplify educational inequality, because homophily bias (the tendency to associate with peers who are similar to oneself) implies that socio-economic differences across families are reproduced in the peer connections children make in schools and neighborhoods \cite{ADSZ2020}.

\citeN{fogli_guerrieri_2018} analyze the relationship between economic inequality and educational inequality in a model with local spillovers in human capital and endogenous residential choices. Rising economic inequality leads to more residential segregation, as richer parents seek out homogeneous neighborhoods with strong peer effects. As a result, educational inequality increases, in turn further widening economic inequality. In a similar setting, \citeN{eckert_kleineberg_2018} show that a policy that equates school funding across neighborhoods has only limited benefits due to the endogenous response of residential segregation. 

As these papers make clear, neighborhoods may affect children through several channels and mechanisms, though empirically distinguishing between them represents a challenge \cite{galster2012mechanism}. One source of evidence on the importance of neighborhoods comes from Moving to Opportunity, a program implemented in five US cities that provided families in public housing with vouchers to move to better neighborhoods. This resulted in an exogenous source of variation in location, and is linked with long-term improvements in outcomes for children who moved \cite{chetty2016effects}. That said, the effects were much smaller than observational differences in outcomes between children across neighborhoods \cite{harding}.

The inputs $E_0$ and $E_1$ depend on the system of education finance. Where public and private schools coexist, richer parents will often prefer high-quality but expensive private schools, whereas children from poorer families attend public schools (\citeNP{crdo03}). If economic inequality between rich and poor families rises, we would expect to see an increase in spending on private schools attended by the rich without a corresponding increase in the quality of public schools. 

Even if most families rely on public school, the system of education finance can still imply a feedback effect from aggregate inequality to educational inequality. In the United States, schools are often financed locally through property taxes, implying that richer neighborhoods have better schools. In such a system, a direct link arises between education spending and the neighborhood effects already discussed.\footnote{\citeN{agrawal2019quantifying} use a structural model to estimate the combined impact of neighborhood and school funding effects, which is revealed to be substantial. Moving from a neighborhood/school combination at the 10th percentile of the distribution to the 90th percentile would increase the probability that a child enrolls in college by 17 percentage points.} When economic inequality rises, so too does inequality between rich and poor neighborhoods, and gaps in school funding will widen. In addition, higher returns to education can incentivize parents to relocate to more expensive neighborhoods with stronger schools (see \citeNP{nech06} for an analysis of the impact of residential sorting on educational inequality). An increase in economic inequality may then result in more residential segregation and, once again, more educational inequality. 

Clearly, policy choices regarding the financing of public schools (e.g., local versus state-level funding) play a large part in determining whether aggregate inequality drives educational inequality through variation in the quality of public schooling. Early formal analyses of these issues is provided by \citeANP{fero98} (\citeyearNP{fero96}, \citeyearNP{fero98}). More recently, \citeN{KoteraSeshadri2017} study the role of school funding in cross-state variation in social mobility in the United States. Building on \citeN{fero03}, they develop a structural model in which funding for public education is decided via majority voting. States where public school funding is decided at the local rather than the state level see more unequal spending on public schools and lower social mobility. 

In a similar vein, \citeN{Zheng2022} develop a life-cycle model of skill acquisition with heterogeneous neighborhoods and endogenous school quality. High house prices prevent low-income households from living in neighborhoods with good schools, which results in low social mobility. In line with the evidence from the Moving to Opportunity program (\citeNP{chetty2016effects}), their model shows that vouchers that give low-income families access to schools in districts with high housing prices can be an effective policy response. The link between  school funding equalization and higher social mobility is supported by the empirical results of \citeN{biasi2022}, who examines the effect of state-level reforms in school financing in the United States and observes that equalized funding increases the social mobility of low-income students.

The relative importance of peers or neighborhoods as compared to parents is difficult to disentangle, as these factors reinforce each other through segregation and the sorting of families. For Denmark, \citeN{Bingleyetal2020} exploit within-family variation in exposure to different neighborhoods, and document that community background accounts for about one-fifth of the sibling correlation in years of schooling.\footnote{These estimates must necessarily be interpreted in the Danish context of extensive redistribution between municipalities and schools. Though, studies comparing sibling correlations with the correlation among unrelated neighbors, such as that by \citeN{SolonPageDuncan2000}, similarly point to a limited contribution of neighborhood factors to educational inequality for other countries.} Moreover, it is difficult to separate the contributions of peers, schools, and neighborhoods. In some cases, institutional factors generate useful variation in specific aspects of the local environment. For example, \citeN{Laliberte2021} take advantage of the fact that school catchment areas in Montreal differ across language groups, and estimate that 50–-70 percent of the benefits of moving to a better area for educational attainment is due to access to better schools.

\subsection{Early versus Late Investments and Dynamic Complementarity}

\label{sec:early_late}

A key feature of our model of skill acquisition is that investments in skills take place in two stages, early and late, governed by the production functions \eqref{eq:ed_tech1} and \eqref{eq:ed_tech2}. This two-stage process allows us to capture the differing role of inputs at different life stages and the dynamic relationship between early and late investments. 

A sizeable literature characterizes the properties of children's skill acquisition process.\footnote{Surveys are provided by \citeNP{HM14} and \citeNP{attanasio15}, and \citeN{Currie2011} summarize the evidence on the impact of early-life influences on children's future outcomes.} In this body of work, findings that are particularly relevant for educational inequality concern the self-productivity and dynamic complementarity of skill acquisition. Self-productivity means that acquiring skills early on facilitates obtaining additional skills at later stages. This feature implies that skill deficits that arise early on in a child's development are difficult to later compensate. Together with evidence that skills are particularly malleable early in life (\citeNP{CH03}, \citeNP{EF05}, \citeNP{HM14}), the self-productivity of skill acquisition points to the important role of skill gaps between young children from different backgrounds in generating overall educational inequality. Conversely, early childhood is likely to be a phase when interventions aimed at compensating deficits are particularly productive. Similarly, dynamic complementarity of skill acquisition means that early investments in skills make later investments more productive. \citeN{CHS10} estimate a general nonlinear technology of skill acquisition and provide evidence for both self-productivity and dynamic complementarity. Self-productivity becomes stronger throughout childhood and, for cognitive skills, complementarity between existing skills and new investments also increases over time.

Early evidence from randomized control trials of high intensity programs targeted at poor, low-educated families in the United States in the 1960s and 1970s indicated that early investment could improve outcomes and reduce educational inequality. Specifically, the Perry pre-school program in Michigan as well as the Abecedarian and CARE programs in North Carolina all demonstrated substantial benefits \cite{elango20164}. The Abecedarian and CARE programs were quite intense, providing full-time childcare from eight weeks of age plus family support, while the Perry pre-school program combined high quality part-time pre-school with home visits. A further factor contributing to success of these programs was arguably the fact that participating children were likely to receive only low-level stimulation at home, making substituting home care with centre-based care more valuable.

Public childcare has become more widespread over recent decades with the share of children enrolled in preschools increasing from 15 to 61 percent between 1970 and 2020 (\citeNP{Worldbank}). Evidence on the success of universal childcare programs is more mixed than for the early targeted programs \cite{blandenrabe}, though most carefully designed studies show that the provision of early childcare is beneficial, especially for children from disadvantaged backgrounds (\citeNP{havnes2015universal}, \citeNP{cornelissen2018benefits}, \citeNP{felfe2018does}, \citeNP{gormley2005promoting}). This is in line with what might be expected, as the level of alternative investment received by children of poorer, less-educated and immigrant backgrounds is likely to be lower.

A few programs, for example \$5-a-day childcare in Quebec and state-funded pre-K in Tennessee, have instead been found to have negative effects  on children's skill acquisition (\citeNP{baker2008universal}, \citeNP{durkin2022effects}).  A possible explanation is that these programs are providing less, or poorer quality, investment than children received previously \cite{haeck2018universal}. This could be due to the quality of the programs themselves and/or to potentially more advantageous home environments in these cases compared to programs that have large beneficial effects. Why some types of public early investment succeed while others do not remains an open question \cite{blandenrabe}. Moreover, public investments may not reach those groups who would most benefit from them. In studying an expansion of early childcare in Germany, \citeN{cornelissen2018benefits} find that children from disadvantaged backgrounds are less likely to attend child care than those from advantaged backgrounds, even though their gains from attendance are higher. 

Incentivizing parents to take parental leave in the first months after birth may improve children’s outcomes due both to mothers' ability to breastfeed and parents' comparative advantage in providing secure attachment. This could reduce educational inequality if poorer parents are less likely to take leave, for example due to credit constraints. Evidence summarized by \citeN{berlinski2019} indicates that paid maternal leave beyond three months has no additional benefit for early childhood development. We know little, however, about the impact of policies that promote paternity leave on child outcomes.

\citeN{carneiro2021intergenerational} investigate the optimal timing of parental income for enhancing children's outcomes as adults and find that income in early and late childhood is more important than income during the middle years, casting some doubt on recent models that emphasize early investment above all else. 

\subsection{Quantifying the Parental Investment Channel}

There is now broad consensus that parental investments play an important role in children's skill acquisition. Parental investments can, however, take different forms and, as outlined in Section~\ref{sec:sources}, there are alternative explanations for why investments vary across families with different backgrounds. Quantifying the contribution of different investment channels to child development represents an ongoing challenge, as does better understanding the relative importance of economic factors, parental skills, and parental beliefs and preferences in generating inequality in investments.

The literature on skill acquisition shows that both cognitive and non-cognitive skills matter for economic outcomes \cite{ADHK11}. Non-cognitive skills, such as perseverance and conscientiousness, are both malleable and important determinants of earnings. In line with the setup developed here, \citeN{CH03}, \citeN{HM07}, and \citeN{CHLM06} show that parents have a crucial impact on the acquisition of both cognitive and non-cognitive skills, in part through transmission of their own skills and in part through their engagement and interaction with the child.\footnote{Sibling correlations are much larger than the square of parent-child correlations, suggesting that the latter capture only part of family influence on children’s cognitive and non-cognitive skills (\citeNP{anger2017cognitive}). In particular, intergenerational correlations might not fully reflect the impact of neighborhood and peer effects (see Section \ref{sec:peersneighborhood}) or latent parental advantages (Section \ref{sec:model_lr}).} \citeN{Deming2017} and \citeN{NybometalSkills2021AEJ} document an increase in the economic returns to social and non-cognitive skills in the United States and Sweden in recent decades. Moreover, the early acquisition of non-cognitive skills may result in higher productivity in acquiring cognitive skills later on. \citeN{blanden2007accounting} argue that children from higher income backgrounds have better non-cognitive skills and that this is an important driver of intergenerational transmission.

The precise mechanisms by which non-cognitive personality traits affect skill acquisition and later life outcomes remain unclear. One possibility is that non-cognitive skills are linked to greater cognitive efforts, in school or the labor market. Little work has been conducted in this regard, partly because effort is rarely observed and instead deduced indirectly as a residual after having accounted for differences in cognitive and non-cognitive skills. In real-effort experiments with fifth graders in Spain, \citeN{apascaritei2021difference} find that self-reported personality scales such as locus of control or conscientiousness are at best weakly associated with real effort.

To assess which parental inputs matter the most, \citeN{DBFW14} estimate a model that distinguishes between inputs of time and money. While both are productive, time inputs generally matter more. The important role of time inputs also implies that direct cash transfers to poor families will have a limited impact on children's skill acquisition, as time inputs don't respond strongly to such transfers and a large fraction of transfers is spent on goods other than educational inputs. Empirical evidence on the causal impact of changes in parental income on child achievement is reviewed by \citeN{BjoerklundJaentti2020} and  \citeN{mogstad2021family}, who conclude that family income does have a direct impact on children’s outcomes, in particular for children from low income or disadvantaged families.\footnote{This is supported by recent evidence from \citeN{troller2022impact}, who find that a regular income boost improves children's brain development in their first year of life.}  

Inequality in parental investment in education can arise from differing beliefs about the productivity of such investment. To this regard, \citeN{RauhBoneva2018} consider parents' perceptions of investment, showing that they generally view investments at different stages of life as substitutes rather than complements and expect higher returns from investments at later compared to early ages. Given that these perceptions are at odds with the empirical evidence, providing better information on returns may foster educational investments, particularly among families at the bottom of the income distribution. A shift in parental beliefs could also affect child beliefs on the returns to education, which are predictive of attainment gaps in higher education \cite{boneva2021can}.

Similarly, differences in parental investments can stem from preferences, either regarding the importance parents attach to children's education and other outcomes, or regarding the cost of investing. \citeN{Kalil20} note that in the United States, more educated mothers spend considerably more time on childcare than mothers with less education, and ask whether part of the reason may be that educated mothers find childcare more enjoyable. Evidence from the Well-Being Modules of the American Time Use Survey does not, however, show significant differences in the enjoyment of childcare by the mother's education. Hence, to the extent that differences in preferences and beliefs matter, these are more likely to relate to the returns to investments in children's skills rather than the cost of investment. Such a channel is supported by evidence that more educated mothers not only spend more time on childcare overall, but also tailor this time more closely to their children's developmental needs \cite{Kalil2012}, and that the overall gap in childcare time between more- and less-educated parents is especially large for education-oriented activities \cite{ramey10,Kaliletal2016,dozi19}. 

A preference channel for intergenerational persistence in educational attainment is also supported by the observation that preferences and aspirations are transmitted from parents to children (see for example \citeNP{dofahusu12}). \citeN{lekfuangfu2020all} show that parental aspirations are strong predictors of the educational aspirations of their children, even conditional on the socio-economic status of the parents and the cognitive and non-cognitive skills of the child. Such intergenerational correlation in aspirations could be due the transmission of information and beliefs about the returns to education, but could also stem from more purposeful actions by parents to shape their children's aspirations according to their own preferences \cite{bisin2011economics}.

\citeN{Caucuttetal2017} use a structural model to distinguish the empirical predictions of alternative channels that can result in unequal parental investments. An important role for either credit constraints or information frictions (i.e., low-income parents being less informed about the returns to parental investments) is supported by the observation that measured returns to parental investment are higher among low-income families. In contrast, if differential investment was primarily driven by intergenerational transmission of ability, we would expect higher investment returns for children from high-income families.

\subsection{Compensating Investments at School}
\label{sec:compensating_investments} 

As discussed, the investments $E_0$ and $E_1$ provided by schools can be a source of educational inequality, for instance if local financing makes the quality of public schools vary with the average income in a neighborhood. Schools can, however, also promote educational equality if they offer equal opportunities to all children regardless of background or even compensate for unequal investments and influences outside of school. Indeed, the belief that public education is the `great leveler' underlies much of the concern over the school closures during the COVID-19 pandemic \cite{ADSZ2020}, an issue we discuss in Section~\ref{sec:covid}. Here, we consider possible ways school systems may reduce inequality, as informed by recent empirical evidence.  

Section~\ref{sec:evidence} documents substantial differences between countries in both the level and inequality of test scores. \citeN{HanushekWoss2011HB} and \citeN{Woessmann2016} explore the extent to which this can be explained by differences in education systems across countries. Their analysis reveals several consistent patterns, with high teacher quality and more hours of instruction shown to be beneficial for outcomes. Perhaps contrary to popular expectation, they find no relationship between the level of school funding and test scores at the national level among OECD countries. Structural factors also matter; early sorting of students into different school tracks increases inequality in educational performance while high-stakes examinations to assess student learning combined with school autonomy leads to better average performance. 

Relying on cross-country differences to identify successful features of school systems has advantages and disadvantages. As noted by \citeN{Woessmann2016}, international variation allows researchers to explore the many possible ways schools can be organized, while within-country comparisons are constrained by the specific institutional setting. Studies targeting a particular country and reform also tend to ignore general equilibrium effects elsewhere in the system. That said, cross-country variation in institutions and policies can be correlated with other important characteristics that are hard to observe, such as cultural differences.  Evidence obtained from the macro analysis of schools and performance should accordingly be complemented with findings from specific natural experiments and randomized control trials (reviewed in \citeNP{fryer2017production}).

The evidence on the importance of school funding for student outcomes is mixed. In contrast to \citeN{HanushekWoss2011HB} and \citeN{Woessmann2016}, several recent studies document moderate effects of school funding based on plausibly exogenous variation in funding levels within countries. In a recent review of studies using US data, \citeN{jackson2021distribution} estimate that greater funding of 1000 dollars per student, sustained for four years, leads to a 3.5 percent of a standard deviation increase in test scores and relatively larger effects on high school graduation and college-going. For the most advantaged students, the effects are smaller. Given that the gaps between students from advantaged and disadvantaged backgrounds in Figure \ref{fig:PISA} are close to one standard deviation, higher funding alone is unlikely to close these gaps, in particular since the average low-income student does not live in an especially low-income district, meaning that distributing more money to poor districts is not a well-targeted transfer (\citeNP{lafortune2018school}).  

School funding is a high level input in the education production function. A given dollar can be spent in different ways, and finding the most cost-effective way to use funds could substantially improve the effectiveness of additional spending.  One obvious possibility is to increase the number of teachers per pupil, i.e., reduce class sizes. Based on randomization and natural experiments, the literature shows that the estimated impacts of money spent on reducing class size are at the upper range of the equivalent estimates from the school funding literature (\citeNP{gibbons2013cep}). A class size reduction of eight students is associated with a 15 percent of a standard deviation improvement in test scores, with some evidence of larger benefits for disadvantaged groups (\citeNP{schanzenbach2020economics}).  An alternative to reducing class size is to provide additional small group teaching. To this regard, \citeN{nickow2020impressive} provide a recent review of the evidence, noting that individual and small group tutoring increases outcomes by 37 percent of a standard deviation over all the available studies and is particularly effective for younger students. Such interventions thus potentially offer an effective means of reducing inequality.

Additional funding could also be spent on increasing the hours of teaching that students receive. This can be done by lengthening the school day, by extending the school term, or by lowering the school starting age. There is evidence that increases along all of these margins can have  modestly beneficial effects. For example, \citeN{huebener2017increased} find that a German reform extending the school week by two hours improved students’ performance in mathematics, science, and reading by slightly more than 0.05 of a standard deviation. Though, as high performing students benefited more, it also increased inequality. \citeN{lavy2020expanding} observes that an additional hour of weekly instruction on mathematics, science, and language improves students’ performance by 0.03 to 0.05 of a standard deviation. In the most disadvantaged schools in southern Italy, \citeN{battistin2016should} document a rise in math scores by almost one third of a standard deviation for one to two hours of additional instruction time, with no effects observed for language. There is some evidence that the number of days spent in school has a greater effect on disadvantaged students, implying that raising instructional time would help close socio-economic gaps. Given that the effects of the number of days of learning are also informative for understanding the potential impact of Covid-19 school closures, we leave a detailed discussion of this literature to Section~\ref{sec:covid}. With regard to school starting age, \citeN{cornelissen2019early} find that an additional month spent in schooling at age 4 increases test scores by 0.06-0.09 of a standard deviation in England, with larger effects for boys with lower social class backgrounds.  \citeN{leuven2010expanding} sees slightly smaller effects, concentrated among students with low-educated parents and those from a minority background.\footnote{The interaction between school starting age and the availability and quality of preschool should also be considered. Namely, if preschool is high quality, this weakens the argument for starting school earlier.}

An alternative approach to improving outcomes for children is to increase the quality of teachers. \citeN{hanushek2011economic} argues that ``no other attribute of schools comes close to having this much influence on student achievement,'' while \citeN{chetty2014measuring} find that a one standard deviation increase in teacher quality in a single grade raises annual earnings by 1.3 percent, which amounts to 39,000 dollars on average in cumulative lifetime income.   \citeN{jackson2018test} and \citeN{liu2021engaging} report even larger effects. Teachers therefore clearly matter, but it is less obvious how their effectiveness be improved. Though scholars have yet to identify observable characteristics that distinguish good teachers (\citeNP{rockoff2004impact}, \citeNP{rivkin2005teachers}), one approach taken in the literature consists of modeling how employment conditions might be altered to select and retain the most effective teachers. \citeN{burgess2019understanding} reaches the conclusion, based on studies that simulate the impact of possible policies \cite{staiger2010searching,rothstein2015teacher}, that teachers should be monitored closely, with the least effective among them being dismissed early on in their careers. Evidence from the Washington IMPACT policy further supports these findings. Specifically, \citeN{dee2015incentives} estimate that this policy improves the quality of teachers both by threatening poor teachers with dismissal and by using performance-related pay (PRP) to reward highly effective teachers. More generally, financial incentives raise both teacher effort and quality by encouraging high-performers to select into and remain in the profession \cite{britton2016teacher,biasi2021labor}.  

Another way to improve teacher performance is to change their pedagogical approach, either through their initial training or by encouraging them to adopt new methods once on the job. Though thus far there have been few opportunities to randomize the contents of teacher training programs,  other interventions do shed light on the features of effective teaching. To this regard, two English programs of are particular interest.  First, the ``literacy hour'' in English primary schools \cite{machin2008literacy} showed that a focused hour a day of teaching children to read (costing but a small amount), could increase student test scores by eight percent of a standard deviation. Second, training teachers to implement a specific pedagogical approach in teaching reading (synthetic phonics) led to sustained gains in reading outcomes for disadvantaged groups as well as long-term reductions in inequality \cite{machin2018changing}. That modifying teaching practices can improve children’s outcomes is supported by the even larger estimated effects of the Success for All \cite{borman2007final} and Reading Recovery \cite{may2014evaluation} programs in the United States, albeit these are small group remedial interventions as opposed to the whole class approaches considered for England.

\citeN{HanushekWoss2011HB} and \citeN{Woessmann2016} note that successful school systems often couple school autonomy with accountability through exit exams. In a previous edition of this Handbook, \citeN{epple2016charter} provide evidence on the beneficial effects of autonomous US charter schools, and explore the puzzle of why such schools are effective.\footnote{ Examples of autonomous schools are charter schools in the United States (including Knowledge is Power of KIP), Swedish free schools, and English academies. All of these schools operate outside local government guidance, and often have distinct pedagogical philosophies.} Arguably, the particular policies associated with charter schools, such as a ``no excuses'' mindset of high expectations of student behavior and strong staff commitment, are crucial to achieving positive long run effects \cite{dobbie2020charter}. \citeN{fryer2014injecting} shows that similar measures have been effective in traditional US public schools, pointing to a bundle of policies in Houston that raised math test scores by 15 to 18 percent of a standard deviation.\footnote{These included increased instructional time, more-effective teachers and administrators, high-dosage tutoring, data-driven instruction, and a culture of high expectations.} Evidence from autonomous English academies meanwhile indicates that while school autonomy is an effective policy prescription in low-performing schools in disadvantaged areas, it has limited effects as a general school reform (\citeNP{eyles2019introduction}, \citeNP{eyles2018academies}).  

The success of school investments in improving outcomes and lessening educational inequality depends to some extent on how parents react. If school inputs and parental investments are substitutes, then increasing inputs through the education system may lead to reduced investment at home. This may partly explain why the effect of increasing school resources on students' achievement is relatively small. There is some support for this hypothesis, with \citeN{greaves2019parental}, \citeN{houtenville2008parental} and  \citeN{fredriksson2016parental} all finding that when parents become aware of school quality improvements they reduce their engagement with children's school work. It is possible that the strength of this effect depends on how easily observable the school investment is to parents. Whereas physical resources (e.g., stationery) are obvious, teaching quality can be more difficult for parents to assess \cite{rabe2019school}. Interestingly, \citeN{greaves2019parental} and \citeN{fredriksson2016parental} find that parents of higher socio-economic status are more likely to respond to increased school investment by reducing their own investment. This could help explain why additional school inputs tend to reduce educational inequality; unlike more advantaged parents, poorer families are less likely to diminish their own investments, perhaps because they were low to begin with.  

The literature thus highlights several fruitful approaches that school policy can act on to close gaps by family background. However, most of the documented effects are fairly small compared to the almost one standard deviation gap in test scores between children from richer and poorer families observed in Section 2. Schools' ability to substantially reduce educational inequality may therefore be limited.\footnote{\citeN{kramarz2015using} conclude that there is more than three times as much variation in outcomes between pupils as there is between schools, and \citeN{agrawal2019quantifying} observe that the share of variation in college attendance associated with schools is 16 percent of the total.}  Nonetheless, public school systems do considerably lessen the extent of inequality compared to what would prevail if all investments were provided or paid for by the family. 

\section{Inequality in Higher Education}
\label{sec:model_high_ed} 

The previous section focused on skill acquisition from early childhood to the end of secondary schooling, the period when parental decisions matter most. However, in high-income economies, educational inequality is also shaped by inequality in higher education. In this phase, parental influences have a less direct impact; educational inequality instead arises from young adults' decisions on whether to work, to go to school, which school and which program to attend, and how much effort to put in their education. In what follows, we provide a simple framework to discuss sources of educational inequality at this stage.

\subsection{Setup}

\label{sec:model_high_ed_setup} 

We consider the education and labor supply decisions of an individual who has already gone through the two periods of childhood and enters adulthood with skills $s_2$. Financial resources are also relevant at this stage, given by assets $a_2$. We can think of $a_2$ as arising from bequests or transfers received from parents. Adult life now unfolds through two further stages. In period 2, the young adult can choose between starting to work immediately and attending college, whereas in period 3 all individuals work. Given that we are focusing on education decisions, we do not model retirement, and hence period 3 is the final period. 

Preferences are given by a standard expected utility function with discounting over the two periods:
\[
U(c_2,c_3)=u(c_2)+\beta u(c_3),
\]
where $c_t$ is consumption and the period utility function $u(\cdot)$ is strictly increasing, strictly concave, and features decreasing absolute risk aversion.

In period 2 (young adulthood), the individual must decide whether to attend college, how much effort (time) $i_2$ to invest in studying, and whether to also work while in college. We consider a simple setting where attending college only affects earnings if college is successfully completed. Let $d_3$ denote whether college was completed, where $d_3=1$ indicates completion and $d_3=0$ lack of completion. Let $e_2\in\{0,1\}$ denote the decision of whether or not to enroll in college. The probability of completing college is given by a function $p_d(e_2,s_2,i_2)$, where $p_d(0,s_2,i_2)=0$ (you cannot graduate if you do not enroll), and where $p_d(1,s_2,i_2)$ is increasing in both arguments and strictly concave in $i_2$.

Adult wages are given by functions $w_2(s_2)$ and $w_3(s_3,d_3)$. We set $s_3=s_2$, hence, basic skills do not evolve throughout adulthood and only college completion matters. This is easily generalized, but the simplified case considered here turns out to be sufficient to illustrate the main tradeoffs and mechanisms of interest.  

The decision problem of the young adult can be written as follows:
\[
\max_{c_2,c_3,a_3,n_2,i_2,e_2\ge 0} E\left\{u(c_2)+\beta u(c_3)\right\}
\]
subject to:
\begin{align}
\label{eq:dyn_bc1} c_2+a_3&=a_2+n_2 w_2(s_2)-T e_2 ,\\
\label{eq:dyn_bc2} c_3&=(1+r) a_3+ w_3(s_3,d_3),\\
\label{eq:dyn_tc1} i_2+n_2&=1.
\end{align}
Constraints \eqref{eq:dyn_bc1}--\eqref{eq:dyn_bc2} are the budget constraints, where $T$ is the tuition that must be paid if college is attended ($e_2=1$). Constraint \eqref{eq:dyn_tc1} is the time constraint in the young adulthood period, where time is split between working $n_2$ and putting effort into college $i_2$. In the old adulthood period, everyone supplies one unit of labor. The constraints are written for the case where there is no insurance available for the uncertainty regarding college completion; we consider the case of insurance below.

\subsection{The Impact of Financial Constraints on Educational Inequality}

Our analysis focuses on how educational outcomes are related to initial skills $s_2$ and initial assets $a_2$. Specifically, we ask: under which conditions will poorer individuals with fewer assets end up with less education, even if they have a lot of skills? 

One potential source for such attainment gaps are financial constraints.\footnote{Following \citeN{Becker75}, the economic literature emphasizes the role of financial investments and constraints. The sociological literature instead stresses the influence of values or norms, as well as differences in parental preferences regarding the value of higher education (e.g., cultural reproduction theory). Structural models such as \citeN{BelleyLochner2007} allow for such additional factors through preference shocks (i.e., a consumption value of schooling) that may depend on parental background. } A natural benchmark is provided by an environment with perfect financial markets, where students are able to borrow to finance education and to obtain insurance for the uncertainty of college completion (or more generally, uncertainty about future income). In this case, education decisions depend entirely on skills $s_2$ but not on wealth.

\begin{prp}[Higher Education under Perfect Financial Markets] \label{prop:college_perfect}
If students can borrow and there is actuarially fair insurance for income uncertainty, college attendance, effort in college, and the return to college  depend only on initial skill $s_2$ but not on resources $a_2$.
\end{prp}

\begin{prf}
The availability of insurance can be represented by replacing actual income in the budget constraint \eqref{eq:dyn_bc2} with expected income:
\begin{equation}
\label{eq:dyn_bc3}
c_3=(1+r) a_3+ E\left\{w_3(s_3,d_3)|e_2,s_2,i_2\right\},
\end{equation}
that is, adults who fail to graduate from college are compensated by those who succeed, conditional on the expected probability of success given enrollment, initial skills, and effort in college. The model is then deterministic, and given no borrowing constraints the two budget constraints \eqref{eq:dyn_bc1} and \eqref{eq:dyn_bc3} can be combined into a single present-value budget constraint:
\[
c_2+\frac{c_3}{1+r}=a_2+ (1-i_2) w_2(s_2)-T e_2+\frac{w_3(s_2,0)+p_d(e_2,s_2,i_2) \left(w_3(s_2,1)-w_3(s_2,0)\right) }{1+r},
\]
where on the right-hand side we have expressed expected income in terms of the graduation probability $p_d(e_2,s_2,i_2)$ and we have replaced $n_2$ with $1-i_2$ from the time constraint \eqref{eq:dyn_tc1}. The form of the budget constraint implies that the enrollment decision $e_2$ and education effort $i_2$ will be chosen so as to maximize the right-hand side of the present-value budget constraint, and hence these decisions do not depend on initial assets $a_2$. Specifically, if the young adult enrolls in college and if optimal effort $i^\star_2$ is interior, it will satisfy:
\[
w_2(s_2)=\frac{1}{1+r}\frac{\partial p_d(e_2,s_2,i_2)}{\partial i_2}\left(w_3(s_2,1)-w_3(s_2,0)\right),
\]
which equates the cost of effort in terms of forgone consumption to the present value of expected future income gains. Likewise, given optimal effort $i^\star_2$ conditional on enrolling, the child will enroll if expected gains exceed the cost:
\[
\frac{1}{1+r} p_d(1,s_2,i^\star_2) \left(w_3(s_2,1)-w_3(s_2,0)\right) >i^\star_2 w_2(s_2)+T.
\]
Neither of these conditions depend on assets $a_2$.
\end{prf}

Hence, in the case of perfect financial markets, unequal financial resources among families (captured by $a_2$) will not result in additional educational inequality. Of course, there is still likely to be inequality in skills $s_2$, and as discussed in the previous section this inequality may in part be generated by family socio-economic characteristics and hence be correlated with $a_2$. Functioning financial markets do not remove existing sources of inequality, but neither do they add an additional channel for further increasing inequality. 

Next, we can consider how the outcome changes if financial markets are less-than-perfect. We first consider a setting in which there is no hard borrowing constraint, but there is no insurance available for the uncertainty of college graduation (or more generally, for the uncertainty about earnings conditional on attending college). Such a setting echoes the current institutional setup of college finance in the United States, where education loans are widely available, but repayment is generally not contingent on future income and discharging student debt in bankruptcy is difficult or impossible. The budget constraints for this case are given by \eqref{eq:dyn_bc1}--\eqref{eq:dyn_bc2}. Here we can show that even though borrowing is possible, initial resources still matter for education decisions. 

\begin{prp}[Higher Education without Insurance] \label{prop:college_no_insurance}
If students can borrow but no insurance for income uncertainty arising from college attendance is available, for students who have a sufficiently low probability of graduating (given their skill $s_2$) effort in college (if interior) and the return to college given skill are increasing in their resources $a_2$, and for given skill, students with greater resources are more likely to enroll in college.
\end{prp}

The proof for the proposition is provided in Appendix~\ref{app:proofs}. Here, the effects of resources on college decisions arise from the impact of wealth on risk attitudes in the environment without insurance. As wealth increases, marginal utility becomes flatter and students come closer to simply maximizing the financial return to college, as in the case with insurance. Students with low wealth have higher marginal utility, more curvature in utility, and higher absolute risk aversion. Attending college is a risky decision, and hence poorer, more risk-averse students will be less willing to enroll. If better insurance markets could be provided, educational inequality between students of low and high wealth would decrease. \citeN{KrebsKuhnWright2015} provide a quantitative analysis of this point. 

Note that in our setting, attending college always increases income uncertainty in later life compared to the case of not attending. In reality, there are forces running in different directions. In many countries, workers who never attended college face higher unemployment risk and may be more impacted by structural change affecting specific occupations, such as the decline of routine non-cognitive occupations in manufacturing. If attending college universally lowered future income risk, young adults with low wealth may actually be more likely to enroll in university, because escaping from the uncertainty that comes with little education is more valuable to them. 

Nevertheless, as our model emphasizes, for young adults with little wealth college completion is a key margin---many more students begin college than finish it, and dropout rates are closely related to socio-economic status. If the probability of actually attaining a degree is sufficiently small, attending college is an inherently risky decision, because it involves an up-front cost with an uncertain outcome. 

The impact of wealth on education is even more severe when there are binding borrowing constraints in addition to imperfect insurance markets.

\begin{prp}[Higher Education with Binding Borrowing Constraints] \label{prop:college_borrowing_constraints}
If there is a binding borrowing constraint and no insurance for income uncertainty arising from college attendance is available, effort in college (if interior) and the return to college for given skill $s_2$ is higher for students with more resources $a_2$, and for given skill students with more resources are more likely to enroll in college.
\end{prp}

The proof for the proposition is provided in Appendix~\ref{app:proofs}. Qualitatively, a binding borrowing constraint has the same implications as missing insurance markets, but quantitatively, the same effects are further amplified. When students with fewer assets face a binding borrowing constraint, their marginal utility of consumption when young is high, which increases the utility cost of paying tuition $T$ to attend college. Similarly, the opportunity cost of putting effort $i_2$ into college is also high. Hence, all else equal a student with fewer assets will put less effort into college and instead work more while also studying, which lowers the probability of graduating. These implications align with evidence indicating that students from low-wealth families are more likely to work during college, attain fewer credits per semester, and ultimately have a lower probability of graduating.

\subsection{Evidence on the Importance of Borrowing Constraints}

The role of borrowing constraints in generating a link between family income and college attendance was observed as early as \citeN{Becker75}.\footnote{See \citeN{LochnerMonge2012} for a recent survey of the role of credit constraints in education decisions.} \citeN{CarneiroHeckman2002} examine the role of ability and credit constraints for college decisions in the United States using the National Longitudinal Survey of Youth 1979 (NLSY79), which covers a cohort that attended college in the early 1980s. They argue that, once ability is controlled for, financial constraints play a small role in college attendance for this cohort. \citeN{CameronTaber2004} come to the same conclusion in a study that compares different methods for assessing the importance of borrowing constraints. \citeN{KeaneWolpin2001} also build a structural model of the education decisions of the NLSY79 cohort, focusing in particular on the role of parental transfers. They estimate that financial constraints are tight, but that nevertheless, relaxing borrowing constraints would have little impact on educational attainment. This is largely because parental transfers and part-time work enable students to attend college even without access to loans. 

According to \citeN{CarneiroHeckman2002}, even though family resources do not matter much conditional on children's skills, financial constraints may still be important for earlier skill investments, reflected in skills acquired by the end of high school. In a structural model of skill acquisition over multiple life stages matched to data from 1990, \citeN{RestucciaUrrutia2004} similarly conclude that financial constraints are binding primarily during early education but much less so for college education. In our model, these results can arise if financial constraints are loose or if the cost of tuition $T$ is low, so that ability-dependent returns (and possibly preferences) rather than financial considerations drive attendance decisions. 

The work of \citeN{KeaneWolpin2001}, \citeN{CarneiroHeckman2002}, and \citeN{CameronTaber2004} focuses on the 1980s. The cost of attending college has since risen substantially in the United States and other countries. \citeN{BelleyLochner2007} reexamine the importance of borrowing constraints using the NLSY 1997 cohort, which attended college around the turn of the 21st century. They show that by the 2000s, family income had become a much stronger determinant of college attendance, consistent with a greater role of financial constraints driven by the higher cost of university.

\citeN{Lochneretal2014} ask whether the cost of attending college and tuition-support policies can explain why the gradient between family income and college attendance is substantially steeper in the United States compared to Canada. They find, however, that financial support for attending college is actually more generous for students from low-income families in the United States, and that factors other than the net costs of attending college must play a role. One possibility is that students from low-income families are not fully aware of the financial aid they can apply for, which would amplify the role of financial constraints.

\citeN{LochnerMonge2011} argue that to account for the empirical relationships between family income, student ability, and college attendance, endogenous borrowing constraints need to be taken into account, meaning that borrowing limits depend on the ability to pay and therefore increase with education. They document that among children from low-income families, college attendance is strongly increasing in ability. The model developed in Section~\ref{sec:model_high_ed_setup} does generate a positive ability-enrollment relationship even with a fixed borrowing constraint, because high-ability students have a higher return to college and are therefore more willing to endure temporarily limited resources to attend. Still, \citeN{LochnerMonge2011} contend that this effect is too small to account for the observed relationship, and hence that endogenous responses to credit limits are necessary to explain the data. They build a model matched to US data that accounts for both government student loan programs and private lending, and claim that this setting can account for a variety of features of the changing relationships between ability, financial resources, and college attendance in the United States.

Additional evidence on the role of borrowing constraints in higher education in the United States is provided by \citeN{Brownetal2011}. Using data from the Health and Retirement Study, they show that children from low-income families who are offered more financial aid for college achieve higher educational attainment, which is consistent with binding financial constraints.

\citeN{Abbottetal2019} examine the implications of college financial aid policies within a general equilibrium model that accounts for parental transfers, government financial aid policies, borrowing constraints, and working during college as an additional source of funding. In the estimated model matched to US evidence, an expansion of financial aid increases college attainment and improves welfare, indicating an important role of binding financial constraints. The estimated model also implies that there are sizeable differences in the return to college by students' ability. As a result, an ability-tested expansion of financial aid is even more effective than a general expansion. 

Studies by \citeN{kaufmann2014understanding}, \citeN{Solis2017}, and \citeN{Caceres-Delpiano:2018aa} on Mexico and Chile further suggest that credit constraints are more important in middle-income countries. Specifically, financial constraints are likely to be binding for a larger share of families, and private financial markets for funding education investment are less developed. Accordingly, lowering attendance costs or providing better access to credit would lead to substantial increases in college enrollment in these settings.

\subsection{Policy Implications}

Our model of higher education decisions under financial constraints has implications for potential policy interventions. For example, providing student loans would lower the impact of financial resources on college attendance and the probability of graduating. Nevertheless, student loans would not completely eliminate educational inequality between children from families with more and less resources. This is clearly the case if borrowing constraints are relaxed but continue to bind on the margin, as in Proposition~\ref{prop:college_borrowing_constraints}. Here, a binding constraint does not mean that a student is unable to borrow enough money to afford tuition. Rather, even if hitting the borrowing constraint merely results in lower consumption (and hence high marginal utility) during the college phase, the incentive to attend college and to put in high effort will be lower among students with fewer financial resources. Furthermore, even if the borrowing constraint is fully eliminated through generous student loans, it still matters that attending college is a risky decision with uncertain returns, which can also introduce an impact of wealth on college attendance and returns (Proposition~\ref{prop:college_no_insurance}). The risk of attending college is  accentuated if student loans need to be repaid regardless of success in college.

From the perspective of college as a risky investment, a ``graduate tax'' that makes loan repayment income-contingent can lower the importance of the risk exposure channel. These benefits must be weighed against disincentive effects for effort during college and later in working life. In particular, higher effective income tax rates may reduce work and study incentives by lowering the realized returns to college. The higher education funding system in England has some features of a graduate tax. Students only repay their loans once their income reaches a particular threshold, and repayments are fixed as a share of salary. Any loans outstanding after 30 years are written off by the state \cite{murphy2019end}.  This is an expensive system, with the state predicted to pay almost 43 percent of the value of the loans made, owing to low interest rates and non-repayment  (\citeNP{crawford2014estimating}). Nevertheless, it does mitigate risk for low income students. Indeed, it appears that the combination of rising fees and the switch to income-contingent loans in England has had no appreciable impact on inequalities in higher education participation (\citeNP{murphy2019end} and \citeNP{azmat2017higher}).

Our results on the impact of financial resources on college completion relate to recent work on socio-economic gaps in college graduation rates. \citeN{BaileyDynarski2011} document a growing disparity in the United States between students from richer and poorer families in terms of college entrance, persistence, and graduation. \citeN{HendricksLeukhina2017} use detailed transcript data to model students' progress throughout college, and show that high dropout rates are associated with slow progress in college from the very start (i.e., few attempted and completed credits). Both initial skills and financial resources can influence progress throughout college, although if students are able to predict their graduation prospects well, financial resources play a minor role. \citeN{Stange2012} and \citeN{Tracher2015} point out that the possibility of dropping out of college early, while introducing risk, can also benefit students with lower ex-ante graduation probabilities by allowing them to learn about their prospects before committing to studying for the entire duration.  

The longitudinal data used in Section 2 enables us examine determinants of non-completion of college by age 25 for those enrolled in bachelor programs at age 20. By far the largest impact of family background is found in the United States, where students with highly educated parents are 14.5 percentage points more likely to complete college than students with less educated parents. Family background also has a significant impact in England (a five percentage point difference in completion rates) but not in Australia.\footnote{The sampled students were still too young to measure non-completion in Germany.} High school test scores give an indication of graduation probability based on ability. Conditioning on this ability measure eliminates the impact of family background in England but it remains important in the United States, with those from highly educated families still nine percentage points less likely to drop out compared to children of similar ability but from less-educated families. In line with the structural models discussed above, these basic facts are consistent with an important role of financial constraints in educational inequality at the college level in the United States. 

For ease of exposition, we have considered investment in skills during the early years (Section~\ref{sec:model}) separately from our analysis of higher education here. Yet, in models of education over the entire life cycle, interactions between these stages naturally arise. For example, \citeN{caucutt_lochner_2017} find that while policies that relax financial constraints at a single stage of the life cycle have only moderate effects, eliminating financial constraints throughout has a large impact on the educational investments of poorer families and on social mobility. Underlying this result is the complementarity of investments at different stages, as discussed in the previous section. The same complementarity also implies that policies that relax borrowing constraints in higher education have larger effects if they are anticipated. Specifically, if poorer parents expect that financial constraints will not stand in the way of their children receiving a college education, they will perceive higher returns to education earlier in childhood and increase their investments. In the language of the model, the skills $s_2$ at the beginning of adulthood will increase if financial constraints are relaxed.

\citeN{BlandinHerrington2021} develop a structural model comprising both early investments in children and college attendance, and they use this model to account for recent trends in college attendance and completion in the United States. They document that college attendance has risen over time for children from richer and poorer families alike. However, consistent with the evidence discussed above, a gap in completing college conditional on attending has emerged. From 1995 to 2015 the probability of completing college conditional on attending rose by more than 10 percentage points among children with at least one college-educated parent raised in a two-parent home, but actually decreased for children raised by a single non-college-educated parent. As in the model developed in Section~\ref{sec:model}, parental investments respond to aggregate conditions, and in particular to the return to college education. Differences in pre-college investments turn out to play a key role in generating socio-economic differences in college completion. As the return to college increases over time, richer families raise their investments in their children's skill more than poorer parents do, which results in a widening disparity in college completion rates. For these reasons, and in line with the work of \citeN{herrington2015}, \citeN{lee_seshadri_2019}, and 
\citeN{daruich2018} discussed above, policies that support earlier parental investments in children's skills turn out to be more effective than later interventions such as tuition subsidies. 

Beyond college attendance and completion, there is also considerable heterogeneity in outcomes conditional on completing college, relating to issues such as choice of college major and post-college career and occupation choices. \citeN{Altonjietal2012} consider these dimensions in a model that captures parental investments at the high school level, college education, and later labor market outcomes. Given that passing rates and average grades vary across majors, pre-college investments once again play an important role. In particular, less well-prepared students may choose ``safer'' majors, even if they ultimately yield a lower financial return.

\section{Educational Inequality across Multiple Generations}
\label{sec:model_lr}

We have reviewed some of the key mechanisms that shape the transmission of educational advantages from parents to their children. However, social mobility can be analyzed not just from one generation to the next, but also across multiple generations. A key finding from the studies reviewed in Section \ref{subsec:Sec2multigenerational} is that educational mobility across multiple generations is lower, and perhaps much lower, than a naive extrapolation from conventional parent-child measures would suggest. This pattern is not only compelling in its own right, but is informative about the underlying mechanisms that generate persistence in educational inequality. Multigenerational estimates of social mobility have accordingly become the subject of debate not only in economics, but also in demographic and sociological studies. 

The literature has focused on two broad classes of explanations to rationalize the evidence on high multigenerational persistence. First, educational status or its determinants might be mismeasured, which would bias estimates of intergenerational persistence from parent to child downward. Mismeasurement may occur because the outcome observed in the data is an imperfect proxy of educational achievement (e.g., years of schooling may not reflect the quality of schooling), or because the parents’ education is only one of many determinants of child education that vary across families (e.g., \citeNP{Clark2014book}, \citeNP{BraunStuhler2018}).

In Section~\ref{subsec:Sec2multigenerational}, we documented multigenerational transmission using regressions of a child's years of schooling on the average years of schooling of the parents and grandparents (see Equation~\eqref{eq:reg3g}). According to the mismeasurement interpretation, the tendency of the coefficient $\beta_{gp}$ on grandparent schooling in such regressions to be positive, as shown in \Cref{fig:3gScatter}, reflects an omitted variable problem, with grandparents’ schooling serving as a proxy for unobserved attributes of the parent generation.\footnote{Consistent with this interpretation, the coefficient $\beta_{gp}$ on grandparent's education tends to be smaller when controlling for both paternal and maternal education (e.g., \citeNP{WarrenHauser1997}, \citeNP{chiang2015grandparents}, \citeNP{BraunStuhler2018}, \citeNP{Engzell2020aa}).} Classical measurement error or misreporting in survey data would yield similar implications (\citeNP{Solon201413}).  

A second, contrasting interpretation is that grandparents or extended family may have an independent causal effect on child education, over and above their indirect influence via the parent generation \cite{Mare2011}. According to this interpretation, a positive coefficient $\beta_{gp}$ on grandparents’ schooling in \Cref{fig:3gScatter} may represent true grandparent effects. One strategy to distinguish this hypothesis from alternative interpretations is to study whether the size of the coefficient $\beta_{gp}$ varies with exposure of grandchildren to their grandparents. In their review of the literature, \citeN{AndersonSheppardMonden2017} note that the coefficient $\beta_{gp}$ does not appear to vary systematically with the likelihood of contact between grandparent and grandchild. However, there are exceptions \cite{ZengXie2014}, and a direct effect of grandparents on their grandchildren is possible through means that do not require contact, such as financial transfers \cite{hallsten2017grand}.

While these interpretations differ, they both imply that researchers can gain a deeper understanding of intergenerational processes by extending their analysis to more distant kin. Focusing on the omitted variable interpretation, we first consider a setting of multi-generational transmission in which we make a distinction between the latent (unobserved) skills and the observed educational outcomes of a given generation. One useful insight is that the transmission of such latent advantages can be indirectly assessed by considering how educational inequalities persist in the extended family beyond the core parent-child relationship. We then consider the potential contribution of assortative mating to educational inequality, and how this contribution can be quantified.  

\subsection{Single- versus Multigenerational Transmission}

Consider a setting in which the skill of the member of a family alive  at time $t$ is given by $S_t$. The true skills $S_t$ are unobserved; observed instead is an educational outcome $Y_t$, such as years of schooling, which is related to true skills by the equation:
\[
Y_t=\alpha S_t+\epsilon_t,
\]
where $\epsilon_t$ is an i.i.d. random variable and $\alpha$ can be interpreted as the effect of skills on educational attainment. Let us assume that skills evolve over generations according to:
\[
S_{t+1}=\lambda S_t+\nu_t,
\]
where $\nu_t$ is an i.i.d. random variable and $0<\lambda<1$ captures the intergenerational persistence of skills. For simplicity, we standardize the variances of $Y_t$ and $S_t$ to one, such that $\alpha$ and $\lambda$ can be interpreted as correlations. In this model, the parent-child correlation in schooling is given by
\[
\rho_{1}=\frac{\text{Cov}(Y_{t+1},Y_{t})}{\text{Var}(Y_{t})}=\alpha^2 \text{Cov}(S_{t+1},S_{t}) = \alpha^2 \lambda ,
\]
while the three-generation (grandparent-child) correlation is given by 
\[
\rho_{2}=\frac{\text{Cov}(Y_{t+2},Y_{t})}{\text{Var}(Y_{t})}=\alpha^2 \text{Cov}(S_{t+2},S_{t}) = \alpha^2 \lambda^2.
\]
More generally, the correlation $k$ generations ahead is equal to $\alpha^2 \lambda^k$. The three-generation correlation is therefore larger than a simple iteration of the parent-child correlation would suggest, 
\[
\rho_{2} > \rho_{1}^2 ,
\]
as long as skills $S_t$ are not a perfect predictor of schooling $Y_t$, that is, as long as $\alpha<1$. As shown in \citeN{BraunStuhler2018}, this observation $\rho_{2} > \rho_{1}^2$ is simply the flip side of the observation that the coefficient $\beta_{gp}$ on grandparents' education in the child-parent-grandparent regression (\ref{eq:reg3g}) tends to be positive (see \Cref{fig:3gScatter}).

In richer models, the inequality could also invert, with the square of the parent-child correlation $\rho_{1}$ being greater than the grandparent-child correlation $\rho_{2}$.\footnote{Indeed, a famous prediction from \citeN{BT79} is that the coefficient on grandparent income in a child-parent-grandparent regression should be negative, which necessarily holds in a simplified version of their model \cite{Solon201413}. This prediction is, however, at odds with the recent multigenerational evidence.} A robust conclusion is that parent-child correlations may not be very informative about the extent to which educational or socio-economic advantages persist across multiple generations. Indeed, if $\lambda$ is sufficiently large, multigenerational persistence can be high even if the parent-child correlation $\rho_1$ itself is only modest in size. In the extreme case of full persistence,  $\lambda=1$, we would observe no further regression to the mean after the first generation, so that $\rho_{1}=\rho_{2}=...=\rho_{k}$.

Multigenerational estimates are then useful for two distinct reasons. First, they offer direct evidence on the persistence of educational advantages in the long run, on which parent-child correlations are less informative. As such, they provide insights on an important descriptive aspect of inequality, independently of what the underlying transmission mechanisms may be. Second, they may provide indirect information on more hidden aspects of the intergenerational transmission process between parents and children, such as the latent skills $S_t$ in the model above. Indeed, this simple latent factor model can be identified from linked data on just three generations, as the intergenerational persistence of skills $\lambda$ is identified by the ratio $\rho_2/\rho_1$. Following this approach, \citeN{BraunStuhler2018} find an intergenerational correlation of latent educational advantages of around 0.6 in Germany, nearly 50 percent larger than the parent-child correlation in years of schooling in their samples. \citeN{NeidhoeferStockhausen} find similarly high latent persistence in a comparison of three different data sets from the United States, the United Kingdom, and Germany, as do \citeN{colagrossi2020like} in comparing 28 European countries. Other studies documenting excess persistence of educational inequalities in the sense of $\rho_{2}>\rho_{1}^2$ or $\beta_{gp}>0$ include \citeN{Pfeffer2014}, \citeN{LindahlPalme2014_IGE4Generations}, \citeN{CelhayGallegosChile2015}, \citeN{KroegerThompson2016}, \citeN{song2016diverging}, \citeN{SheppardMonden2018}, \citeN{ferrie2020grandparents}, and \citeN{chuard2021multigenerational}, among others. \citeN{AndersonSheppardMonden2017} provide a systematic review of this literature.

While informative, multigenerational correlations do suffer from a limited comparability of socio-economic measures across generations, as the marginal distributions of educational attainment and other outcomes usually differ across generations. Moreover, the evidence they provide is insufficient to fully characterize richer models of intergenerational transmission. A recent ``dynastic'' or ``distant-kin'' approach addresses these limitations by shifting the attention from multigenerational relations to family relations in the horizontal dimension, such as between cousins in the same generation. Using data from Sweden, \citeN{adermon2021dynastic} document that the average educational attainment of dynasties in the parent generation (including parents’ siblings and cousins, and their siblings and spouses) are highly predictive of child education, suggesting that conventional parent-child correlations underestimate long-run persistence by at least one-third. \citeN{hallsten2020shadow} merge historical and modern data sources to compare socio-economic status across many generations and across distant cousins, showing that outcomes for past generations predict education, occupations, and wealth many generations later. 

\citeN{ColladoOrtunoStuhler2019aa} exploit administrative data to distinguish many different types of kin, which they use to fit a detailed intergenerational and assortative model. They show that distant-kin as well as more traditional measures of educational inequality, such as parent-child and sibling correlations, can be all integrated within a consistent transmission model. In particular, multigenerational estimates can be related to earlier evidence from sibling correlations, which are an omnibus measure of the importance of family and community that reflect parental influences, but also other influences shared by siblings that are orthogonal to parental characteristics \cite{bjorklund2011education}. Siblings share educational advantages to a much a greater extent than what is captured by parent-child correlations, implying that intergenerational correlations represent only a small share of family background effects \cite{BjorklundJaentti2012aa}. Recent multigenerational studies are consistent with this hypothesis, and imply that the unexplained gap between sibling and parent-child correlations is at least partially due to unmeasured family influences (as opposed to influences that are orthogonal to family characteristics, such as neighborhood or peer influences). Moreover, a comparison of distant kin suggests that sibling correlations still understate the importance of family influences, as most of the advantages that siblings share are not reflected in observables such as years of schooling \cite{ColladoOrtunoStuhler2019aa}.

\subsection{What Drives High Multigenerational Persistence?}

What are the mechanisms underlying the high degree of persistence in educational advantages documented in recent studies? Interpreted from the perspective of the model outlined above, a higher degree of persistence implies an intergenerational persistence parameter $\lambda$ closer to 1. However, the model is silent as to what drives this parameter. 

One possibility is that persistence reflects genetic transmission. If ability is in large part determined by genetic endowments that are passed on from generation from generation, a high multigenerational degree of persistence can arise. Yet, genetic transmission necessarily combines the endowments of different families. Because mothers and fathers contribute in approximately equal parts to the genetic endowments of their children, a child's genetic endowment is given by:
\[
S_{t+1}=\lambda \frac{S_{f,t}+S_{m,t}}{2}+\nu_t,
\]
where $S_{f,t}$ and $S_{m,t}$ are the endowments of mother and father, respectively.  For this model to generate a high multi-generation correlation of skills, a $\lambda$ close to one is not sufficient. For example, if the skills of husband and wife are uncorrelated, the persistence of skills across generations is driven by $\lambda/2$, which would imply very low multi-generation correlations after a few generations. 

As a result, for a genetic model to explain high multigenerational correlations there would also have to be high assortative mating in terms of genetic endowments \cite{Clark2021Bell}. For illustration, the endowments of the mother can be described as:
\[
S_{f,t}= \gamma S_{m,t}+\xi_t,
\]
where $0<\gamma<1$ describes the assortativeness of mating and $\xi_t$ is an i.i.d. term. For high intergenerational correlations to arise, both $\lambda$ and $\gamma$ would have to be sufficiently close to one. Little direct evidence on the assortativeness of mating in terms of genetic endowments exists. A standard assumption in quantitative genetics has been to consider that assortment occurs in the phenotype (\citeNP{CrowFelsenstein1968}), which in our context corresponds to education. But because spousal correlations in years of schooling tend to "only" be between 0.4--0.6 (\citeNP{fegu01}), and because genes explain but a part of the variation in education (\citeNP{LeeWedowOkbayKongMaghzianOmeed2018}, \citeNP{young2019solving}), phenotypic assortment necessarily implies a limited extent of genetic assortativeness. Recent work finds that the genetic predictors of education are more correlated than could be explained by phenotypic assortment, due to secondary assortment on a genetically correlated trait that also increases the spousal correlation in genetic predictors of education (\citeNP{robinson2017genetic}). 

While this is an active area of research, we would argue that the assortativeness in genetic advantages is unlikely to be sufficiently high to explain the high multigenerational persistence in educational advantages documented by the recent economic literature. Strong assortativeness would also be required to account for the findings of studies comparing relatives in the horizontal dimension. \citeN{ColladoOrtunoStuhler2019aa} estimate that in Sweden, spousal correlations in latent advantages would have to be around 0.75 to explain the correlation pattern in educational attainment between distant siblings-in-law. This evidence suggests that assortative mating would need to be far stronger than that reflected in the conventional measures, such as the spousal correlation in years of schooling. 

Two alternative explanations for strong multigenerational persistence can be hypothesized. One is that the mechanisms described in \Cref{sec:model} reach across multiple generations. For example, persistence will be higher if grandparents have a direct impact on the education and socialization of their grandchildren. The reach of this mechanism is limited by the fact that there is usually substantial overlap between at most three generations at the same time. A second possibility is that transmission is generation-to-generation, but works in large part through mechanisms that are not genetic in nature but may be subject to high rates of assortative mating. For example, the variable $S_t$ could represent a concept termed ``family culture,'' which summarizes how a given family views its role in society and the degree of ambition of its members. In a class-based society, we can envision that a given family takes its overall place in the hierarchy as given, and will therefore preserve its relative standing over multiple generations. These mechanisms are reinforced if society imposes limits that hinder mobility across classes. Such restrictions were codified in many European societies throughout the aristocratic period (e.g., \citeNP{goni2020assortative}), and exist in much stronger and more persistent form in caste societies. To this regard, \citeN{dozi08b} provide an analysis of the intergenerational transmission of values and attitudes in a class based society, and show that persistent class differences can arise and are reinforced by the different economic circumstances of each class. Less clear, however, are the mechanisms that might explain high multigenerational persistence in contemporary populations in high-income countries, constituting an interesting question for future research.

\section{Educational Inequality in the Covid-19 Pandemic}

\label{sec:covid}

Public health measures in response to the Covid-19 pandemic have led to an unprecedented shock to investments in education. According to UNESCO, 94 percent of the world’s student population was impacted by educational institution closures in the spring of 2020. By May 2021, schools across countries had been fully closed for an average of 17 weeks \cite{UNESCO}. These closures varied widely in length, and were only partially determined by infection rates \cite{OECD2021}. Students in many developing nations, and in some US states, experienced closures that lasted more than a year, whereas there were no school closures at all in Belarus and Burundi. 

\begin{figure}[p]
\centering
\captionsetup[sub]{font=footnotesize}
\begin{subfigure}[]{0.995\linewidth}
    \centering
   \includegraphics[width=\linewidth,height=8cm,trim=2 2 2 2,clip]{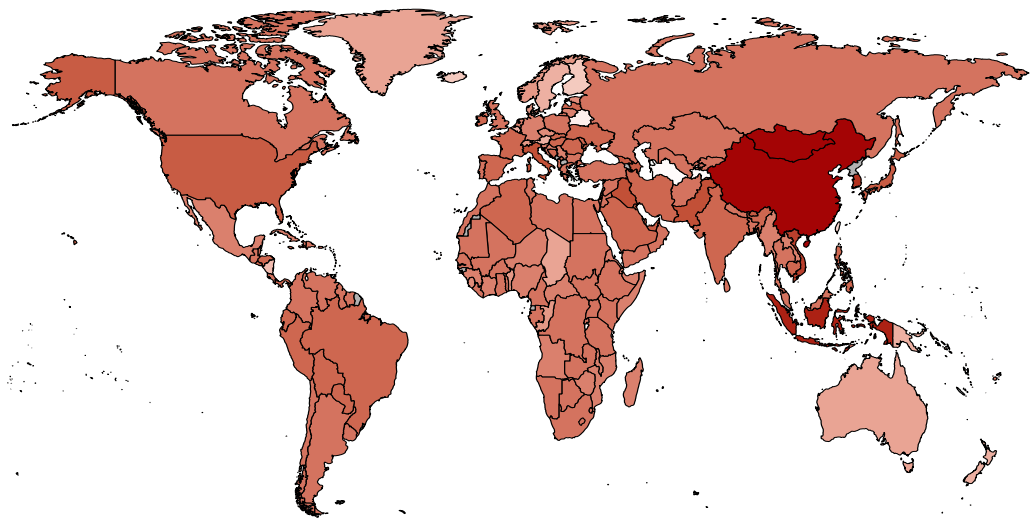}
   \caption{June 1, 2020}
   \label{fig:Map1} 
\end{subfigure}
\hfill
\begin{subfigure}[b]{0.995\linewidth}
    \centering
   \includegraphics[width=\linewidth,height=8cm,trim=2 2 2 2,clip]{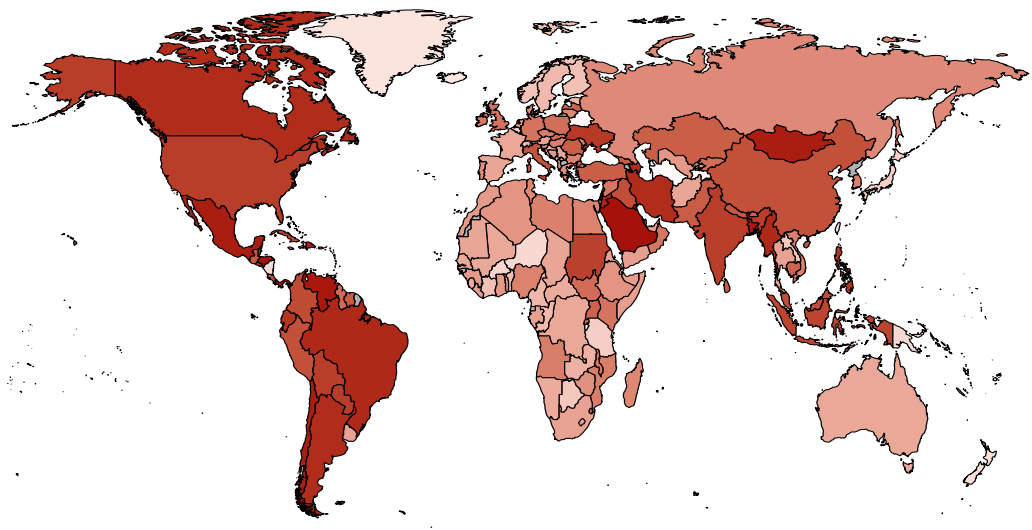}
   \caption{June 1, 2021}
   \label{fig:Map2} 
\end{subfigure}
\centering
\caption[World map]{Number of weeks of school closures at (a) June 1, 2020 and (b) June 1, 2021, based on \emph{Oxford Covid-19 Government Response Tracker, Blavatnik School of Government}. Weekly values assigned as 1 for full closure at all levels, 0.5 for partial closure, and 0 for minor alterations or no measures.}
\label{fig:Map} 
\end{figure}

Figure \ref{fig:Map} illustrates the worldwide distribution of school closures, based on data collected by the Oxford Covid-19 Government Response Tracker (\citeNP{OxCGR2021global}). We compare the accumulated number of weeks of school closure at an early stage of the pandemic (June 1, 2020) and one year later (June 1, 2021), assigning weekly values of 1 for full closure at all levels, 0.5 for partial closure, and 0 for minor alterations or no measures. By June 2020 (Panel (a)), nearly all countries had imposed some form of school closure to halt the progression of the pandemic. Closures were particularly long in China, Indonesia, and some countries in Southern Europe, but of similar intensity in most other countries. Over the following year, there was more variation in school closures across countries, being particularly extensive in the Americas, in some countries in the Middle East, and in Southern Asia (Panel (b)). Schools and universities adjusted to a lack of in-person teaching by shifting to remote learning, albeit with great variation in speed and efficiency.

As noted by \citeN{OxCGR2021global}, while the initial policy response was relatively uniform, later measures were driven by local epidemiological situations and differing political environments, leading to divergent policy responses. Students around the world accordingly experienced highly unequal exposure to school closures.

School closures have been especially controversial in the United States and Canada, where closures have continued much longer than in other developed economies. In the United States, school closures were not nationally mandated and were rather determined at the state or school district level, resulting in substantial variation in children’s experiences, as well as posing a challenge for the systematic collection of data. Evidence from mobile phone locations \cite{parolin2021large} indicates that 9 out of 10 schools were closed in April 2020, falling to 40 percent in September but then rising again to 56 percent in December. \citeN{Halloran_Oster} use school district-level data from 12 US states for the 2020--2021 academic year and show that shares of in-person schooling time varied from 9 percent in Virginia to 98 percent in Florida. At the time of writing in early 2022, the Covid-19 omicron variant is spreading, leading a number of countries and regions to once again close their schools \cite{washingtonpost2022}. 

Economists have expressed concern over the future economic cost of these closures (\citeNP{burgess2020schools}; \citeNP{Psach_covid}; \citeNP{Hanushek_covid}). As emphasized in the model described in Section~\ref{sec:model}, learning is a cumulative process where skills acquired at one life stage foster further learning later on. This dynamic nature of skill acquisition implies that learning losses, once incurred, are subsequently difficult to compensate. Many of the schoolchildren affected by the pandemic are therefore likely to enter adult life with fewer skills and lower educational attainment than they would have otherwise. This loss of human capital will be reflected in lower lifetime earnings at the individual level, and could result in a lower stock of human capital and lower national income at the aggregate level for decades to come. 

\captionsetup[figure]{font=footnotesize,labelfont=footnotesize}

We are particularly interested here in whether school closures increase educational inequality by differentially impacting children of different socio-economic backgrounds. There are two reasons why this might occur. First, the incidence of school closures themselves may vary by social background, for instance when public schools close while private schools attended by richer families stay open. Second, children from disadvantaged backgrounds might experience greater learning loss if their school closes. 

In the model of skill acquisition presented in Section~\ref{sec:model}, children's learning depends on inputs from educational institutions, parental inputs, and neighborhood and peer effects. All of these inputs are affected by school closures in ways that are likely to differ across families. During the closures, the inputs provided by schools and teachers were often delivered via online education. Yet, just how well virtual education can replace in-person schooling depends on factors such as having a reliable internet connection, functioning tablets or laptops, and a quiet work environment; all of which are more likely to be met in higher-income families. Parents also play an important role, and richer parents may not only be more capable of assisting their children in making up for lost time but also more likely to work from home, where they can help if need be.\footnote{\citeN{fiorini2014allocation} analyze time-use data in Australia and suggest that among a set of different child activities, time spent with parents on educational activities is most effective at increasing cognitive skills.} Social networks that exist outside of school are likely to matter for peer and neighborhood effects, and residential stratification by income might once again put lower-income families at a disadvantage.

School closures are not the only mechanism via which the pandemic may impact educational inequality. The macroeconomic effects of the pandemic could decrease parents’ income and educational investments, reduce public spending on schooling, or affect the returns and incentives to acquiring education. The full impact of all of these changes on educational inequality will gradually emerge over the next few years, but much can already be learned from pre-pandemic evidence on the consequences of school closures, from new evidence on the pandemic already collected, and from structural modeling of the long-run impact of the pandemic. 

\subsection{The Effect of School Closures: Pre-pandemic Evidence}\label{PrepandemicEvidence}

Given that direct evidence on the impact of pandemic school closures on skill acquisition and educational inequality takes time to accumulate, researchers have considered other sources of variation in school access or instruction time in an effort to make first assessments of the likely impact of the crisis. Table~\ref{table:3} summarizes the evidence and reports the implied effect of a 12-week school closure on standardized test scores. 

Initially, pandemic school closures were anticipated to last two to three months, about the time it took to contain the initial Covid-19 outbreak in China with social distancing measures. This time span allowed researchers to draw parallels with the so-called summer learning loss---or the fact that children’s knowledge and skills degrade over the summer school break. However, estimates of the magnitude of the loss vary widely between studies, from 0.06 to 0.6 of a standard deviation in skills for a 12-week closure (\citeNP{kuhfeld2020projecting}, \citeNP{cooper1996effects}, \citeNP{McCombs_et_al2014}). There is greater agreement over variation in learning loss across subjects. In particular, summer learning loss has been found to be less severe for reading skills, where children have more opportunity for independent practice, compared to mathematics or spelling (\citeNP{shinwell2017investigation}, \citeNP{paechter2015effects}). A limitation of applying the summer-learning-loss literature to the pandemic is that the impacts are measured over a short period of time (from just before to just after summer) and hence are not necessarily informative for long-term outcomes.

Other studies exploit variation in instructional time between students in different locations or cohorts, using research designs that can be informative about long-term effects. \citeN{carlsson2015effect}, for example, consider variations in the length of preparation time available for male Swedish students before taking a cognitive test. Even among 18-year-olds (for whom we might expect investment effects to be weaker) an extra ten days of school instruction increases scores on crystallized intelligence tests (synonyms and technical comprehension tests) by approximately 0.01 of a standard deviation. Assuming linearity, this implies effects of around 0.06 of a standard deviation for a 12-week closure. \citeN{fitzpatrick2011difference} estimate stronger impacts for the early years of schooling, implying a drop of around 0.3 of a standard deviation in math scores and 0.4 in reading scores as the result of a 12-week school closure. Meanwhile, \citeN{lavy2015differences} exploits differences in instruction time spent on different subjects across countries.  As noted by \citeN{burgess2020schools}, Lavy's results would predict a 0.06 of a standard deviation fall in math scores as a result of a 12-week school closure. \citeN{pischke2007impact} studies the effect of shortened school years that resulted from education reforms in Germany. While he does observe short-run impacts on test scores and grade repetition, long-run effects on earnings or employment are not found. Another approach is to consider variation in effective instruction time caused by differences in the age at school entry. For example, exploiting regional variation in school entry rules in England, \citeN{cornelissen2019early} see large effects of the effective length of the first school year on cognitive and non-cognitive outcomes at the end of the first school year, implying a fall of around 0.2 of a standard deviation in test scores as a result of a 12-week school closure. The impacts on cognitive outcomes largely disappear by age 11, while the effect on non-cognitive skills is more persistent. 

A further set of papers investigates the consequences of unanticipated shocks to instruction time. For example, \citeN{belot2010teacher} document that a seven-month teacher strike in Belgium, which ended with six weeks of uninterrupted school closures, led to five percent lower educational attainment, although the estimates are imprecise. For Argentina, \citeN{jaume2019long} find long-run effects of a strike on earnings and intergenerational effects on education. \citeN{marcotte2008unscheduled} observe measurably lower performance from just five days of lost school instruction due to snow days in Maryland, with the effect being larger for younger children. Developing this approach further, \citeN{hansen2011school} considers snow days in Maryland, Minnesota, and Colorado, and find effects between 0.05 and 0.15 of a standard deviation for a five-day absence for children in third, fifth, and eighth grade, though no consistent evidence that these effects are larger for younger children. \citeN{goodman2014flaking} finds no impact of snow days resulting in school closures in Massachusetts, but does document effects of individual absences. Similarly, \citeN{aucejo2016assessing} consider variations in instruction time that affect entire cohorts due to variations in the date of a test and absences that are idiosyncratic to specific children. They arrive at similar conclusions, suggesting that individual absences have more adverse effects than do shocks affecting an entire class. \citeN{goodman2014flaking} attributes this pattern to coordination problems; teachers find it more difficult to manage situations where children are learning at different rates. Further evidence of the detrimental effect of student absences is provided by \citeN{gershenson2017student} and \citeN{liu2021short}. \citeN{cattan2022} use Swedish historical and administrative data to show that in addition to reducing academic performance, absences have a negative long-run impact on labor income over the life cycle.

Some of the studies based on pre-pandemic evidence compare outcomes across children of varying socio-economic backgrounds. \citeN{lavy2015differences}, for instance, finds that disadvantaged children are more strongly affected by cross-country differences in instruction time. The effects of early education found by \citeN{cornelissen2019early} are particularly large for English boys from low socioeconomic backgrounds, though in the German data, \citeN{passaretta2021does} do not observe differences in the effect of the first year of schooling by socioeconomic background. Results in the summer learning literature are also mixed. \citeN{kuhfeld2020projecting} notes that while there is substantial variation in the extent of summer learning loss across students in the United States, this variation is not strongly related to family background. \citeN{paechter2015effects} document that children from more educated parents in Austria experience less summer learning loss in mathematics, but that there is little variation in effect size by family background in spelling or reading. A possible reason for these contrasting findings is that most children take a break from learning during the regular scheduled summer vacation irrespective of their family background. Interruptions of the regular school year may lead to different behavioral responses. In line with this interpretation, a study on teacher strikes in Canada by \citeN{johnson2011strikes} suggests that the children of less educated families were particularly affected. Similarly, \citeN{marcotte2008unscheduled} and \citeN{goodman2014flaking} both find that individual absences related to snow days have a more negative impact on achievement in schools with higher poverty rates, and \citeN{aucejo2016assessing} and \citeN{gershenson2017student} show that individual absences have significantly worse effects on the outcomes of low-income students. These studies support the notion that unexpected reductions in instruction time increase educational inequality between children of lower socio-economic backgrounds and those from well-off families.

\vspace{.2cm}

 \begin{scriptsize}    
    \begin{longtable}[ht!]
    					{l L{.11\textwidth}
                      L{.11\textwidth}
                      L{.12\textwidth}
                      L{.17\textwidth}
                      L{.17\textwidth}
                      L{.17\textwidth}} 
\caption{Evidence on the Effect of School Closures on Educational Outcomes\label{table:3}}\\
      \toprule      &
      \multicolumn{1}{C{.11\textwidth}}{\textbf{Paper}}    &
      \multicolumn{1}{C{.11\textwidth}}{\textbf{Variation}}    &
      \multicolumn{1}{C{.12\textwidth}}{\textbf{Country and Age Group}}    &
      \multicolumn{1}{C{.17\textwidth}}{\textbf{Implied effect 12-week closure on tests in SD}}    &
      \multicolumn{1}{C{.17\textwidth}}{\textbf{Other effects}}    &
      \multicolumn{1}{C{.17\textwidth}}{\textbf{Differences by family background}}    \\ \midrule
     
  &\RaggedRight \citeN{kuhfeld2020projecting} &   \RaggedRight   Summer learning loss   &\RaggedRight  US, grades 3 to 8 &\RaggedRight  Between 0.06 and 0.6 &\RaggedRight  &\RaggedRight No.  \\ \\
 
  &\RaggedRight \citeN{cooper1996effects}  &\RaggedRight   Summer learning loss     &\RaggedRight  US, grades 1 to 8 &\RaggedRight 0.13   &\RaggedRight Summer learning loss increases with students’ grade level. &\RaggedRight  Higher-income students gain on reading recognition while lower-income students lose.  \\\\
  
&\RaggedRight  \citeN{McCombs_et_al2014}  &\RaggedRight    Summer learning loss    &\RaggedRight US, grade 3 &\RaggedRight 0.22-0.26 on mathematics  &\RaggedRight Non-significant effect on reading or socio-emotional tests. &\RaggedRight No.   \\ \\

&\RaggedRight  \citeN{shinwell2017investigation} &\RaggedRight   Summer learning loss  &\RaggedRight  Scotland and North Eastern England, 5-10-year-olds &\RaggedRight  0.34 on spelling &\RaggedRight Non-significant effect on reading. &\RaggedRight The study sample is limited to children from low socioeconomic status (SES) areas. \\\\

&\RaggedRight  \citeN{paechter2015effects} &\RaggedRight   Summer learning loss  &\RaggedRight  Austria, Between 10 and 12 years old &\RaggedRight  0.73 on arithmetic problem solving, 0.38 on spelling &\RaggedRight Summer learning loss in arithmetic problem solving and spelling but gains in reading. &\RaggedRight  Greater learning loss among children from less-educated parents in mathematics, but not other subjects.  \\\\

&\RaggedRight  \citeN{carlsson2015effect} &\RaggedRight   Timing of tests  &\RaggedRight Sweden, 18-year-old males  &\RaggedRight  0.07 on synonym tests and 0.05 on technical comprehension test&\RaggedRight Test scores on fluid intelligence (spatial and logic) do not increase with additional school days. &\RaggedRight  No. Similar effects based on parental education and father’s earnings.  \\\\

&\RaggedRight \citeN{fitzpatrick2011difference} &\RaggedRight  Timing of tests  &\RaggedRight  US, kindergarten and grade 1 &\RaggedRight 0.3 in mathematics and 0.4 in reading  &\RaggedRight  &\RaggedRight No significant differences.   \\\\

&\RaggedRight  \citeN{lavy2015differences} &\RaggedRight   Instruction time across subjects    &\RaggedRight International (50 countries), 15-year-olds (PISA)  &\RaggedRight 0.06 on mathematics, science or language  &\RaggedRight Lower impact in developing countries (0.025 SD). &\RaggedRight Instruction time has greater effect for students from lower-educated backgrounds.   \\\\

&\RaggedRight  \citeN{pischke2007impact} &\RaggedRight    Short school years   &\RaggedRight Germany, grades 1 to 4   &\RaggedRight  20-50\% more grade repetitions, 10\% fewer students attend intermediate track &\RaggedRight No long-term effects on earnings and employment. &\RaggedRight Not measured.    \\\\

&\RaggedRight \citeN{cornelissen2019early} &\RaggedRight  School entry age  &\RaggedRight  England, reception year (age 4) &\RaggedRight 0.17-0.25 on cognitive (at age 5), 0.17-0.19 on non-cognitive (at age 5)  &\RaggedRight  Effect on cognitive outcomes largely disappears by age 11. &\RaggedRight Larger effects for boys from low socioeconomic background.  \\\\

&\RaggedRight  \citeN{belot2010teacher} &\RaggedRight Teachers' strikes  &\RaggedRight Belgium, high school students  &\RaggedRight Cohort affected by seven-month strike, reduced educational attainment by about 5 percent  &\RaggedRight Reallocation of students from university studies to higher vocational education.  &\RaggedRight   Not measured. \\ \\
         
&\RaggedRight  \citeN{jaume2019long} &\RaggedRight   Teachers' strikes  &\RaggedRight  Argentina, primary school students  &\RaggedRight Lowers high school completion, college completion, and years of schooling for males (females) by 3.2\% (2.5\%), 8.7\% (1.9\%), and 1.4\% (1.4\%)  &\RaggedRight Lower wages and increased likelihood of being unemployed. Males experience occupational downgrading while females increase home production. &\RaggedRight  Not measured. But there are intergenerational effects: children of individuals exposed to strikes suffer negative education effects as well.    \\\\

&\RaggedRight  \citeN{marcotte2008unscheduled} &\RaggedRight Weather (snow days) and timing of tests   &\RaggedRight  Maryland, grades 3, 5 and 8 &\RaggedRight  Grade 3 pass rate in reading (math) exam decreases by 0.49\% (0.53\%) with each day of unscheduled closure &\RaggedRight Smaller impacts for students in grades 5 and 8.   &\RaggedRight 50\% greater impact in schools with highest compared to lowest share of low-income students.   \\\\

 &\RaggedRight  \citeN{hansen2011school} &\RaggedRight  Weather (snow days) and timing of tests    &\RaggedRight Colorado, Maryland and Minnesota, grades 3, 5 and 8  &\RaggedRight Between 0.6 and 1.8   &\RaggedRight  &\RaggedRight  Not measured.  \\ \\
 
 & \RaggedRight  \citeN{goodman2014flaking} &\RaggedRight  Weather (snow days) &\RaggedRight Massachusetts, grades 3 to 8 and 10  &\RaggedRight  0.48 in mathematics and English &\RaggedRight Large negative effect for individual absences of student or student's peers. No effect of school closure. &\RaggedRight  Effects slightly more negative in schools with higher poverty rate.  \\\\

  &\RaggedRight  \citeN{aucejo2016assessing} &\RaggedRight   Individual absences and timing of tests    &\RaggedRight  North Carolina, grades 3 to 5 &\RaggedRight  0.10 in mathematics and 0.05 in reading  &\RaggedRight  Absenteeism is more detrimental in higher grades and for low performing students. &\RaggedRight   Missing school days has more negative effects on test scores for low income students. \\\\
 
   &\RaggedRight \citeN{cattan2022} &\RaggedRight Primary school absences &\RaggedRight Sweden, grades 1 to 4 &\RaggedRight  0.38 &\RaggedRight Long-term effect on lifetime earnings. Ten annual days of absence in elementary school decrease income by 1-2\%. &\RaggedRight No significant difference in impacts between children of agricultural workers and production or service workers. \\

&\RaggedRight  \citeN{gershenson2017student} &\RaggedRight   Individual absences    &\RaggedRight  North Carolina and US survey, kindergarten and grades 1, 3, 4, 5 and 8 &\RaggedRight 0.12–0.42 in mathematics and 0.12-0.24 in reading  &\RaggedRight Unexcused absences are 2-3 times more harmful than excused absences. &\RaggedRight  Absences decrease the reading achievement of low-income students more (by 25\%).  \\\\

  &\RaggedRight \citeN{liu2021short} &\RaggedRight  Individual absences   &\RaggedRight  California, grades 2 to 11 &\RaggedRight   0.18–0.24 in mathematics and ELA (English Language Arts) tests  &\RaggedRight Ten absences in 9th grade reduce both the probability of on-time graduation and ever enrolling in college by 2\%.  &\RaggedRight Not measured.   \\\\

  &\RaggedRight \citeN{passaretta2021does} &\RaggedRight Differences in testing period.  &\RaggedRight Germany, kindergarten cohort (age 4-5) &\RaggedRight  0.38 in mathematics, 0.30 in science and 0.12 in vocabulary &\RaggedRight &\RaggedRight No effect of 1st-grade schooling on socioeconomic gaps in any skill domain. Schooling is beneficial for all children equally. \\

  &\RaggedRight \citeN{johnson2011strikes} &\RaggedRight Teachers’ strikes and work-to-rule campaigns  &\RaggedRight Ontario, grades 3 to 6 &\RaggedRight  Reduce pass rates by 1.8-4.6\% on mathematics tests and 0.8-2.5\% on reading tests &\RaggedRight Negative effects are stronger on reading tests for grade 3 students compared to grade 6 students, but less so on writing and mathematics tests. &\RaggedRight Largest reductions in results are found at schools where children enter with other disadvantages, as measured by level of parental education. \\

\bottomrule
\end{longtable}
\end{scriptsize}

\subsection{Pandemic Evidence: Inequality in the Incidence of School Closures}

The pre-pandemic evidence suggests that Covid-19 could increase educational inequality via two channels: a greater incidence of school closures in low-income neighborhoods, and a greater learning loss conditional on closure among disadvantaged students. Indeed, early work on the pandemic supports both channels. \citeN{parolin2021large} show that school closures in the United States in the fall of 2020 were more common for students from ethnic minorities. School closures were also more widespread in institutions with lower third grade math scores, more homeless students, more students with limited English proficiency, and a larger share of students eligible for free or subsidised lunch. \citeN{Halloran_Oster} confirm this picture in a sample of 12 states, documenting that districts with a greater share of black students and a higher share of students receiving free lunches offered less in-person schooling.  

In the United Kingdom, school closures are determined at the national level, but local mitigation procedures led to varying incidence, as groups of children were required to isolate if a positive case was detected in their ``bubble.'' \citeN{eyles2021schools} show that in the fall of 2020, these localized measures led to nine days of missed schooling in the poorest areas of the United Kingdom compared to only two days in the most affluent municipalities. This evidence suggests that variation in the incidence of school closures could exacerbate educational inequalities. 

\subsection{Pandemic Evidence: Test Scores}

Some alarming direct evidence is emerging on the impact of school closures in the early phases of the pandemic (see Table~\ref{table:4} for a summary). \citeN{engzell2021learning} find that in the Netherlands, eight weeks of online rather than in-person learning led to 0.08 of a standard deviation lower test scores for students aged eight to eleven. Notably, this is precisely the effect size we would expect if test scores increase by 0.4 of a standard deviation for each academic year of in-person schooling (as found by \citeNP{bloom2008performance} and \citeNP{azevedo2021simulating}) and online learning leads to no learning gains at all. The impact is 40 percent larger among those in the least educated homes, suggesting that the pandemic not only increased educational inequality, but that disadvantaged children's skills actually deteriorated.  

\citeN{tomasik2021educational} analyze improvement in student skills in German-speaking Switzerland over the initial eight-week school closure starting in March 2020, compared to the eight weeks just prior.  On average, primary school pupils learned half as much under distance learning, and there was more inequality in their progression. In particular, those with higher ability going into the pandemic saw stronger effects. Students in secondary school learned at the same speed as before. Though these estimates may be affected by seasonal variation, at face value they suggest less severe impacts than indicated by \citeN{engzell2021learning}. 

\citeN{maldonado2020effect} provide evidence on 5th graders in Flemish Belgium who experienced seven weeks of school closure, partially replaced with online teaching. This period led to reduced test scores, equivalent to 0.19 of a standard deviation in math and 0.29 of a standard deviation in Dutch compared to earlier cohorts. Students performed worse than if they had simply retained their initial knowledge, suggesting a slide in skills. The authors observe weak effects on inequality, with no differences across schools by initial average test scores or by the school’s social mix for math outcomes, and only slightly larger effects for poorer schools in Dutch. However, measuring social background at the school level is less accurate than at the family level. In addition, this data is not based on a student-level panel and may be affected by attrition. 

\citeN{Kuhfeld2020eval} and \citeN{Kuhfeld2021} use the results of aptitude tests for students in grades 3--8 in US public schools to compare the outcomes of students affected by the pandemic with those of the previous cohort.\footnote{The identifying assumption here is that any difference between cohorts is generated by the pandemic, which is not necessarily the case. An alternative research design is to make use of exogenous variation in access to school during the pandemic. \citeN{BlandenSDQ} use differences in eligibility for an early return to school by grade in England and find that being out of school has substantial negative mental health effects. Such effects are likely to both partly explain and compound the test score effects we focus on here.} Differences between cohorts are greater in the spring of 2021 than in the fall of 2020, larger for math than reading, and stronger among younger children and those in high poverty schools. It is difficult to ascertain how much of these effects are a consequence of skill losses due to the pandemic and how much is due to additional sources of variation between cohorts. Using the same between-cohort design, other studies examine outcomes in the fall of 2020 in New South Wales, Australia \cite{gore2021impact} and in Baden-W\"{u}rttemberg, Germany \cite{schult2021did}. \citeN{gore2021impact} observe that third-grade children in the most disadvantaged schools are two months behind their pre-pandemic achievement in math, with no significant effects in reading or math for fourth-grade students. \citeN{schult2021did} document that fifth-grade students are slightly behind across the board, with the largest impact among low-achieving students in math.  

The results of \citeN{engzell2021learning} and \citeN{maldonado2020effect} suggest that school closures during the pandemic had greater effects on test scores than one might have expected based on extrapolations from prior evidence. Children's learning may have been affected not just by the school closures themselves but also by other effects of the crisis, such as the disruption of peer interactions or increased anxiety during the pandemic. Improved virtual instruction might have helped reduce learning losses as the pandemic wore on, but uneven engagement has the potential to worsen inequalities even further. \citeN{Kuhfeld2021} point to ``pandemic fatigue''  as an explanation for why they observe more learning loss in the spring of 2021 compared to six months previously, and cite evidence that students were more likely to report not liking school in the winter of 2021 compared to the start of the academic year. 

Using panel data for several cohorts of US students from the third to eighth grade, \citeN{Halloran_Oster} show that proficiency rates in English and math were on average 14 percentage points lower during the pandemic. By associating variation in time spent in different learning modes over the 2020/2021 academic year (remote, hybrid, in-person) with  district-level information on test scores, the authors conclude that this gap would have only been four percentage points if schools had remained open throughout the period, though the effects are likely to be downward biased due to missing data.  The authors see larger effects in districts with more students of color and a greater number of students eligible for free lunch. These findings are based on a phase during the pandemic when hybrid and online study was well established. Differential impacts could therefore be attributable to uneven engagement with online schooling, an issue we turn to next.

\subsection{Pandemic Evidence: Inequalities in Home Learning}

Underlying the impact of school closures on overall learning and educational inequality are several distinct mechanisms. First, the availability and quality of virtual learning offered by schools is clearly important. Second, there may be differences in parents' ability to support virtual learning and to compensate for the lost investments from schools. Third, the work put in by the students themselves matters as well. These mechanisms also interact in that the efforts of parents and children may respond to the inputs provided by schools, as discussed in \Cref{sec:compensating_investments}. A growing body of work has begun to uncover the importance of each of these factors during the pandemic.  

\citeN{clark2021compensating} consider evidence from three schools in China and show that children who have access to online learning through their school do 0.22 of a standard deviation better on tests that follow the end of a seven-week period of school closures. While effects by family background are not reported, the authors find that effective online learning is especially beneficial for low achievers. This observation suggests that inequalities in access to online learning may in part drive inequalities in the impact of school closure.
 
 \citeN{bacher2021inequality} use information from internet searches for digital learning resources in the United States before and during the pandemic. Searches doubled in April 2020 compared to the pre-pandemic period, and were 20--40 percent higher in high-income compared to low-income areas. This implies that uneven take up of online learning may be a contributor to the unequal impact of the pandemic.\footnote{The direction of these effects may not, however, be universal. \citeN{Amer-Mestre2021} document that in the early stages of the pandemic in Italy, searches for online learning resources increased more in regions with lower academic performance.} 
 
 \citeN{andrew2020inequalities} survey parents in the first period of English school closures and find that primary school students in the tenth percentile of the family income distribution did about 35 minutes less learning per day than those from median-income families, and 1 hour and 10 minutes less that a child from a family in the 90th percentile of the income distribution. That richer children spent more time on learning in England’s first lockdown is confirmed in \citeN{Delbonoreport}, with students in the top quartile of household earnings spending an additional 20 minutes on homework.  Similarly, \citeN{grewenig2021covid} and \citeN{werner2021legacy} find that during school closures, low-achieving students in Germany disproportionately replace learning time with less productive activities, such as playing video games. \citeN{chetty2020did} reports evidence on student effort from an online math program used by a representative sample of schools. They observe that while students from the richest quartile of neighborhoods quickly recovered to their prior level of progress, students from the bottom half were completing at least 40 percent fewer lessons in April 2020 than before schools closed.
 
 Gaps in measureable parent support are narrower than gaps in student effort. \citeN{Delbonoreport} observe no differences in the time spent by parents in supporting children's learning by parental education level. \citeN{bansak2021covid} instead find that in the United States, parents with a college education spent 2.2 hours a week more time on home learning than parents who are high school drop outs. Evidence from the United States and the United Kingdom suggests that inputs from both parents and children rise with schooling inputs (\citeNP{bansak2021covid}, \citeNP{Delbonoreport}) while in Germany, daily online instruction substantially increases student learning time \cite{werner2021legacy}. Effective distance learning provision for more vulnerable groups has a multiplicative effect on reducing the unequal impact of school closures.\footnote{ In addition to differences in time spent, parents may vary in their capacity to help due to their own skills or confidence.  \citeN{bol2020inequality} documents a large gap in the extent to which parents feel capable of supporting their child, with lower-educated mothers and fathers feeling less sure of themselves, even when their children are still in primary school.} 

 \begin{scriptsize}    
    \begin{longtable}[ht!]
    					{l L{.12\textwidth}
                      L{.13\textwidth}
                      L{.13\textwidth}
                      L{.21\textwidth}
                      L{.30\textwidth}} 
\caption{Covid-19 Pandemic Evidence on Educational Outcomes\label{table:4}}\\
      \toprule      &
      \multicolumn{1}{C{.12\textwidth}}{\textbf{Paper}}    &
      \multicolumn{1}{C{.13\textwidth}}{\textbf{Variation}}    &
      \multicolumn{1}{C{.13\textwidth}}{\textbf{Country and Age group}}    &
      \multicolumn{1}{C{.21\textwidth}}{\textbf{Implied effect 12-week closure on tests in SD}} &
      \multicolumn{1}{C{.30\textwidth}}{\textbf{Differences by family background}}    \\ \midrule

  &\RaggedRight \citeN{engzell2021learning} &   \RaggedRight  Pandemic, compared with previous year &\RaggedRight   Netherlands, grades 4 to 7 &\RaggedRight  0.12 in maths, spelling, and reading &\RaggedRight Effects 60\% larger among those with low-educated parents. \\

  &\RaggedRight \citeN{tomasik2021educational} &   \RaggedRight Pandemic, comparing learning to previous 8 weeks &\RaggedRight Switzerland, primary and secondary students  &\RaggedRight Primary school pupils learned more than twice as fast during in-person as during distance learning; no significant difference for secondary school students &\RaggedRight Greater heterogeneity in learning progress during distance learning among primary school pupils, but not among secondary school pupils. \\

  &\RaggedRight \citeN{maldonado2020effect} &\RaggedRight  Pandemic, compared with previous year &\RaggedRight  Belgium, grade 6  &\RaggedRight   0.25 in maths and 0.39 in Dutch  &\RaggedRight Learning loss increases in most indicators for socio-economic status.  \\ 

&\RaggedRight \citeN{Kuhfeld2020eval}  &\RaggedRight Fall 2020 compared with same grades in Fall 2018 &\RaggedRight US grades 3-8 &\RaggedRight   No overall effect on reading, 5-10 percentile decline in math performance. Larger in grades 3-6 than grades 6-8  &\RaggedRight    \\

&\RaggedRight \citeN{Kuhfeld2021} &\RaggedRight Spring 2021 compared with same grades in spring 2019 &\RaggedRight US grades 3-8 &\RaggedRight   4-6 percentile decline in reading, 7-11 percentile decline in math. Larger in grades 3-5 &\RaggedRight Effects greater for Blacks, Latinos and Native American and Alaskans, and greater for younger children in high poverty schools. \\

  &\RaggedRight \citeN{gore2021impact} &   \RaggedRight Pandemic, compared with previous year &\RaggedRight  Australia (New South Wales), grades 3 and 4 &\RaggedRight No significant differences between 2019 and 2020 in mathematics or reading tests &\RaggedRight Lower achievement growth in mathematics for grade 3 students in least advantaged schools.\\
 
  &\RaggedRight \citeN{schult2021did} &   \RaggedRight Pandemic, compared with previous year &\RaggedRight Germany (Baden-Württemberg), grade 5 &\RaggedRight 0.11 for reading comprehension, 0.14 for operations, 0.05 for numbers comprehension &\RaggedRight Math competencies of low-achieving students and those with lower socio-cultural capital particularly affected.\\
 
 &\RaggedRight \citeN{Halloran_Oster} &   \RaggedRight Pandemic, compared with previous year &\RaggedRight  US, grades 3 to 8 &\RaggedRight Overall decline in students’ 2021 test scores in maths (14.2 percentage points) and English language arts (ELA) (6.3 percentage points) &\RaggedRight Districts with a larger share of Black and Hispanic students or students receiving free lunch experience a greater decline in ELA. \\
 
  &\RaggedRight \citeN{clark2021compensating} &\RaggedRight  Online education &\RaggedRight China (Guangxi Province), grade 9 &\RaggedRight 0.34 increase for students that had access to online learning compared to those that did not. Higher impact (0.45) for students that had online lessons from external high-quality teachers  &\RaggedRight  Low achievers benefit the most from teacher quality. Students with a computer benefited more than those who used a smartphone. \\

&\RaggedRight  \citeN{bacher2021inequality}  &\RaggedRight  Online education (internet searches)  &\RaggedRight  US, K-12 students   &\RaggedRight     &\RaggedRight  Searches for school-centered (parent-centered) online resources 39\% (24\%) higher in high-income areas than in low-income areas.   \\

  &\RaggedRight \citeN{Amer-Mestre2021} &   \RaggedRight Online education &\RaggedRight Italy, primary and secondary students &\RaggedRight  &\RaggedRight Searches for e-learning tools increased more in regions with previously low academic performance than higher-performing regions. \\

&\RaggedRight \citeN{andrew2020inequalities} &\RaggedRight  Pandemic, children time use &\RaggedRight  UK, reception year to grade 10  &\RaggedRight     &\RaggedRight  Primary school students in bottom decile of family income spent 70 minutes less learning per day than those in top decile. Poorer children attended schools with less active home-learning support and had less resources such as computers or dedicated study space. \\

&\RaggedRight   \citeN{Delbonoreport}  &\RaggedRight  Pandemic, children time use   &\RaggedRight  UK, primary and secondary students   &\RaggedRight     &\RaggedRight  Children’s time spent on schoolwork was lower in disadvantaged families, but parental time spent on home schooling did not differ by indicators of socio-economic background.   \\

&\RaggedRight  \citeN{grewenig2021covid} &\RaggedRight  Pandemic, children time use  &\RaggedRight  Germany, primary and high school students &\RaggedRight  &\RaggedRight Daily learning reduction was significantly larger for low-achievers (4.1 hours) than high-achievers (3.7 hours), but not larger for children of low-educated parents.  \\

&\RaggedRight  \citeN{werner2021legacy} &\RaggedRight Pandemic, children and parent time use &\RaggedRight Germany, primary and high school students&\RaggedRight &\RaggedRight Children’s learning time decreased severely during the first school closures, particularly for low-achieving students.\\

&\RaggedRight  \citeN{chetty2020did} &\RaggedRight Online education   &\RaggedRight  US, K-12 students   &\RaggedRight     &\RaggedRight  Children in high-income areas temporarily learned less but recovered to baseline levels, while children in low-income areas remained 50\% below baseline through the end of the school year.   \\

&\RaggedRight  \citeN{bansak2021covid} &\RaggedRight Pandemic, children and parent time use   &\RaggedRight   US, K-12 students &\RaggedRight  &\RaggedRight Parental time helping children positively associated with parental education. College-educated parents spent 2.2h more per week compared to those without a high school degree. In addition, less-educated households were much more likely to experience computer or internet access problems. \\

\bottomrule
\end{longtable}
\end{scriptsize}

\subsection{Structural Estimates of the Long-term Impact of the Pandemic}

Given that the pandemic is ongoing, the empirical literature has so far been able to quantify only a subset of potential channels that may affect educational inequality, and findings on long-run impacts will take time to materialize. A number of papers leverage structural modeling that is disciplined by both current and pre-pandemic data to assess the combined impact of different channels as well as the potential long-run repercussions of the crisis.
  
\citeN{ADSZ2021} use a model along the lines of Section~\ref{sec:model} to assess the potential impact of the pandemic on educational inequality among US high school students. Educational achievements depend on school inputs, parental effort, parenting style, and on peer effects. As in \citeN{AS18}, the formation of peer groups is endogenous and subject to possible interventions of parents. School closures during the Covid-19 pandemic affect learning through three channels. First, there is a decline in the overall efficiency of skill accumulation because remote learning is less effective than in-person instruction. Second, parental time inputs become more important during remote learning, as parents have to replace some inputs that are usually provided by teachers. Parents' ability to provide these inputs depends on time constraints: parents who are able to work from home during the pandemic have an easier time helping their children with school work than do essential workers who must work outside the home. Third, peer effects and peer-group formation is also disrupted during the pandemic, as during closures children lose contact with some existing friends and new peer interactions are restricted to the local neighborhood rather than schools. 

\citeN{ADSZ2021} assess the contribution of these channels to educational inequality. The model estimation relies on a combination of empirical findings specific to the pandemic (such as lost learning time during school closures and variation in parents' time budgets depending on whether they can work from home) and pre-pandemic evidence on issues for which contemporary evidence is not yet available (such as peer effects). All three channels are found to contribute to a widening of educational inequality. While all parents increase their time investments during the pandemic, the ability of low-income parents to respond is hampered by the fact they are much less likely to have jobs that can be done from home (\citeANP{adbogora20} \citeyearNP{adbogora20}, \citeyearNP{adbogora20b}). Hence, inequality in parental input increases between high- and low-income neighborhoods. Inequality in peer effects also rises, in part because children from low-income neighborhoods lose the ability to meet more high-ability peers at school, and in part because the effect of losing any peer connection on learning is worse for children already struggling in school. The impact on educational inequality is large: students from poor neighborhoods suffer a learning loss close to half a standard deviation, whereas the skill accumulation of children from rich neighborhoods is barely affected. 

\citeN{Fuchs_et_al_2020} focus on the long-run implications of school closures for educational attainment and children's future earnings. Similar to \citeN{ADSZ2021}, they build their analysis on a human capital production function with time and monetary investments by parents and public investments provided by schools. The model is calibrated to the US economy. They find large effects: among children aged 4-14 during the pandemic, a school closure lasting six months increases the share without a high school degree by seven percent and reduces the share of children with a college degree by 3.2 percent. These changes diminish the average lifetime earnings of the affected children by about one percent. The effects are largest for younger children, an implication that follows from two features of the skill accumulation technology emphasized in Section~\ref{sec:model}, namely self-productivity (investments today increase human capital tomorrow) and dynamic complementarity (investments today increase the marginal productivity of further investments tomorrow). 

Turning to educational inequality, \citeN{Fuchs_et_al_2020} consider the role of both the financial resources and the education of parents. Poorer children are predicted to suffer more from school closures for two reasons. First, a larger share of their educational investments come from schools rather than from parents. Second, richer parents increase their investments by more in response to school closures. Lower college attendance rates are one key mechanism for the overall loss in welfare. This implies that the welfare loss is non-monotonic in parental education, being strongest for children with high school educated parents (who have sizeable college completion rates) rather than for those whose parents did not complete high school (whose attendance rates were low even before the pandemic). Intuitively, students who are unlikely to attend college anyways have less to lose from school closures than those whose participation decision is more marginal. The welfare loss is, however, monotonic in the financial resources of parents, as financially constrained parents find it hard to increase investments in response to the reduced governmental investment associated with school closures.

\citeN{fuchs2021fiscal} extend their earlier work to account for the empirical distribution of school closures as observed in the early stages of the pandemic. In the United States, secondary and public schools were closed for longer periods than elementary and private schools, respectively. Extending their earlier life cycle model to include the choice of parents between public and private school options, the authors predict that the earnings- and welfare losses will be largest for children who started public secondary schools at the onset of the crisis. Welfare losses are smaller for children from richer families, who are more likely to send their children to private school. The authors further suggest that a policy intervention to extend schooling (by shortening the summer breaks in future years) would raise tax contributions sufficiently to be self-financing. 

\citeN{JangYum2020} also study the impact of school closures in a dynastic overlapping-generations calibrated to the US economy. Unlike \citeN{Fuchs_et_al_2020}, they account for general-equilibrium effects. The authors find that school closures that last for one year reduce the lifetime income of the affected cohorts by about one percent, compared to a one percent reduction for a half-year closure as estimated by \citeN{Fuchs_et_al_2020}. General equilibrium effects have a substantial impact on the aggregate impact: the decline in the supply of human capital by the affected cohorts raises the return to education, which provides additional incentives for parents to make up for at least some of the learning losses through higher investments. In the setting of \citeN{JangYum2020}, school closures have only a small impact on cross-sectional inequality. Nevertheless, there is a sizeable effect on intergenerational mobility. On average, a one-year school closure would lower the probability of children born into the bottom income quintile to move up to the top quintile by two to three percent and increase the rank correlation in income by 0.4 to 0.9 percent. In contrast to the findings of \citeN{Fuchs_et_al_2020}, these effects turn out to be larger for older children, who have less time to compensate for learning losses through greater time investments.  

\citeN{Alon_et_al_2022} use a structural model to assess the macroeconomic impact of pandemic school closures in low-income countries. One reason for different outcomes compared to high-income countries is the extent of the learning loss itself. Survey evidence shows that considerably fewer children continued learning activities during school closures in low-income countries, with particularly large reductions in sub-Saharan Africa. Limited education funding and less access to communications technology implies that few children had access to virtual lessons during school closures. Many children essentially received no education at all during prolonged school closures, so that the total learning loss is likely to be severe. Moreover, beyond the size of the learning loss, a given learning loss is likely to have a greater long-run economic impact in low-income countries. This is partly due to demographic reasons. Low-income countries have much younger populations than do high-income countries, which means that cohorts of children finishing school are large compared to the adult labor force. Hence, a given reduction in human capital for children finishing school has a strong impact on overall human capital in the economy. The effect is additionally amplified as, in the lowest-income countries, older cohorts usually received little schooling. This further increases the share of total human capital accounted for by recent graduates, and hence heightens the aggregate economic impact of learning loss of children who will enter the labor market in the coming years. 
 
The findings in this section can be considered in light of the educational inequality in PISA test scores shown in Figure 1. The gap between children in the top and bottom quintile of family background is roughly one standard deviation.  Results from \citeN{engzell2021learning} for an eight-week school closure imply that this gap would further increase by around 0.05 of a standard deviation.\footnote{Engzell et al.'s comparison is based on the 4 percent most educationally disadvantaged homes, for whom effects are 60 percent larger than for the overall population.} This could easily rise to 0.10 as schools in many countries were closed for twice as long, with potentially even larger effects in places such as the United States with particularly long closure periods.
  
The effect sizes found by \citeN{engzell2021learning} show the impact of the pandemic in the short term. Incorporating time constraints and peer effects, \citeN{ADSZ2021} indicate that skill inequality could increase by 0.5 of a standard deviation, or to 1.5 times its current high level. Dramatic policy action would be needed to close such a gap. While the mechanisms explored in the structural studies vary, there is a general consensus that pandemic school closures are likely to increase educational inequality, with long-term consequences for the educational attainment, lifetime income, and social mobility of the affected cohorts.  Notably, additional unexplored mechanisms could lead to even more substantial inequalities. For example, it is assumed that all children are affected by a school closure of the same duration, whereas the evidence discussed above shows that (at least in the United States and the United Kingdom) children from poorer areas missed more days of school compared to those in affluent neighborhoods (\citeNP{parolin2021large},  \citeNP{eyles2021schools}). Other factors that may further increase educational inequality include the role of family structure (e.g., single parents, who are often poorer and perhaps less able to provide support) and access to learning technology (e.g., the availability of a reliable internet connection, a functioning laptop or tablet, and a quiet place to work).

Commentators on both sides of the Atlantic have called for policy action to help offset the damaging effects of school closures (\citeNP{Language}; \citeNP{Burgess2020}; \citeNP{sibieta2021}). These primarily focus on what schools can do once children return. A wealth of research evaluates school-level interventions, much of it reviewed in \Cref{sec:compensating_investments}. Limiting the additional educational inequality caused by the pandemic requires policies able to target the most affected groups and successfully aid struggling students. Proposed interventions include increased school funding, providing small group instruction, and lengthening the school day or year. All of these have potential, with targeted small group instruction shown to be especially fruitful. Evidence suggests that additional days spent at school raise test scores for poorer students, but the likelihood of diminishing returns means the optimal length of the post-pandemic school year is unclear. 

Overall, it has now become quite clear that the Covid-19 pandemic has had a major negative impact on many children's learning and is likely to have substantially increased educational inequality within the affected cohorts. Tracing the effect of this shock over the following years and contributing to the design of effective policy responses represents an important challenge for future research. At the same time, the crisis provides an opportunity to learn more generally about the sources and consequences of educational inequality. The pandemic introduced large, previously unanticipated changes in various inputs in children's skill acquisition in a way that varied substantially across countries, schools, and families. We expect that the growing body of work on the consequences of this shock will advance our understanding of educational inequality both during the pandemic and beyond.

\section{Outlook and Conclusions}

\label{sec:conc}

The literature in recent years has made tremendous strides in measuring educational inequality, understanding its sources, working out its long-run implications, and examining policy options. We have documented that educational inequality between children from different family backgrounds is pervasive and manifests both in test scores and educational attainment. At the aggregate level, we observe a ``Great Gatsby Curve,'' whereby more unequal countries also experience lower social mobility. To show that educational inequality contributes to this relationship, we provide evidence for an ``Educational Great Gatsby Curve,'' meaning that in more unequal countries the intergenerational correlation between parents and children in educational attainment is higher. We also review research documenting that educational inequality is more persistent across generations than captured by such intergenerational correlations, and discuss the potential mechanisms underlying this finding.

We show how structural models of skill acquisition and education decisions describe the role of and interactions between different factors contributing to educational inequality, including parental investments in children's education, public inputs, and peer and neighborhood effects. One advantage of a structural approach is that it allows for counterfactual policy analysis, including assessments of the impact of interventions ranging from expanded early childhood education to student loans for higher education on educational inequality. 

Despite substantial progress in research on educational inequality, many challenges remain. Understanding educational inequality is a complex undertaking. As we have outlined, there are several channels that potentially contribute to inequality, and a variety of interactions between them. Distinguishing the contribution of different channels to overall inequality is further complicated by the fact that the scope for experimental and quasi-experimental evidence on this issue is limited. Some drivers of educational inequality are difficult or impossible to control via randomized evaluations, and even where randomized experiments are possible, results can be difficult to interpret. For example, the introduction of a new preschool program may have a direct effect on the enrolled children, but might also lead to changes in parents' investments. How parents react, in turn, may differ depending on the economic, institutional, and cultural setting in which the intervention takes place. 

Given these limitations, greater understanding of educational inequality will likely emerge from combining alternative approaches, including structural modeling and a variety of sources of empirical evidence. We conclude by outlining several particularly important outstanding issues that future research might address. 

{\bf Productivity of different inputs:} While there is plenty of evidence that investments in children's skills are productive, there is still much to be learned about what kind of interactions with children have the highest returns. Little experimental evidence exists on this issue, especially relative to parental investments. Observational data often does not permit fine-grained distinctions between different kinds of parental inputs, and observational associations (e.g., children who routinely eat dinner with their parents do well) are likely not causal. In recent years, a number of experimental intervention studies have been carried out that allow for different treatments (such as nutrition assistance for children versus information sessions for parents), emphasizing the benefit of talking with young children. Much more detailed research in this direction is needed to shed greater light on the effects of different parental inputs.

{\bf Sources of multigenerational persistence:} As outlined above, the recent literature has established that rates of social mobility over multiple generations are low compared to what a simple extrapolation of single-generation correlations would suggest. One interpretation of this observation is that the potential impact of educational inequality on social mobility is even larger than apparent at first sight, making policies that push back against educational inequality even more desirable. Yet, whether this is the correct conclusion crucially depends on the sources of low multigenerational mobility. One possibility is that there is a direct influence across multiple generations, for example through childcare or funding for other investments provided by grandparents. In this case, policies that provide similar investments for children who receive less support from their grandparents would be expected to increase long-run mobility. Conversely, if low multi-generation mobility is linked to genetic transmission within families, policy interventions may be less effective. A third possibility is that the persistence of status across generations is related to the transmission of values, attitudes, or preferences within families, which would have yet other implications for how persistence responds to policies and changes in the economic environment. Sorting out these possible mechanisms should be a high priority for research on multigenerational persistence.

{\bf Reconciling opposing trends in educational inequality:} We have discussed mechanisms that suggest that the rise in overall economic inequality observed in a number of high-income economies in recent decades should result in higher educational inequality. There is, indeed, clear evidence that parental inputs of money and time have become more unequal across the income scale in different countries (\citeNP{ramey10}; \citeNP{KornrichFurstenberg2012}; \citeNP{corak2013income}; \citeNP{SchneiderHastingsBriola2018}; \citeNP{dozi19}). Nevertheless, measures of educational inequality based on attainment or test scores often suggest little change in overall educational inequality. There are three main possibilities for reconciling the opposing trends in the inequality of educational inputs and outcomes. First, it may take time for the change in inequality in inputs to be fully reflected in outcomes, in which case we should expect educational inequality to rise in the near future. Second, the change in inequality in inputs does push towards higher educational inequality, but may have been offset by a more equal provision of other inputs (such as public schooling inputs). Third, the parental inputs that have been increasing quickly among richer parents are relatively unproductive or run into strongly diminishing returns.\footnote{This channel would be consistent with the observation that marginal returns for investing in children's skills in early childhood are higher for children with lower initial skills \cite{AW16}.}  Which of these channels is dominant has important implications for future trends in overall inequality, social mobility, and the desirability of potential policy responses.

Our survey has focused on evidence on educational inequality in high-income economies. Yet, the majority of the world's children live in low- and middle-income countries where, given generally lower living standards, the consequences of inequality can be even more marked.\footnote{See \citeN{attanasio2021} for a recent survey of research on early childhood development with a focus on developing countries.} While most of the issues discussed here apply to these settings, additional factors to consider include the role of child labor, nutrition, and the varying quality of formal education. Much work remains to be done to examine the causes and consequences of educational inequality on a global scale.

\small

\bibliography{handbook}

\appendix

\section{Proofs for Propositions}

\begin{prfof}{\bf Proposition~\ref{prop:college_no_insurance}:} Let $c_{3,1}$ denote consumption in period 3 if graduating from college, and $c_{3,0}$ if not graduating:
\begin{align*}
c_{3,0}&= (1+r) a_3+ w_3(s_3,0), \\
c_{3,1}&= (1+r) a_3+ w_3(s_3,1).
\end{align*}
Conditional on attending college, the optimality condition for savings $a_3$ is given by:
\begin{equation}\label{eq:sav_opt}
u'(c_2)=\beta (1+r)\left[p_d(1,s_2,i_2)u'(c_{3,1})+
(1-p_d(1,s_2,i_2))u'(c_{3,0})
\right].
\end{equation}
The optimality condition for effort in college $i_2$ is:
\[
w_2u'(c_2)=\beta \frac{\partial p_d(1,s_2,i_2)}{\partial i_2} 
\left[u(c_{3,1})-u(c_{3,0})\right],
\]
which can also be written as:
\begin{equation}\label{eq:opt_i}
w_2u'(c_2)=\beta \frac{\partial p_d(1,s_2,i_2)}{\partial i_2} 
\int_{c_{3,0}}^{c_{3,1}} u'(c)\;\text{d}c.
\end{equation}
In the limit where the probability of succeeding in college approaches zero, the savings condition~\eqref{eq:sav_opt} is:
\[
u'(c_2)=\beta (1+r) u'(c_{3,0}).
\]
Using this to replace $u'(c_2)$ in \eqref{eq:opt_i} gives:
\[
w_2\beta (1+r) u'(c_{3,0})=\beta \frac{\partial p_d(1,s_2,i_2)}{\partial i_2} 
\int_{c_{3,0}}^{c_{3,1}} u'(c)\;\text{d}c
\]
or:
\[
w_2 (1+r) =\frac{\partial p_d(1,s_2,i_2)}{\partial i_2} 
\int_{c_{3,0}}^{c_{3,1}} \frac{u'(c)}{u'(c_{3,0})}\;\text{d}c.
\]
The only term that varies by assets $a_3$ is the integral. Making the dependence of consumption $c_{3,0}$ and $c_{3,1}$ on assets explicit, this integral can be written as:
\[
\int_{0}^{c_{3,1}-c_{3,0}} \frac{u'((1+r) a_3+ w_3(s_3,0)+x)}{u'((1+r) a_3+ w_3(s_3,0))}\;\text{d}x.
\]
For a given $x$, the derivative of the integrand with respect to assets $a_3$ is given by:
\[
(1+r)\frac{u''(c_{3,0}+x)u'(c_{3,0})-u''(c_{3,0})u'(c_{3,0}+x)}{\left(u'(c_{3,0})\right)^2}.
\]
This term is positive if we have:
\[
-\frac{u''(c_{3,0}+x)}{u'(c_{3,0}+x)}< -\frac{u''(c_{3,0})}{u'(c_{3,0})},
\]
which is satisfied because we assume that the utility function exhibits decreasing absolute risk aversion. The integral is therefore increasing in assets $a_3$. Moreover, $a_3$ is increasing in initial wealth $a_2$, implying that effort $i_2$ is also increasing in wealth. Intuitively, the risky investment in college education has a higher return for otherwise identical students with more assets because marginal utility diminishes more slowly and hence the return to college is worth relatively more to them. 

The expected return to attending college is given by:
\[
p_d(1,s_2,i_2)(w_3(s_2,1)-w_3(s_2,0)).
\]
Hence, conditional on skills $s_2$, the return is increasing in effort $i_2$, so that if richer students put in more effort they also get a higher return.

Lastly, for given skills and given effort, attending college is a risky investment with an expected return that does not depend on wealth. Given this same investment opportunity, decreasing absolute risk aversion implies that attending college is more attractive to students with higher wealth. Moreover, given that poorer students put in less effort conditional on attending, they also experience lower returns to attending college, which lowers the likelihood of attending even more.
\end{prfof}

\begin{prfof}{\bf Proposition~\ref{prop:college_borrowing_constraints}:} With a binding borrowing constraint we have $a_3=0$ and the budget constraints \eqref{eq:dyn_bc1} and \eqref{eq:dyn_bc2} read:
\begin{align*}
c_2&=a_2+n_2 w_2(s_2)-T e_2 ,\\
c_3&=w_3(s_3,d_3).
\end{align*}
Conditional on enrolling in college ($e_2=1$) the first-order condition governing the choice of effort in college $i_2$ is given by:
\[
u'(a_2+ (1-i_2)\,w_2(s_2)-T)=
\beta \frac{\partial p_d(1,s_2,i_2)}{\partial i_2} 
\left[u(w_3(s_3,1))-u(w_3(s_3,0))\right].
\]
Here the marginal benefit of putting effort on the right-hand side does not depend on assets $a_2$, but the marginal cost of effort on the left-hand side is decreasing in assets, so that students with more resources will put in higher effort $i_2$, and conversely students with fewer resources will spend more time on working during college ($n_2$ is decreasing in $a_2$). Higher effort for wealthier students also implies that the expected return from attending college (conditional on skill $s_2$) is increasing in wealth. Lastly, a student will decide to attend college ($e_2=1$) if:
\[
u(a_2+\,w_2(s_2))-u(a_2+ (1-i_2)\,w_2(s_2)-T)\le 
\beta p_d(1,s_2,i_2) \left[u(w_3(s_3,1))-u(w_3(s_3,0))\right]
\]
Here the right-hand side is increasing in assets $a_2$ (because effort $i_2$ increases in assets) and the left-hand side is decreasing in assets (because of diminishing marginal utility), implying that students with more assets $a_2$ are more likely to attend (put differently, the threshold for skill $s_2$ above which a student attends college is decreasing in assets $a_2$).
\end{prfof}

\label{app:proofs}

\end{document}